\documentclass[aps,prd,nofootinbib,twocolumn,superscriptaddress,preprintnumbers,letterpaper]{revtex4}

\usepackage{amssymb}
\usepackage{amsmath}
\usepackage{epsfig}
\usepackage[colorlinks=true
,urlcolor=blue
,anchorcolor=blue
,citecolor=blue
,filecolor=blue
,linkcolor=red
,menucolor=blue
,linktocpage=true
,pdfproducer=medialab
,pdfa=true
]{hyperref}
\usepackage{comment}
\usepackage{color}
\usepackage{cleveref}
\usepackage{multirow}
\usepackage{capt-of}
\usepackage{siunitx}
\usepackage{graphicx}
\usepackage{gensymb}
\usepackage[caption=false]{subfig}
\usepackage{slashed}
\usepackage{tabularx}
\usepackage{blindtext}
\usepackage[utf8]{inputenc}

\preprint{MIT-CTP/4996}

\begin{document}

\title{
Too Hot, Too Cold or Just Right? Implications of a 21-cm Signal for Dark Matter Annihilation and Decay
}

\author{Hongwan Liu}
\email{hongwan@mit.edu}
\affiliation{Center for Theoretical Physics, Massachusetts Institute of Technology, Cambridge, MA 02139, U.S.A.}

\author{Tracy R. Slatyer}
\email{tslatyer@mit.edu}
\affiliation{Center for Theoretical Physics, Massachusetts Institute of Technology, Cambridge, MA 02139, U.S.A.}

\begin{abstract} 
Measurements of the temperature of the baryons at the end of the cosmic dark ages can potentially set very precise constraints on energy injection from exotic sources, such as annihilation or decay of the dark matter. However, additional effects that lower the gas temperature can substantially weaken the expected constraints on exotic energy injection, whereas additional radiation backgrounds can conceal the effect of an increased gas temperature in measurements of the 21-cm hyperfine transition of neutral hydrogen. Motivated in part by recent claims of a detection of 21-cm absorption from $z\sim17$ by the EDGES experiment, we derive the constraints on dark matter annihilation and decay that can be placed in the presence of extra radiation backgrounds or effects that modify the gas temperature, such as DM-baryon scattering and early baryon-photon decoupling. We find that if the EDGES observation is confirmed, then constraints on light DM decaying or annihilating to electrons will in most scenarios be stronger than existing state-of-the-art limits from the cosmic microwave background, potentially by several orders of magnitude. More generally, our results allow mapping any future measurement of the global 21-cm signal into constraints on dark matter annihilation and decay, within the broad range of scenarios we consider.
\end{abstract}


\maketitle


\section{Introduction}
\label{sec:Introduction}

Between thermal decoupling of baryons from the CMB and star formation at $z \sim 20$, the universe can function as a sensitive calorimeter for exotic sources of energy injection. In the conventional $\Lambda$CDM model, the matter temperature $T_m$ and hydrogen ionization fraction as a function of redshift are simple and well-understood \cite{AliHaimoud:2010dx,Chluba:2010ca}. After recombination, Compton scattering between the residual free electrons and cosmic microwave background (CMB) photons keeps $T_m$ at the CMB temperature $T_\text{CMB}$ down to $z \sim 150$. Subsequently, the energy transfer rate becomes too small to prevent thermal decoupling, and soon after, $T_m$ is determined solely by adiabatic expansion, evolving as $T_m(z) \propto (1+z)^2$. Deviations in temperature from this well-understood standard history are thus a strong indication of new sources of heating or cooling in the universe. 

The recent measurement of an absorption profile at 78 MHz in the sky-averaged spectrum by the Experiment to Detect the Global Epoch of reionization Signature (EDGES) Collaboration \cite{Bowman:2018yin} opens a new window into the cosmic dark ages, shedding new light on the ionization and thermal history at precisely this period of interest. Radiation with a frequency near the hyperfine transition of hydrogen illuminates neutral hydrogen clouds during this epoch, and gets absorbed as they redshift into the transition frequency. The brightness temperature of the 21-cm hydrogen absorption line relative to the background radiation is given by \cite{Zaldarriaga:2003du}
\begin{multline}
    T_{21}(z) \approx x_\text{HI}(z) \left(\frac{0.15}{\Omega_m}\right)^{1/2} \left(\frac{\Omega_b h}{0.02}\right) \\
    \times  \left(\frac{1+z}{10}\right)^{1/2} \left[1 - \frac{T_R(z)}{T_S(z)}\right] \SI{23}{mK},
    \label{eqn:T_21}
\end{multline}
where $x_\text{HI}$ is the neutral hydrogen fraction, $\Omega_m$ and $\Omega_b$ are the matter and baryon energy density as a fraction of the critical density, $h$ is the Hubble parameter today in units of \SI{100}{km\, s^{-1} Mpc^{-1}}, and $T_R(z)$ is the effective temperature of the background 21-cm radiation at redshift $z$. $T_S(z)$, the spin temperature, determines the ratio of neutral hydrogen in the higher-energy spin-triplet state to the lower-energy spin-singlet state. 

The expected value of $T_S$ as a function of redshift has been studied extensively (see e.g. \cite{Furlanetto:2006jb} for a review). At $z \sim 30$, we expect $T_S = T_R$, with the radiation temperature commonly assumed to be $T_\text{CMB}$. Once the first stars start forming at $z \sim 20$ and begin to emit UV radiation, downward transitions from the spin-triplet to the spin-singlet state through the Wouthuysen-Field effect \cite{1952AJ.....57R..31W,1959ApJ...129..536F,1959ApJ...129..551F} start to occur, driving the spin temperature toward $T_m$. The combination of the background 21-cm radiation, UV radiation from stars and collisional hyperfine excitation/de-excitation ensures that well before reionization,
\begin{alignat}{1}
    T_m \lesssim T_S \lesssim T_R.
    \label{eqn:spin_temperature_bound}
\end{alignat}
A measurement of a negative $T_{21}(z)$ at this time indicates that $T_S$ lies below $T_R$, and also sets an upper bound on $T_m$ if $T_R$ is known.

The EDGES collaboration measured a strong 21-cm absorption trough in the redshift range $14 < z < 20$, reporting a value of $T_{21}$ at $z \sim 17.2$ of  $T_{21} = -500^{+200}_{-500}\text{ mK}$ \cite{Barkana:2018lgd}, with 99\% confidence limits specified. This result, together with Eq.~(\ref{eqn:spin_temperature_bound}), sets the following constraint on the matter and radiation temperature at $z = 17.2$ at the 99\% confidence level:
\begin{alignat}{1}
    \frac{T_m}{T_R}(z = 17.2) \lesssim 0.105.
    \label{eqn:T_m_T_R_ratio}
\end{alignat}
Precise calculations of the temperature evolution after recombination assuming the $\Lambda$CDM model \cite{AliHaimoud:2010dx,Chluba:2010ca} give $T_m(z = 17.2) \sim \SI{7}{K}$; however, assuming $T_R = T_\text{CMB}$ in Eq.~(\ref{eqn:T_m_T_R_ratio}), we obtain $T_m \lesssim \SI{5.2}{K}$, which lies well below the expected value.

Since the publication of the EDGES result, this discrepancy has been explained by either a colder-than-expected gas temperature or an additional source of 21-cm photons at $z \sim 20$. In both cases, the effect is to reduce the expected value of the ratio $T_m/T_R$. Models with interactions between baryons and cold dark matter (DM) with a Rutherford-like cross section have been explored~\cite{Barkana:2018lgd} as a mechanism to cool the gas, particularly in the context of millicharged DM models~\cite{Munoz:2018pzp,Berlin:2018sjs,Fraser:2018acy,Barkana:2018qrx}. These models have been shown to be highly constrained, with millicharged DM likely to only make up a subdominant component of DM. Modifications to the redshift of thermal decoupling of baryons from the CMB can also result in a cooler-than-expected gas temperature. Such a scenario can occur due to an imbalance between the proton and electron number densities~\cite{Falkowski:2018qdj} or early dark energy~\cite{Hill:2018lfx} (although the latter scenario appears difficult to reconcile with other observations). The possibility that interacting dark energy or other effects could modify the evolution of the Hubble parameter and change the 21-cm brightness temperature was proposed in~\cite{Costa:2018aoy}, but the change to the Hubble parameter required at $z \lesssim 20$ is large. Finally, models which inject additional 21-cm radiation through light DM decays~\cite{Fraser:2018acy,Pospelov:2018kdh} or radio emission from black holes~\cite{Gong:2018sos,Ewall-Wice:2018bzf} have been studied as a means of raising $T_R$.

In any model of DM, the annihilation and decay rates into Standard Model (SM) particles are important quantities to understand. Models with DM-baryon scattering are likely to imply the existence of DM annihilation to SM particles by crossing symmetry, and these annihilation processes could potentially set the relic abundance of DM via thermal freezeout at early times. Even if DM-baryon scattering does not occur or is not strong enough to markedly affect the matter temperature, new constraints on annihilation and decay can be set using the information on the thermal history provided by 21-cm measurements of this epoch.

Previous studies \cite{Poulin:2016anj,Furlanetto:2006wp,Lopez-Honorez:2016sur,Valdes:2007cu,Evoli:2014pva} have explored such constraints under the assumption that there are no other modifications to the conventional thermal history. However, any attempt to explain the EDGES result mandates the presence of additional effects, and such modifications could also be present even if the EDGES result is not confirmed.

In this paper, we will study the implications that a confirmed 21-cm absorption measurement from $z \sim 20$ would have for DM annihilation and decay, in conjunction with three general mechanisms that could deepen an absorption signal: (i) non-standard recombination histories; (ii) baryon-DM scattering; and (iii) an additional source of 21-cm photons at $z \sim 20$. We will use the EDGES result as a benchmark; if it is confirmed, the forecast limits in this work can be applied as constraints on the DM parameter space.

Throughout this paper, all algebraic expressions will be written in natural units with $\hbar = c = k_B = 1$, and we adopt cosmological parameters that are equal to the Planck 2015 TT,TE,EE+lowP central values \cite{Ade:2015xua}.

\section{Ionization and Thermal History}
\label{sec:IGMHistory}

We first review the standard thermal and ionization history, and the modifications necessary to include DM energy injection.

\begin{figure*}[t]
    \subfloat[]{
        \label{fig:example_temp_histories}
        \includegraphics[scale=0.38]{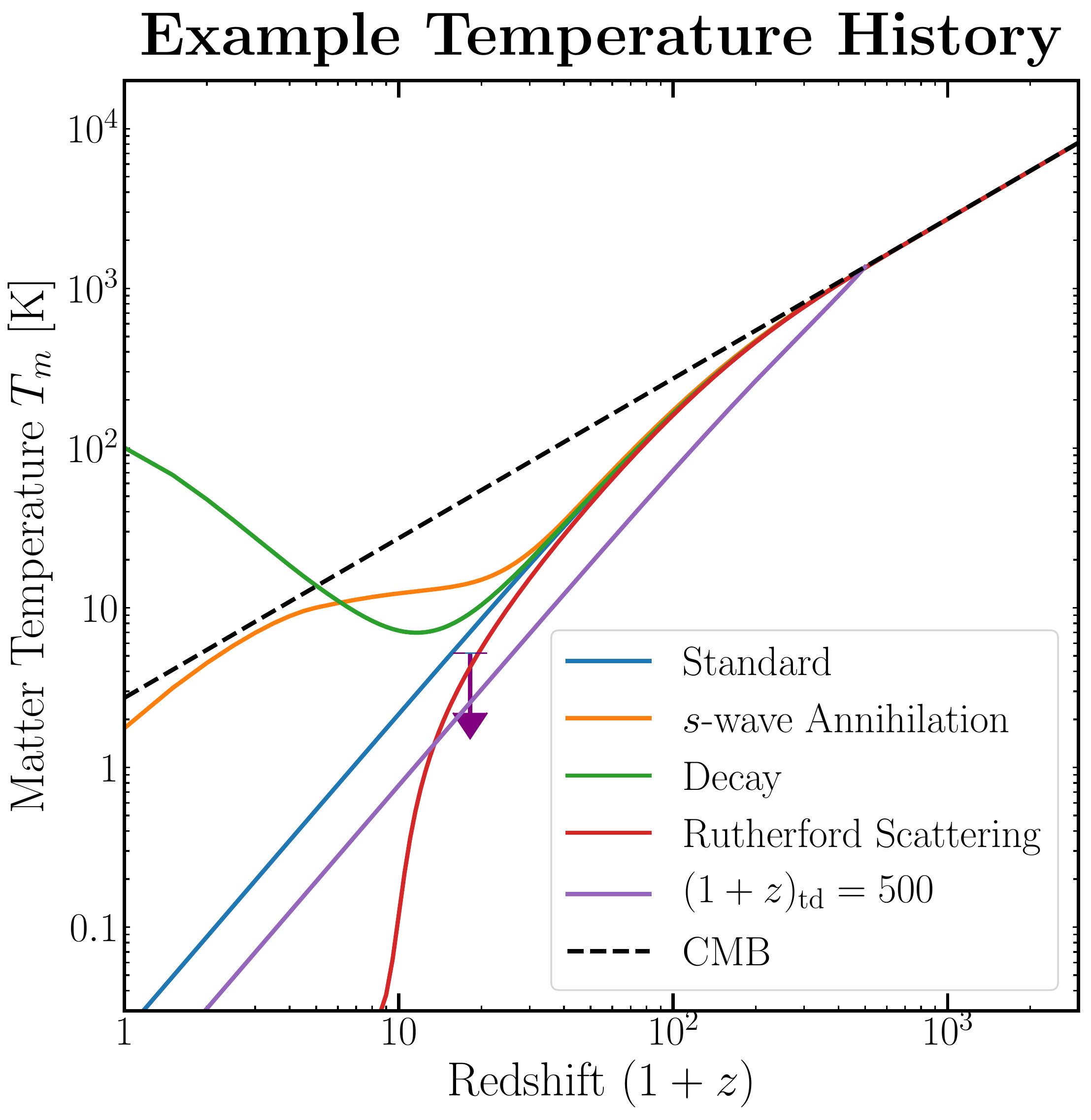}
    }
    \hfil
    \subfloat[]{
        \label{fig:example_ion_histories}
        \includegraphics[scale=0.38]{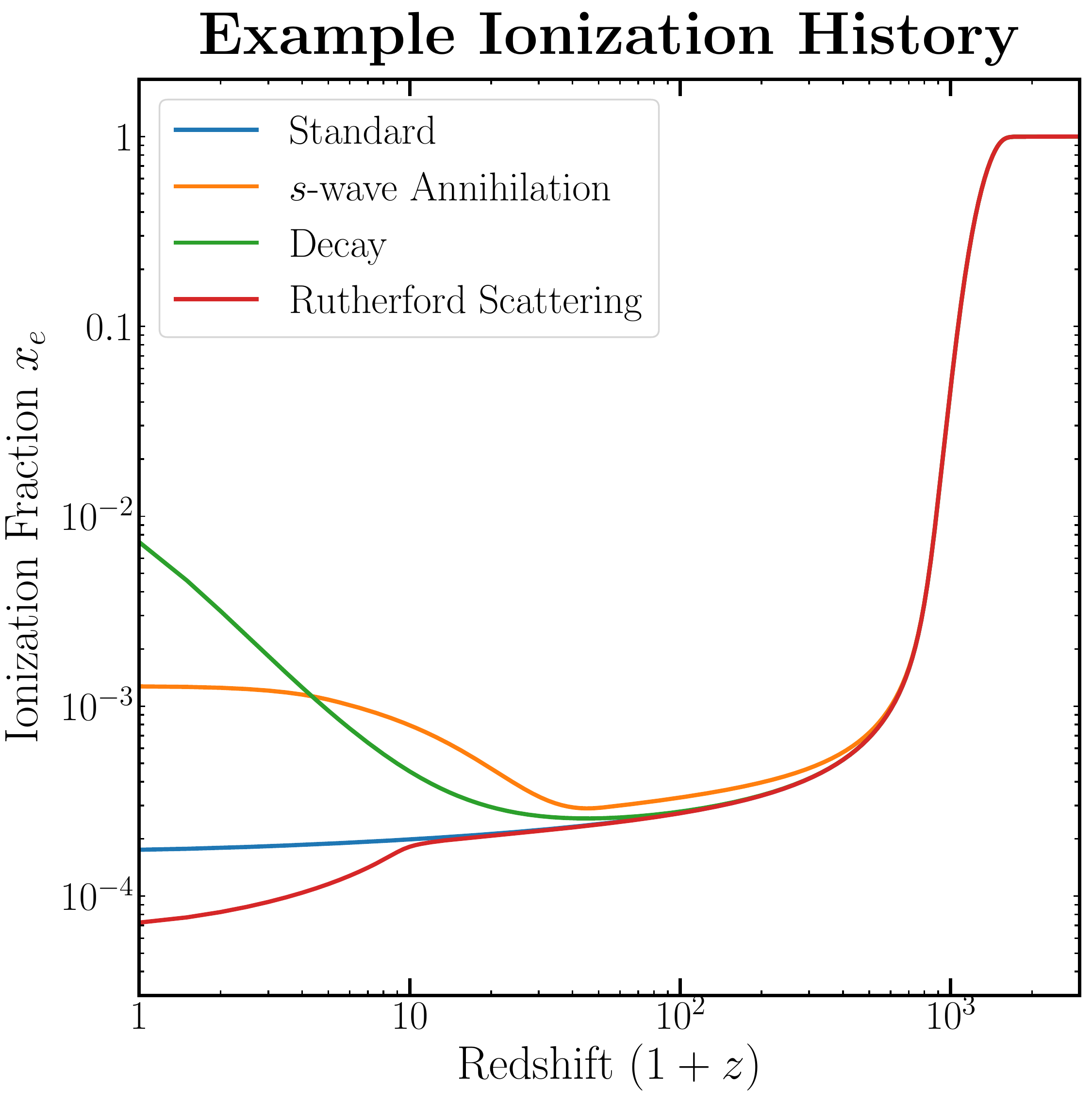}
    }
  \caption{Example thermal (left) and ionization (right) histories. The standard history (blue), with DM $s$-wave annihilation (orange), with DM decays (green) and with DM-baryon Rutherford scattering (red) are shown. The thermal history with an earlier redshift of thermal decoupling at $(1+z)_\text{td} = 500$ (light purple) is also shown, with the CMB temperature (black, dashed) plotted for reference. The purple arrow indicates the EDGES temperature limit. 
  }
  \label{fig:example_histories}
\end{figure*}

The conventional understanding of the thermal and ionization history of the universe during and after recombination is well-approximated by the three-level atom (TLA) model first derived in \cite{Peebles:1968ja,Zeldovich:1969en} (a review of the model is given in e.g. \cite{AliHaimoud:2010dx}). In this model, the evolution of the matter temperature $T_m$ and the ionization fraction are:
\begin{alignat}{1}
    \dot{T}_m^{(0)}  &= -2HT_m + \Gamma_C \left(T_\text{CMB} - T_m \right), \label{eqn:Tm} \\
    \dot{x}_e^{(0)} &= -\mathcal{C}\left[n_\text{H} x_e^2 \alpha_\text{B} - 4(1-x_e) \beta_\text{B} e^{-E_{21}/T_\text{CMB}}\right],
    \label{eqn:xe}
\end{alignat}
where $H$ is the Hubble parameter, $n_\text{H}$ is the total number density of hydrogen (both neutral and ionized), $x_e \equiv n_e/n_\text{H}$ where $n_e$ is the number density of free electrons, and $E_{21} = \SI{10.2}{eV}$ is the Lyman-$\alpha$ transition energy. $\alpha_\text{B}$ and $\beta_\text{B}$ are case-B recombination and photoionization coefficients respectively, and $\mathcal{C}$ is the Peebles-C factor \cite{Peebles:1968ja,AliHaimoud:2010dx} that represents the probability of a hydrogen atom in the $n=2$ state decaying to the ground state before photoionization can occur. $\Gamma_C$ is the Compton scattering rate, given by
\begin{alignat}{1}
    \Gamma_C = \frac{x_e}{1 + f_\text{He} + x_e} \frac{8 \sigma_T a_r T_\text{CMB}^4}{3 m_e},
    \label{eqn:compton_rate}
\end{alignat}
where $\sigma_T$ is the Thomson cross section, $a_r$ is the radiation constant, $m_e$ is the electron mass and $f_\text{He} \equiv n_\text{He}/n_\text{H}$ is the relative abundance of helium nuclei by number.

DM annihilation or decay into SM particles injects energy into the universe, leading to additional ionization, excitation and heating of the gas. Given a velocity-averaged $s$-wave annihilation cross section $\langle \sigma v \rangle$ or a decay lifetime $\tau$ (assumed to be much longer than the age of the universe), the rate of energy injection for these processes is given by
\begin{alignat}{1}
    \left(\frac{dE}{dV \, dt}\right)^\text{inj} = \begin{cases}
        f_{\chi,\text{ann}}^2 \rho_{\chi,0}^2 (1+z)^6 \frac{\langle \sigma v \rangle}{m_\chi}, & \text{annihilation}, \\
        f_{\chi,\text{dec}} \rho_{\chi,0}(1+z)^3 \frac{1}{\tau}, & \text{decay},
    \end{cases}
    \label{eqn:energy_injection}
\end{alignat}
where $\rho_{\chi,0}$ is the mass density of DM today, and $m_\chi$ is the DM mass. $f_{\chi,\text{ann}}$ and $f_{\chi,\text{dec}}$ are the fraction by mass of the DM that annihilates or decays respectively; our constraints are presented with these quantities set to 1 unless explicitly stated otherwise, and a straightforward rescaling of $\langle \sigma v \rangle$ or $\tau$ will give the limits for other values of these fractions. An additional factor of $1/2$ is required if the annihilation occurs between distinct particles (e.g. Dirac fermions), but we will assume throughout this paper that the two particles in an annihilation are indistinguishable unless otherwise stated: constraints for distinguishable particles can be obtained by a simple rescaling of $\langle \sigma v \rangle$ by a factor of 2.

The effect of the injected energy on the thermal and ionization history may not be instantaneous, and can in fact be significantly delayed: these effects are captured by a deposition efficiency $f_c(z)$ into some energy deposition channel $c$ (ionization, excitation or heating), which is defined as
\begin{alignat}{1}
    \left(\frac{dE}{dV \, dt}\right)^\text{dep}_c = f_c(z) \left(\frac{dE}{dV\, dt}\right)^\text{inj}. \label{eqn:f_c_definition}
\end{alignat}
Values for $f_c(z)$ into the relevant channels have been previously computed in \cite{Slatyer:2015kla}, and were extended to include structure formation boosts to the annihilation rate in \cite{Liu:2016cnk}. We note that $f_c(z)$ is computed assuming the standard ionization history, whereas new sources of energy injection or cooling will no doubt change the ionization history. However, the atomic processes that determine these deposition efficiencies remain relatively unaffected by small changes in the ionization fraction. This is a valid assumption in this study, where we will ultimately constrain DM energy injection rates small enough that $x_e \lesssim 0.1$ at $z \sim 20$. 

The resulting modifications to the temperature and ionization history for DM energy injection are
\begin{alignat}{1}
    \dot{T}_m^\chi &= \frac{2 f_\text{heat}(z)}{3(1 + f_\text{He} + x_e)n_\text{H}} \left(\frac{dE}{dV \, dt}\right)^\text{inj}, \nonumber \\ 
    \dot{x}_e^\chi &= \left[\frac{f_\text{ion}(z)}{\mathcal{R} n_\text{H}} + \frac{(1-\mathcal{C})f_\text{exc}(z)}{0.75 \mathcal{R} n_\text{H}} \right] \left(\frac{dE}{dV \, dt}\right)^\text{inj},
    \label{eqn:TLA_DM}
\end{alignat}
where $\mathcal{R} = \SI{13.6}{eV}$ is the ionization potential of hydrogen. The contribution from excitation to ionization is given by the number of excitation events (each event deposits an energy of $E_{21} = 0.75 \mathcal{R}$), multiplied by the probability of an excited hydrogen atom getting ionized, given by $1-C$.

Fig.~\ref{fig:example_histories} shows a number of representative ionization and thermal histories, including the standard history obtained by integrating Eq.~(\ref{eqn:Tm}) and~(\ref{eqn:xe}), as well as two examples with DM $s$-wave annihilation and decay, which includes Eq.~(\ref{eqn:TLA_DM}). The EDGES upper limit on the matter temperature if we take $T_R = T_\text{CMB}$ in Eq.~(\ref{eqn:T_m_T_R_ratio}) is also indicated. We have also included two of the new interactions that we will examine later: Rutherford-like interactions between the dark sector and hydrogen, as well as a temperature history with early decoupling of the photon and baryon temperatures. Throughout the paper, no star-formation or reionization models are included in this analysis: excluding these effects, which would only raise the matter temperature near $z \sim 20$, leads to annihilation cross section or decay lifetime limits that are less constraining and thus conservative. The impact of $s$-wave annihilation becomes significantly enhanced beginning at $z \sim 40$ due to structure formation, which greatly increases the local DM density. We discuss the systematics associated with structure formation in Appendix~\ref{app:systematics}. 

The authors of~\cite{Venumadhav:2018uwn} have recently pointed out that Lyman-$\alpha$ radiation at $z\sim 20$ is able to mediate a transfer of energy from the 21-cm CMB photons to the thermal motion of the gas, providing an additional and significant source of heating during this epoch. Although the inclusion of this effect would ultimately be important in setting precise DM annihilation and decay constraints, we neglect this effect in this paper, and leave a proper treatment of this process to future work. This is consistent with our omission of the process of reionization, and leads to limits that are less constraining than they would be in a more complete treatment.

\section{Additional 21-cm Sources}
\label{sec:additional_sources}

A large absorption trough can be explained by the existence of an additional 21-cm source, which would raise the effective radiation temperature at $z \sim 20$ above the CMB temperature. If $T_R$ is large enough so that Eq.~(\ref{eqn:T_m_T_R_ratio}) is satisfied with $T_m \gtrsim \SI{7}{K}$, no additional sources of cooling are required to explain the EDGES result. For $T_m = \SI{7}{K}$, we require $T_R = \SI{67}{K}$, compared to the CMB temperature at this redshift, $T_\text{CMB} = \SI{50}{K}$. 

\begin{figure*}[t]
    \subfloat[]{
        \label{fig:source_elec_decay}
        \includegraphics[scale=0.34]{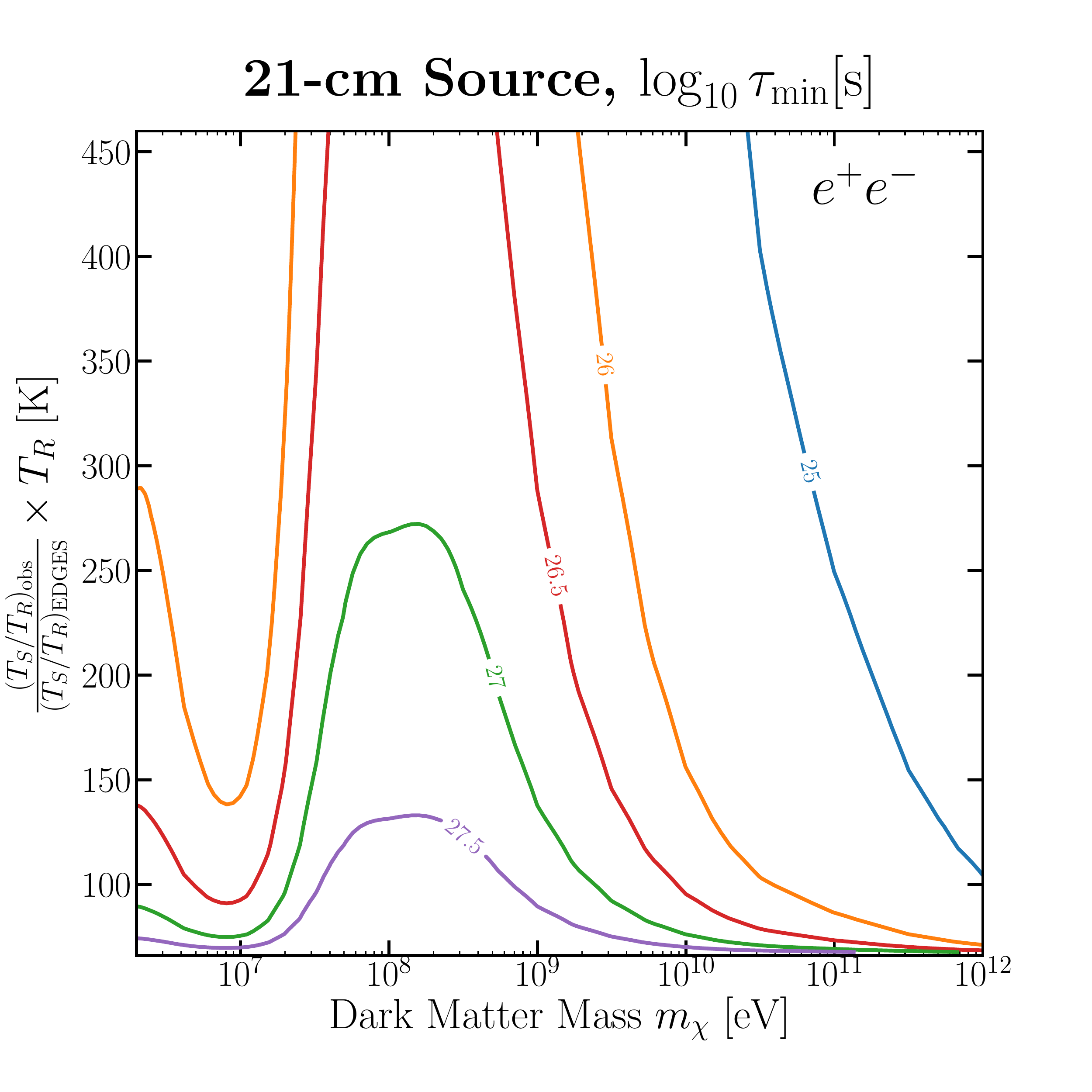}
    }
    \hfil
    \subfloat[]{
        \label{fig:source_phot_decay}
        \includegraphics[scale=0.34]{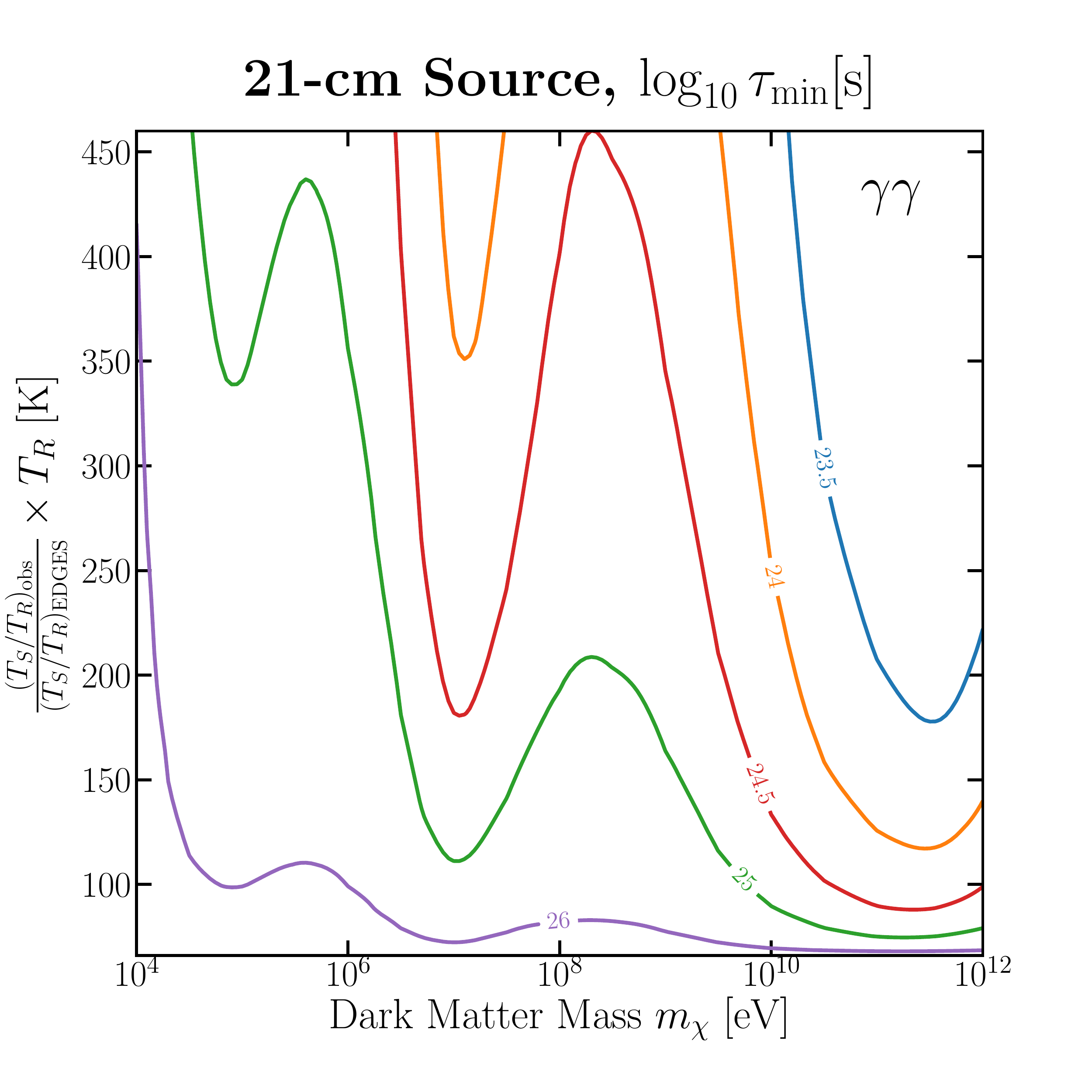}
    }
  \caption{Decay lifetime constraints with an additional 21-cm source with $\chi \to e^+e^-$ (left) and $\chi \to \gamma \gamma$ (right), as a function of $m_\chi$ and $(T_S/T_R)_\text{obs}/(T_S/T_R)_\text{EDGES} \times T_R$. Contour lines of constant minimum $\log_{10}\tau$ (in seconds) are shown. 
  }
  \label{fig:source_decay}
\end{figure*}

\begin{figure*}[t]
    \subfloat[]{
        \label{fig:source_elec_swave}
        \includegraphics[scale=0.34]{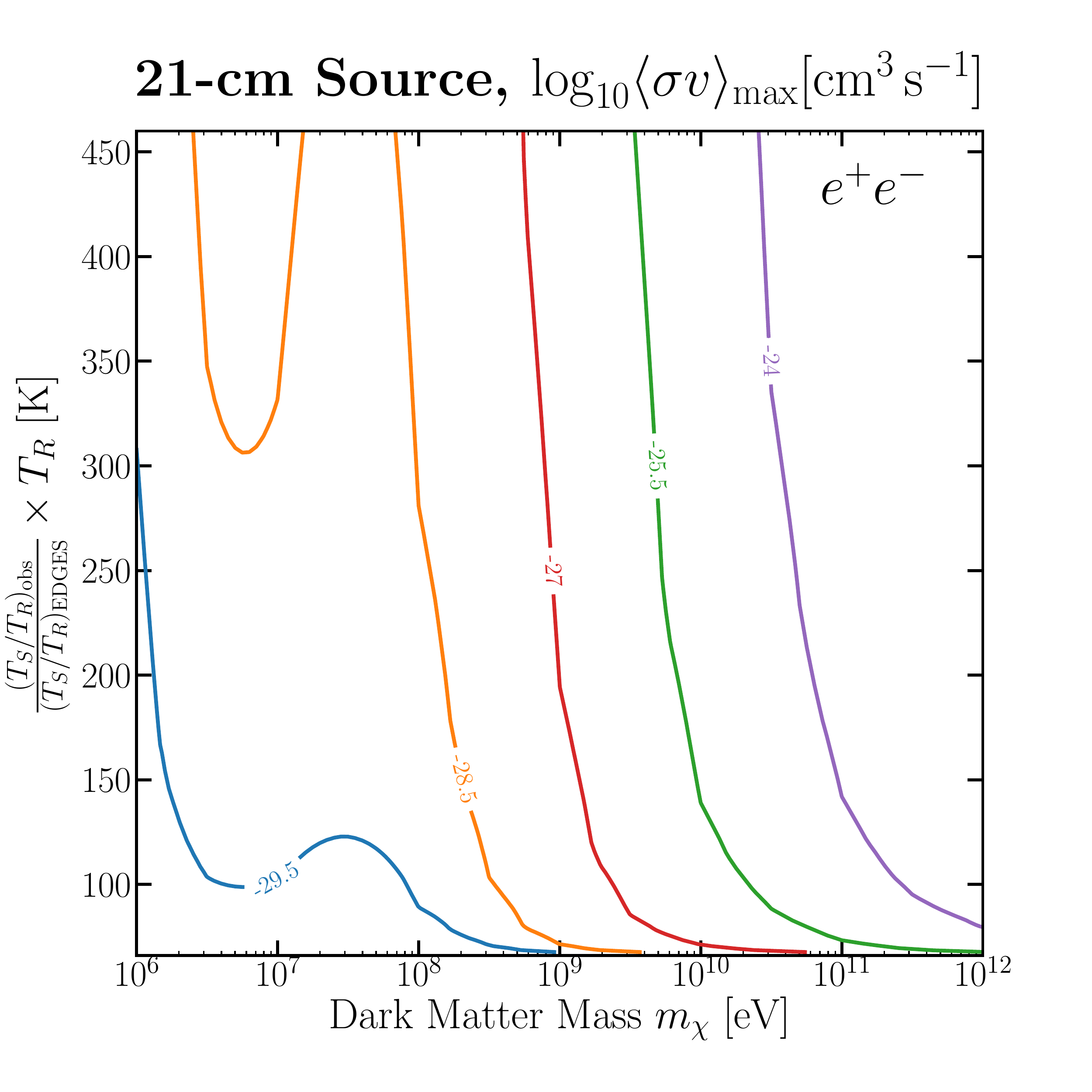}
    }
    \hfil
    \subfloat[]{
        \label{fig:source_phot_swave}
        \includegraphics[scale=0.34]{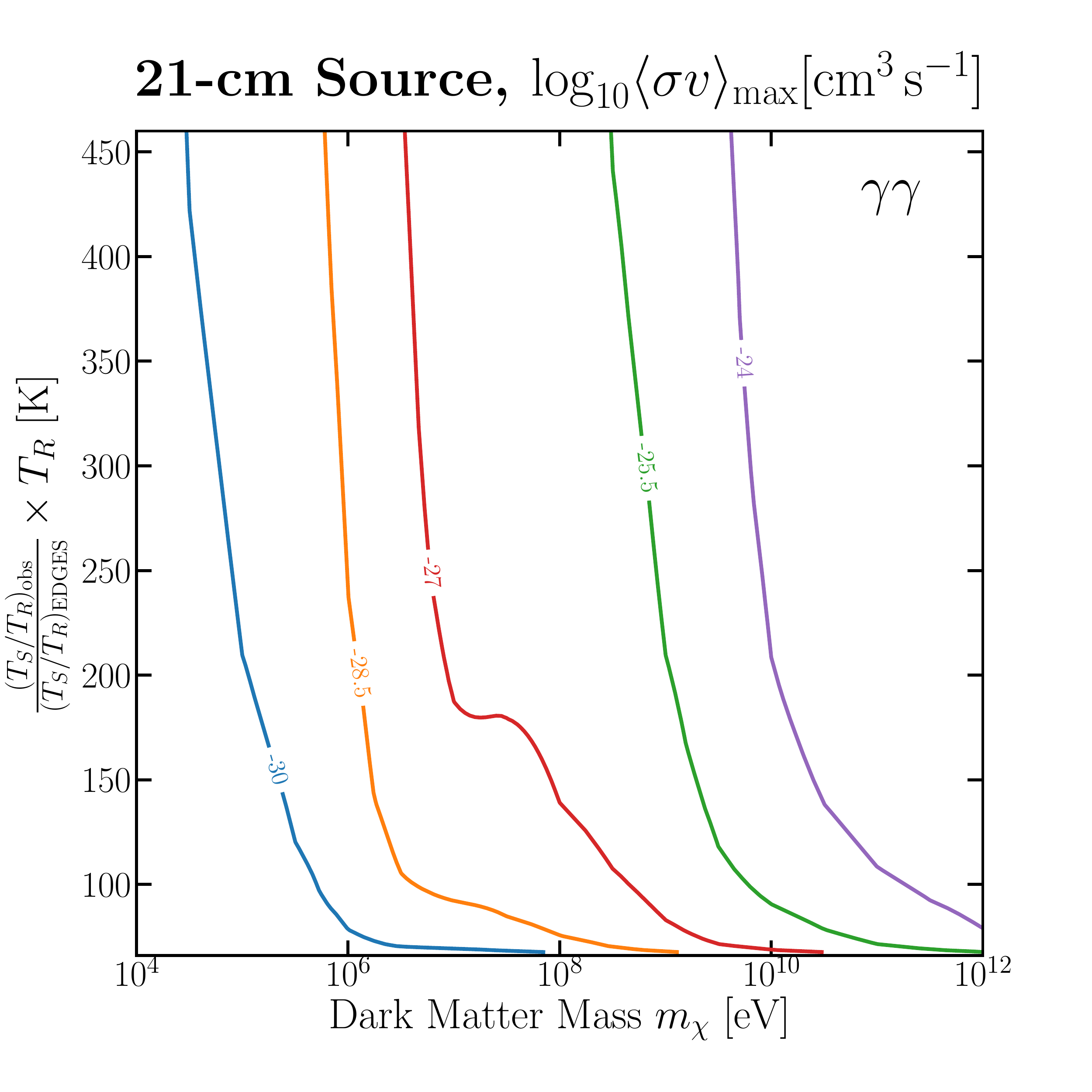}
    }
  \caption{Annihilation cross section constraints with an additional 21-cm source with $\chi \chi \to e^+e^-$ (left) and $\chi \chi \to \gamma \gamma$ (right), as a function of $m_\chi$ and $(T_S/T_R)_\text{obs}/(T_S/T_R)_\text{EDGES} \times T_R$. Contour lines of constant maximum $\log_{10} \langle \sigma v \rangle$ (in $\text{ cm}^3 \text{ s}^{-1}$) are shown. The green contour corresponds to the canonical relic abundance cross section of $\SI{3e-26}{cm^3 \, s^{-1}}$.
  }
  \label{fig:source_swave}
\end{figure*}

This required value of $T_R$ is large: since $T_R \gg \nu_{21}$, where $\nu_{21} = \SI{1.42}{GHz}$ is the hyperfine transition frequency, the effective temperature is directly proportional to the intensity, and so the additional 21-cm source must have approximately 35\% of the intensity of the CMB at this frequency. Distortions to the CMB from measurements of the energy spectrum today are constrained at the level of approximately one part in $10^4$ to $10^5$ \cite{Chluba:2014sma}, and so this large contribution to the photon spectrum must cut off before $\sim$10 GHz. Models where such a strong photon emission comes directly from DM decay or de-excitation run into stringent electroweak precision constraints \cite{Fraser:2018acy}, but models where DM decays into dark photons which oscillate into $21$-cm photons are viable, and can provide an order one or larger contribution to the intensity compared to the CMB \cite{Pospelov:2018kdh}. This large temperature also cannot be explained by uncertainties in the radio emission from astrophysical sources \cite{Barkana:2018lgd}, but may be possible to explain with optimistic black hole formation rates \cite{Ewall-Wice:2018bzf}.

We shall therefore set constraints on DM annihilation and decay as a function of $T_R$, assuming that $1.3 T_\text{CMB} \lesssim T_R \lesssim 10 T_\text{CMB}$ at $z = 17.2$. If we take the EDGES 99\% confidence limit on $T_m$ shown in Eq.~(\ref{eqn:T_m_T_R_ratio}), this corresponds to gas temperatures in the range $\SI{7}{K} \lesssim T_m \lesssim \SI{50}{K}$ at the same redshift. Lower values of $T_R$ lead to values of $T_m$ that are below the standard matter temperature at this redshift, in which case no additional heating would be allowed.

The evolution equations we solve are
\begin{alignat}{1}
    \dot{T}_m &= \dot{T}_m^{(0)} + \dot{T}_m^\chi, \nonumber \\
    \dot{x}_e &= \dot{x}_e^{(0)} + \dot{x}_e^\chi.
\end{alignat}
Note that the CMB temperature used in these equations remains unchanged, as we do not expect significant modifications to the overall energy density of the CMB. Since the evolution equations are essentially the same as the TLA with DM annihilation or decay, these constraints are related to those derived in~\cite{DAmico:2018sxd}, but are broadly applicable to measurements with $T_{21} \lesssim -\SI{200}{mK}$, including the EDGES measurement. We also use the structure formation prescription described in \cite{Slatyer:2012yq,Liu:2016cnk}, with the boost factor included in $f_c(z)$, to account for any delayed deposition of energy.

Figs.~\ref{fig:source_decay} and~\ref{fig:source_swave} show the constraints on the minimum decay lifetime and maximum annihilation cross section with an additional source of 21-cm radiation. The limits are presented as a function of $m_\chi$ and the ratio $(T_S/T_R)_\text{obs}/(T_S/T_R)_\text{EDGES} \times T_R$, with $(T_S/T_R)_\text{EDGES} = 0.105$ as given in Eq.~(\ref{eqn:T_m_T_R_ratio}); these limits can be rescaled if future 21-cm measurements alter or improve the measurement of $(T_S/T_R)_\text{obs}$. The constraints found in~\cite{DAmico:2018sxd} for $T_{21} = \SI{100}{mK}$ and \SI{50}{mK} are equivalent to the constraints obtained with $(T_S/T_R)_\text{obs}= 0.26$ and 0.41 respectively, and setting $T_R = T_\text{CMB}$ at $z = 17.2$. Zoomed-in versions of these plots for lower temperatures are shown in Figs.~\ref{fig:source_decay_zoom}-\ref{fig:source_swave_zoom}. 

For a given measurement of $T_{21}$, Eq.~(\ref{eqn:T_m_T_R_ratio}) permits a higher matter temperature for larger values of $T_R$, which weakens the constraints that can be set. Taking the observed EDGES measurement of $T_S/T_R$, a radiation temperature of $T_R \sim \SI{100}{K}$ constrains the decay lifetime for $\chi \to e^+e^-$ to more than $10^{25}$ s across all DM masses, which is significantly stronger than the existing Planck power spectrum limits \cite{Slatyer:2016qyl}. Cross section constraints similarly strengthen considerably with respect to the Planck limits for $T_R < \SI{100}{K}$.

\section{Non-Standard Recombination}
\label{sec:non_standard_recombination}

Thermal decoupling occurs when the Compton scattering rate becomes comparable to the adiabatic cooling rate, marking the point where the matter temperature transitions from $T_m \propto (1+z)$ to $T_m \propto (1+z)^2$. The standard redshift of thermal decoupling without additional sources of heating or cooling $(1+z)_{\text{td},0}$ is therefore obtained by setting $2HT_m = \Gamma_C T_m$, giving
\begin{alignat}{1}
    (1+z)_{\text{td},0} \approx \left[\frac{45 m_e H_0 \sqrt{\Omega_m}}{4 \pi^2 \sigma_T x_e T_{\gamma,0}^4}\right]^{2/5}.
    \label{eqn:td_redshift}
\end{alignat}
Substituting a value of $x_e = 3 \times 10^{-4}$, a typical value for $x_e$ during the dark ages, we get $(1+z)_{\text{td},0} \approx 155$.

In non-standard models of recombination, the ionization history can be altered in such a way that the $x_e$ evolution equation Eq.~(\ref{eqn:xe}) is modified while leaving the temperature evolution unchanged; this can happen, for example, if the background radiation at energies on the order of the ionization potential for hydrogen deviates significantly from a blackbody distribution during recombination \cite{DeBernardis:2008tk}. 
Another example will be discussed in Sec.~\ref{sec:rutherford_cooling}: if a small fraction of DM couples strongly to baryons, it can act as an additional heat sink and likewise modify the ionization and thermal history. Other mechanisms for early thermal decoupling, which have been recently proposed to explain the EDGES measurement, include the influence of early dark energy ~\cite{Hill:2018lfx}, or charge sequestration~\cite{Falkowski:2018qdj}, where the number density of protons and electrons are unequal owing to the presence of an additional dark charged species. In this work, we remain agnostic as to the cause of early decoupling, parametrizing it by the modified redshift of decoupling.

While the ionization history in such a situation would be model-dependent, once thermal decoupling occurs, the evolution of the thermal history without DM energy injection is completely specified by $\dot{T}_m = -2 H T_m$. The full evolution equation that we will thus solve is
\begin{alignat}{1}
    \dot{T}_m = -2 H T_m + \dot{T}_m^\chi
\end{alignat}
starting from the redshift of thermal decoupling. In reality, the non-zero value of $x_e$ would still provide some additional Compton heating, but limits set by ignoring this effect are less constraining and thus conservative.

The effect of such modifications on the thermal history can therefore be parametrized by the redshift of decoupling $(1+z)_{\text{td}}$. An earlier redshift of decoupling, occurring when the condition specified in Eq.~(\ref{eqn:td_redshift}), results in a lower temperature at later times: The EDGES result can be explained, for example, by a modified ionization and thermal history of this sort \cite{Bowman:2018yin}. Without considering specific models for increasing $(1+z)_\text{td}$, we can set constraints on DM energy injection processes as a function of this quantity, as long as heating from these processes are unimportant relative to adiabatic expansion prior to thermal decoupling.

\begin{figure*}[hp]
    \subfloat[]{
        \label{fig:recomb_elec_decay}
        \includegraphics[scale=0.34]{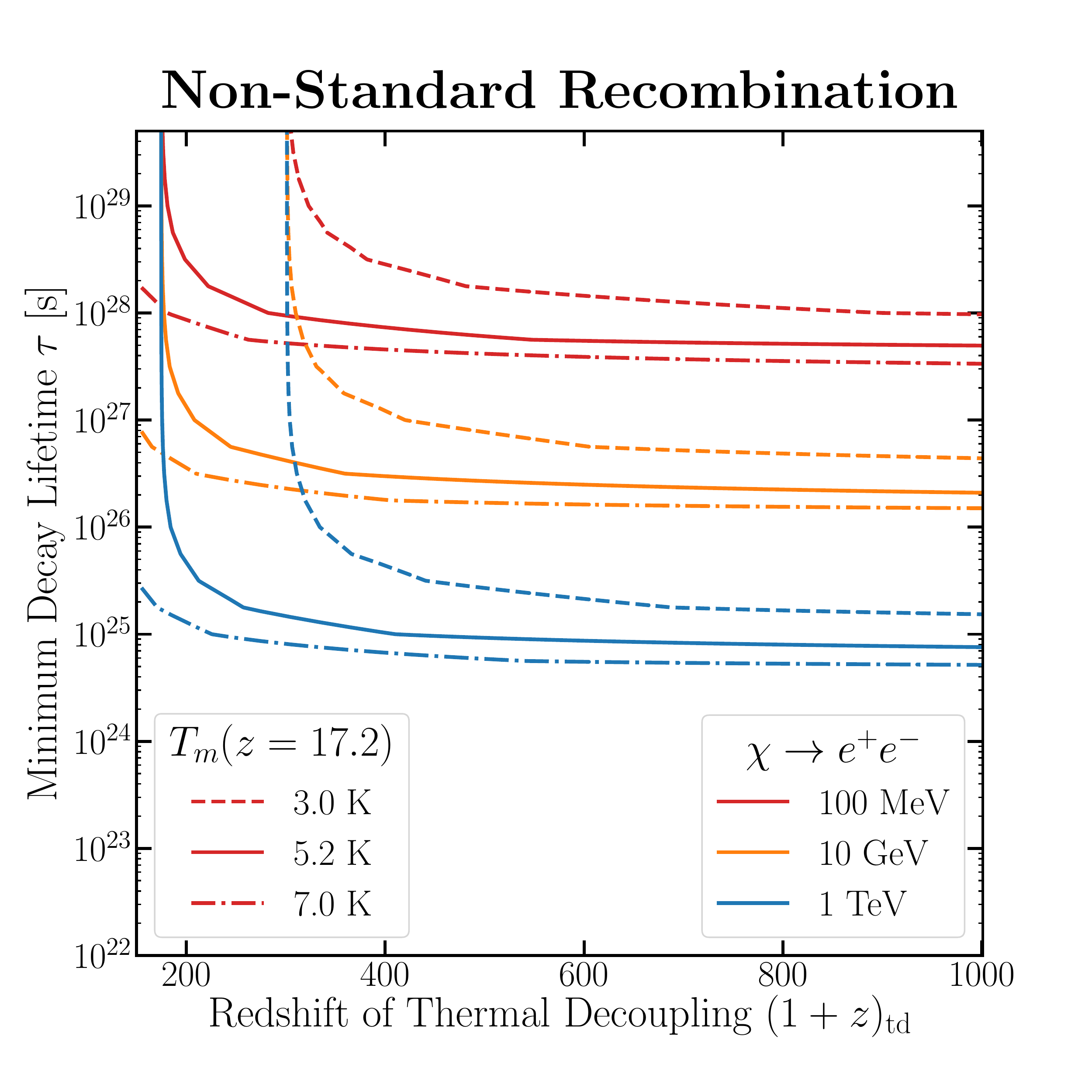}
    }
    \hfil
    \subfloat[]{
        \label{fig:recomb_phot_decay}
        \includegraphics[scale=0.34]{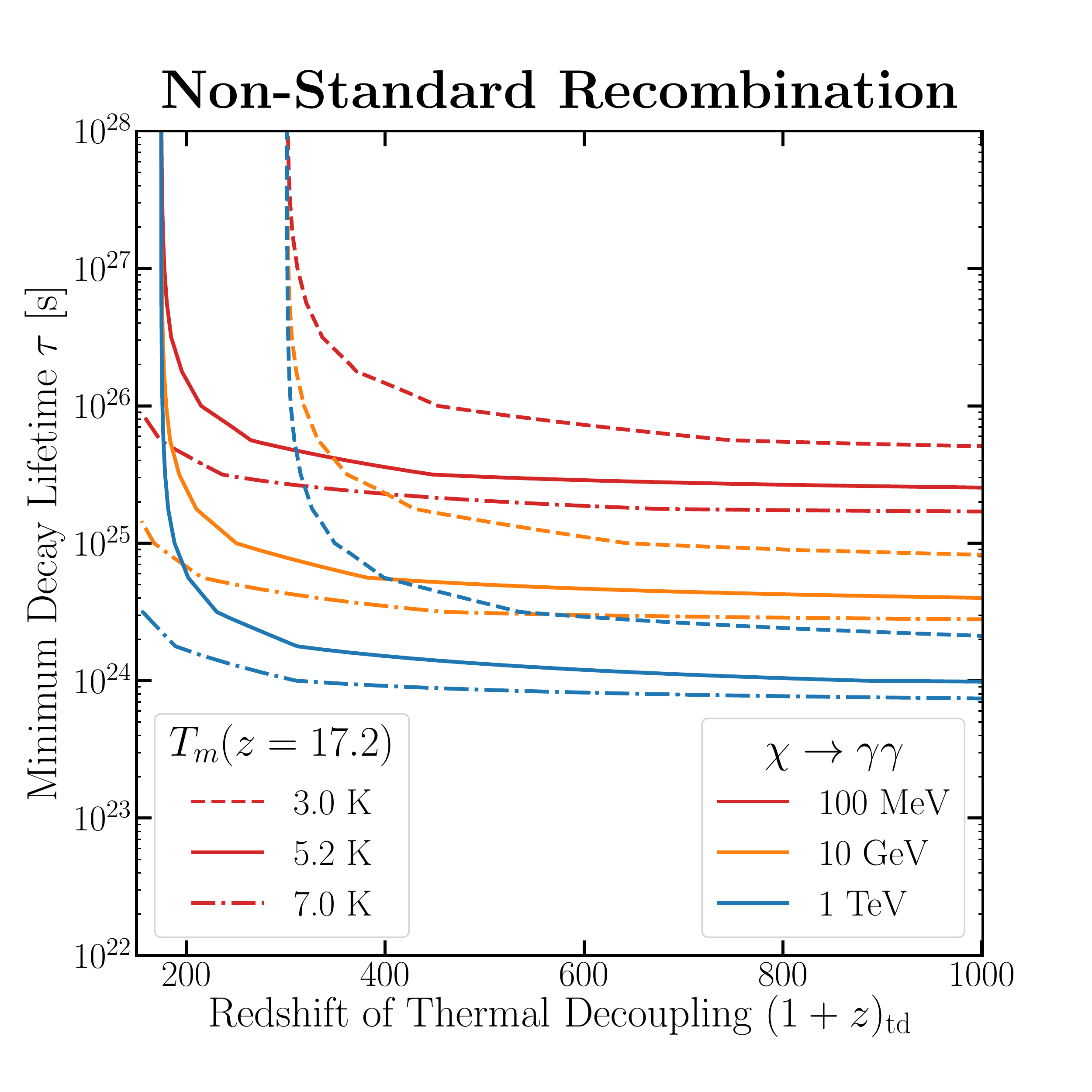}
    }
  \caption{Decay lifetime constraints for non-standard recombination as a function of the redshift of thermal decoupling, with $\chi \to e^+e^-$ (left) and $\chi \to \gamma \gamma$. In both plots, we show the limits for 100 MeV (red), 10 GeV (orange) and 1 TeV (blue) DM, assuming a measured upper limit on the matter temperature at $z = 17.2$ of 3.0 K (dashed), 5.2 K (solid) and 7.0 K (dot-dashed). The 5.2K value corresponds to the EDGES limit.
  }
  \label{fig:recomb_decay}
\end{figure*}

Fig.~\ref{fig:recomb_decay} shows the constraints set on the decay lifetime of a DM particle $\chi$ decaying to $e^+e^-$ and $\gamma \gamma$ respectively as a function of $(1+z)_\text{td}$, for different possible values of $T_m$ at $z = 17.2$. For temperatures below 7 K, the temperature for standard recombination, thermal decoupling must occur at a sufficiently high redshift before adiabatic cooling can bring $T_m$ to that value. For the EDGES value of 5.2 K, this corresponds to $(1+z)_\text{td} \sim 175$; additional heating from DM energy injection is only allowed when the thermal decoupling occurs above this value. 

Once $(1+z)_\text{td}$ exceeds the critical value for sufficient cooling, the constraints on the minimum decay lifetime depends only weakly on $(1+z)_\text{td}$. To understand this, note that if the baryon temperature in the absence of heating is well below the observed temperature limit, then the temperature including heating is solely determined by the energy injection rate, and is relatively independent of the baseline baryon temperature without heating and hence $(1+z)_\text{td}$. 

With $T_m(z = 17.2) = $ 5.2 K, the constraints set by this temperature measurement for $m_\chi $ = 100 MeV is $\sim 5 \times 10^{27}$ s for decays to $e^+e^-$ and $\sim 10^{25}$ s for $\gamma \gamma$, which is both at least an order of magnitude stronger than limits set by the Planck CMB power spectrum measurement \cite{Slatyer:2016qyl}. These limits are valid assuming only no additional sources of cooling for the matter temperature after thermal decoupling. 

\begin{figure*}[hp]
    \subfloat[]{
        \label{fig:recomb_elec_swave}
        \includegraphics[scale=0.34]{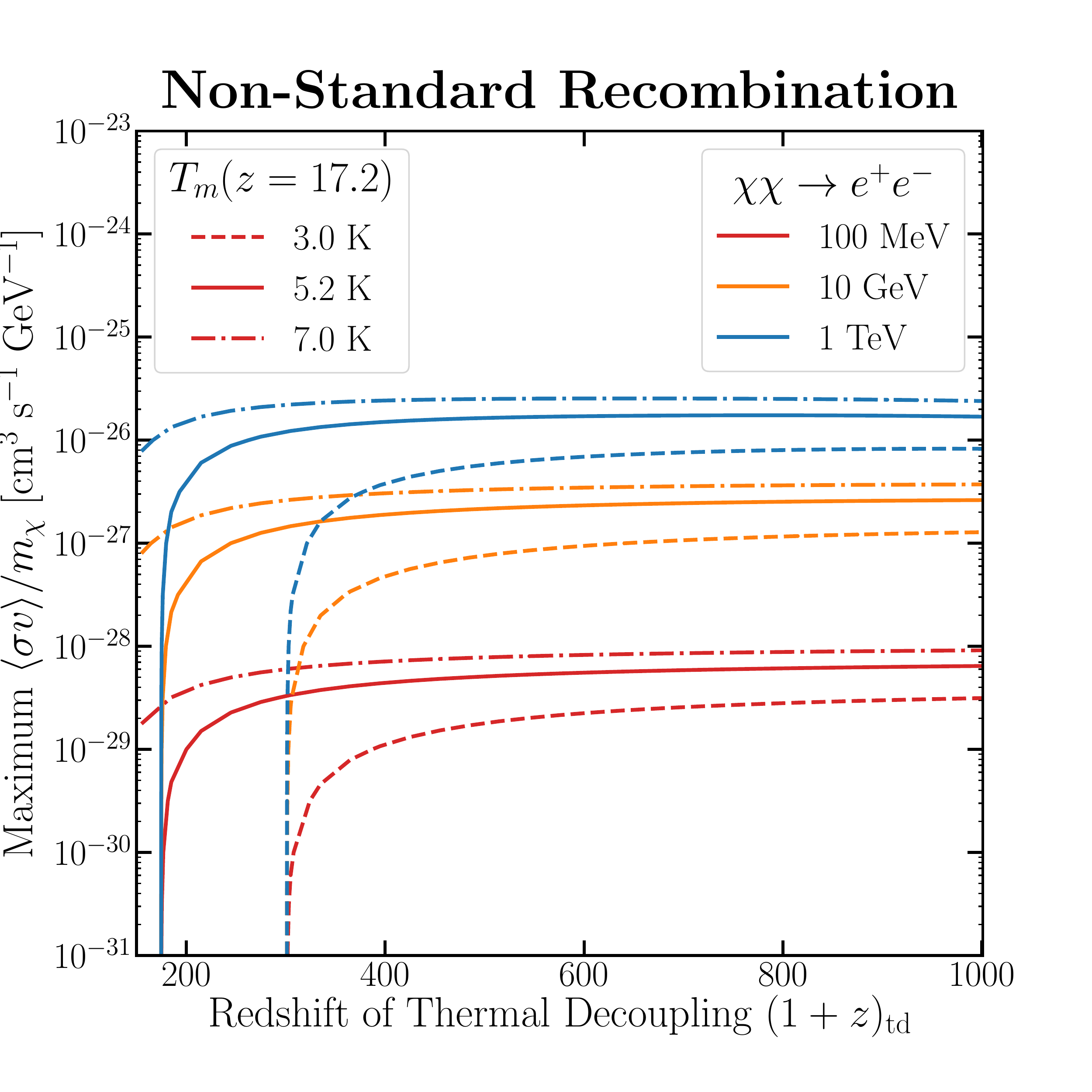}
    }
    \hfil
    \subfloat[]{
        \label{fig:recomb_phot_swave}
        \includegraphics[scale=0.34]{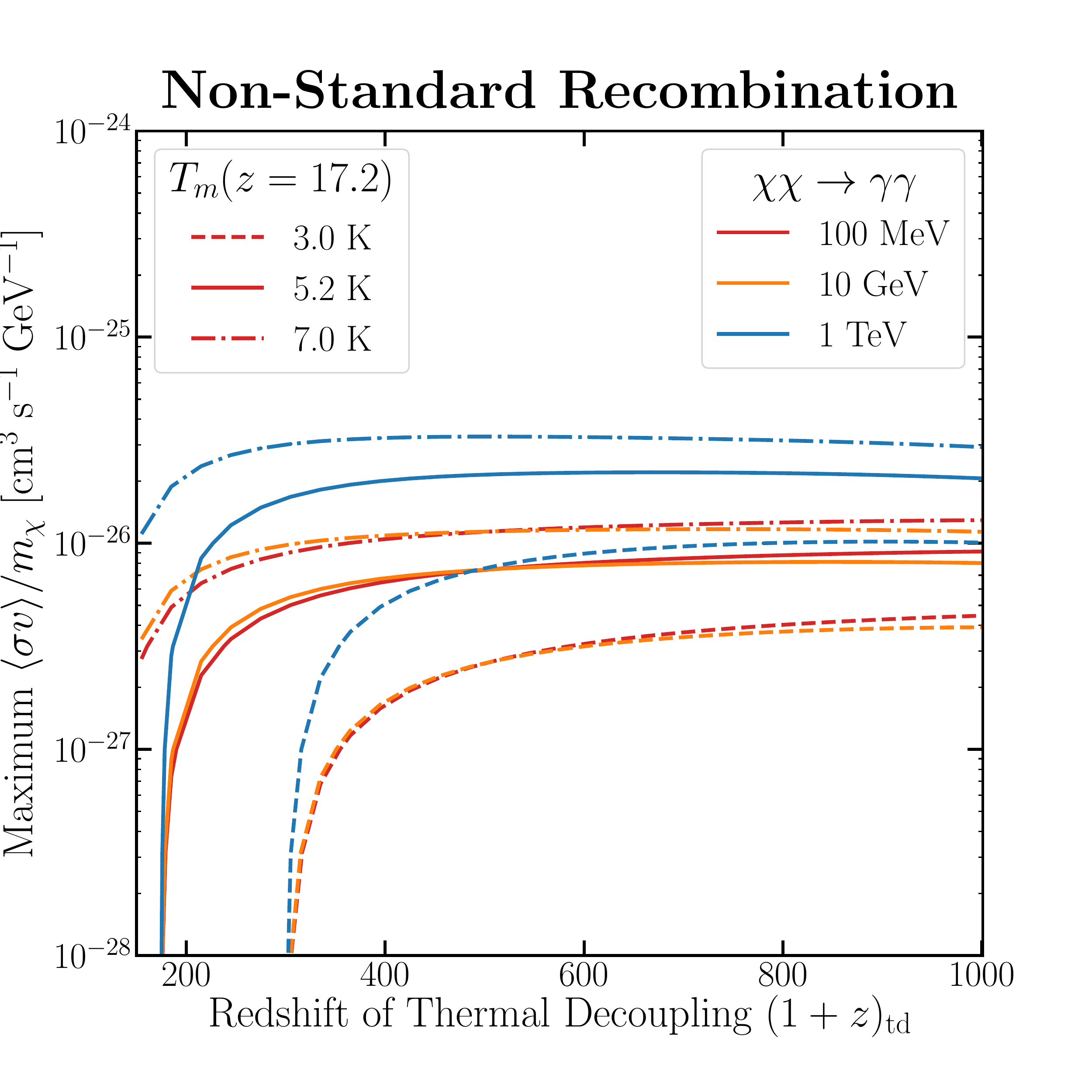}
    }
  \caption{Annihilation cross section constraints for non-standard recombination as a function of the redshift of thermal decoupling, with $\chi \chi \to e^+e^-$ (left) and $\chi \chi \to \gamma \gamma$. In both plots, we show the limits for 100 MeV (red), 10 GeV (orange) and 1 TeV (blue) DM, assuming a measured upper limit on the matter temperature at $z = 17.2$ of 3.0 K (dashed), 5.2 K (solid) and 7.0 K (dot-dashed). The 5.2K value corresponds to the EDGES limit.}
  \label{fig:recomb_swave}
\end{figure*}

Fig.~\ref{fig:recomb_swave} shows a similar plot for the constraints on the annihilation cross section, with the main features of these constraints being similar to the result for decays. The constraints set by $T_m(z = 17.2) = $ 5.2 K are once again stronger than the current Planck constraints \cite{Slatyer:2015jla,Ade:2015xua} by about an order of magnitude, with little dependence on $(1+z)_\text{td}$.

\begin{figure*}[t]
    \subfloat[]{
        \label{fig:recomb_chan_decay}
        \includegraphics[scale=0.34]{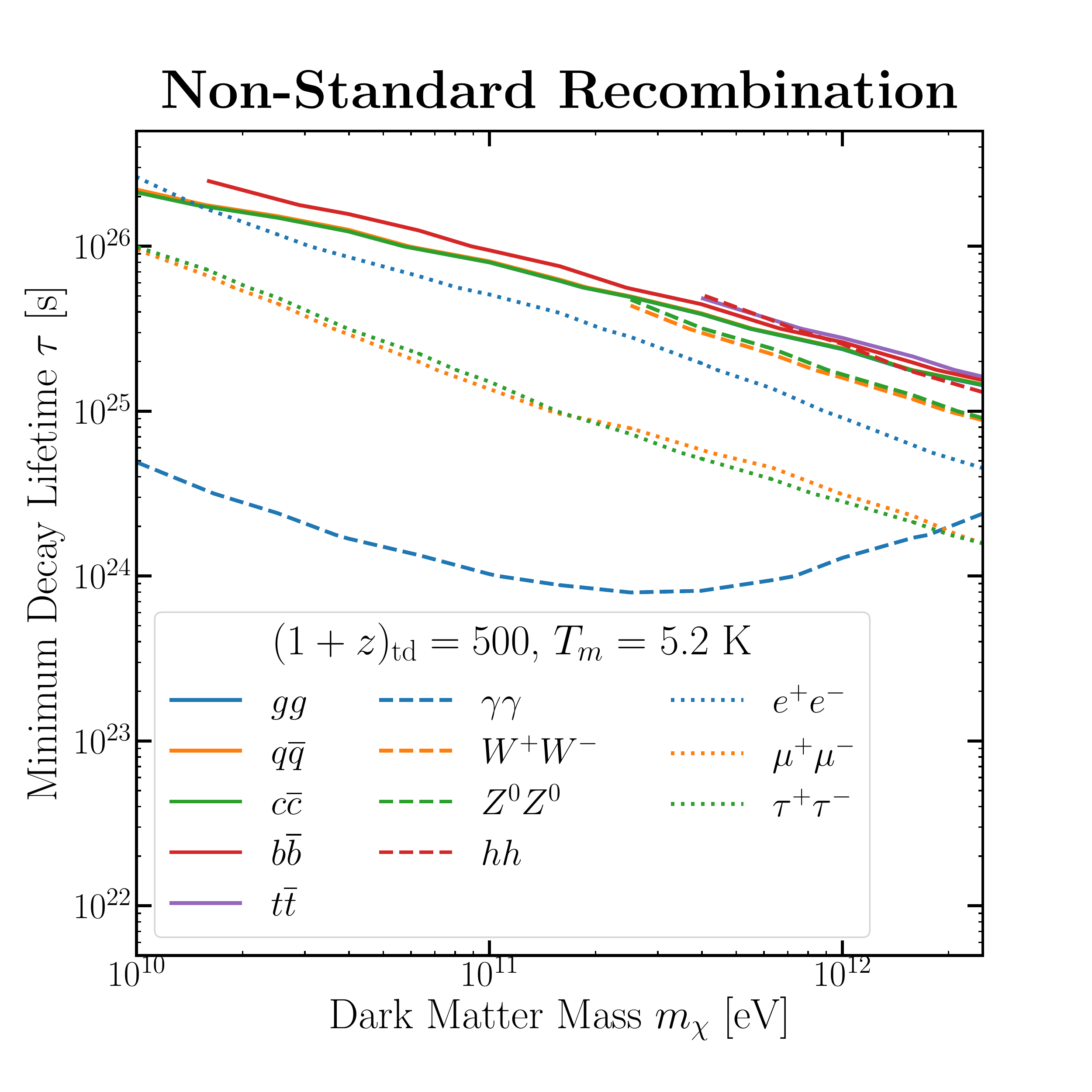}
    }
    \hfil
    \subfloat[]{
        \label{fig:recomb_chan_swave}
        \includegraphics[scale=0.34]{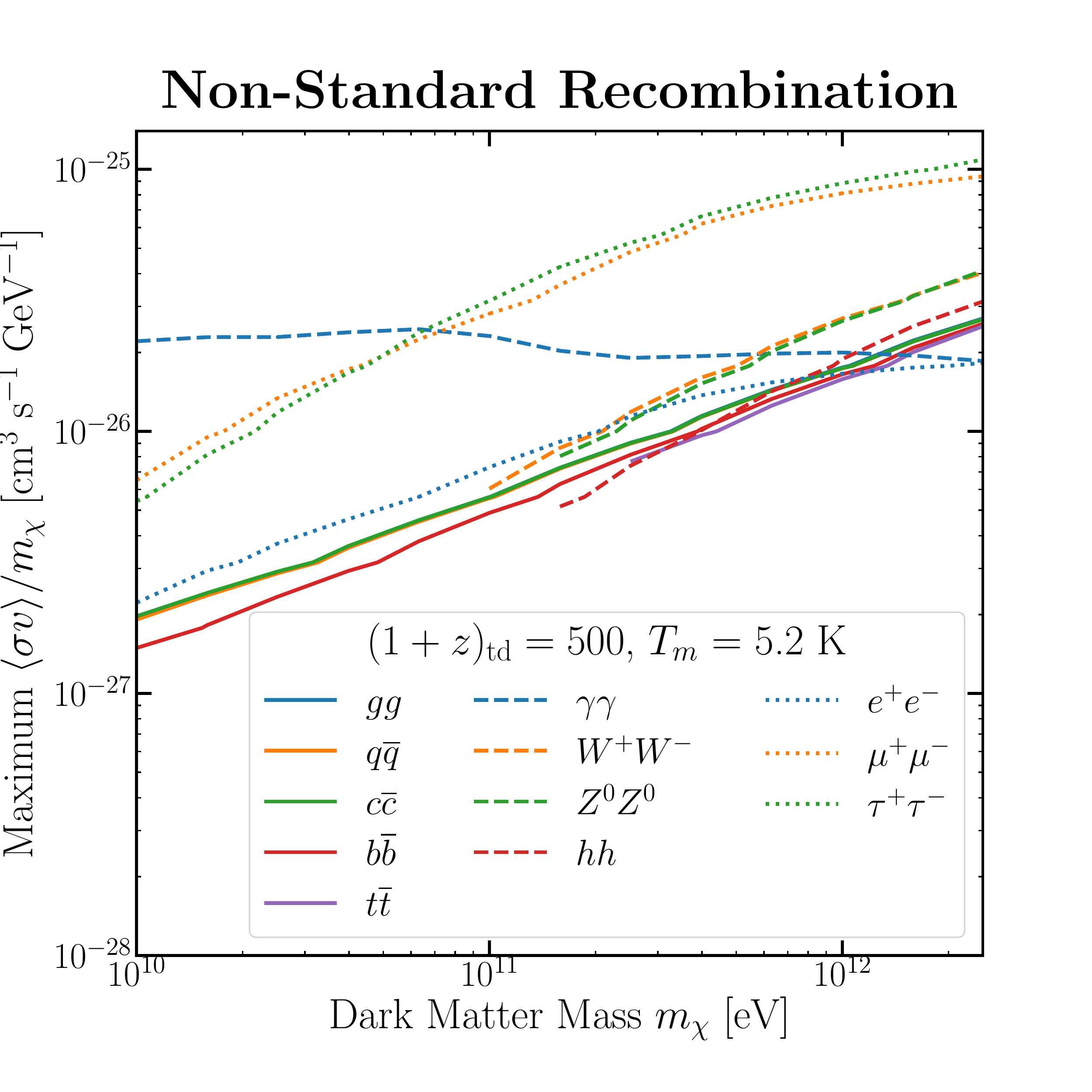}
    }
  \caption{Minimum decay lifetime (left) and maximum annihilation cross section (right) constraints for non-standard recombination as a function of the redshift of thermal decoupling for several SM channels. In both plots, $(1+z)_\text{td} = 500$, and $T_m(z = 17.2) = $ 5.2 K. 
  }
  \label{fig:recomb_chan}
\end{figure*}

Fig.~\ref{fig:recomb_chan} shows the constraints for several decay and annihilation channels into SM particles for the case where $(1+z)_\text{td} = 500$, and $T_m(z = 17.2) = $ 5.2 K. These constraints apply across a large range of $(1+z)_\text{td} \gtrsim 250$ due to the weak dependence on $(1+z)_\text{td}$. To obtain these constraints, electron and photon spectra were computed using the \texttt{PPPC4DMID} \cite{Cirelli:2010xx}, and the corresponding $f_c(z)$ computed by taking an energy-weighted sum of the spectra \cite{Slatyer:2015jla}. The behavior of these limits are set mainly by the ability of the secondary photons and electrons of these decays to heat baryons at $z \sim 20$. The universe between $z \sim 20$ and recombination is mostly transparent to high energy photons, which explains the relatively weak constraints for $\chi \to \gamma \gamma$. The limits of the other channels relative to $\chi \to e^+e^-$ weaken when more neutrinos are produced during the cascade, and strengthen when soft electrons/positrons with energy $\sim \SI{100}{MeV}$ are produced: electrons and positrons in this energy range are particularly effective at depositing their energy into baryons during this epoch \cite{Liu:2016cnk,Slatyer:2016qyl}.

\section{Rutherford Cooling}
\label{sec:rutherford_cooling}

\subsection{Evolution Equations}

To ensure that the matter temperature at $z \sim 17$ satisfies Eq.~(\ref{eqn:T_m_T_R_ratio}) while taking $T_R = T_\text{CMB}$, several groups \cite{Barkana:2018lgd,Munoz:2018pzp,Berlin:2018sjs,Fraser:2018acy,Barkana:2018qrx} have examined the possibility of a new DM-baryon or DM-electron scattering cross section that has a Rutherford-like behavior, i.e. $\sigma = \sigma_0 v^{-4}$. This interaction may occur between only a fraction of DM and the SM. Both the difference in temperature between matter and DM as well as their bulk relative velocity $V_{\chi b}$ from earlier DM clustering can affect $T_m$, which evolves according to \cite{Munoz:2018pzp,Munoz:2015bca} 
\begin{multline}
    \dot{T}_m^c = \sum_j \frac{2}{3(1 + f_\text{He} + x_e)n_\text{H}} \frac{f_{\chi,\text{int}} \rho_\text{DM} \rho_j}{(m_\chi + m_j)^2} \\
    \times  \frac{\sigma_{0,j}}{u_j} \left[\sqrt{\frac{2}{\pi}} \frac{e^{-r_j^2/2}}{u_j^2} (T_\chi - T_m) + m_\chi \frac{F(r_j)}{r_j}\right],
    \label{eqn:T_m_cooling}
\end{multline}
where the sum is over all species $j$ that can interact with the DM: this may be over all baryons \cite{Barkana:2018lgd}, or over free protons and electrons in millicharged DM models \cite{Munoz:2018pzp}. $\rho_\text{DM}$ and $\rho_j$ are the mass densities of all DM and species $j$ respectively, with $f_{\chi,\text{int}}$ being the fraction of DM interacting with the SM by mass. $m_\chi$ and $T_\chi$ is the mass and temperature of the interacting DM respectively, $u_j \equiv (T_m/m_j + T_\chi/m_\chi)^{1/2}$ is the thermal sound speed of the DM-$j$ fluid, and $r_j \equiv V_{\chi b}/u_j$. The function $F(r)$ is
\begin{alignat}{1}
    F(r) \equiv \text{erf} \left(\frac{r}{\sqrt{2}}\right) - \sqrt{\frac{2}{\pi}} e^{-r^2/2} r.
\end{alignat}

To solve for the full evolution, we must also evolve the temperature of the interacting DM \cite{Munoz:2015bca},
\begin{multline}
    \dot{T}_\chi = -2HT_\chi + \sum_j \frac{2}{3} \frac{m_\chi \rho_j}{(m_\chi + m_j)^2} \\
    \times \frac{\sigma_{0,j}}{u_j} \left[\sqrt{\frac{2}{\pi}} \frac{e^{-r_j^2/2}}{u_j^2} (T_m - T_\chi) + m_j \frac{F(r_j)}{r_j}\right],
    \label{eqn:T_chi_cooling}
\end{multline}
as well as the bulk relative velocity \cite{Munoz:2018pzp}
\begin{alignat}{1}
    \dot{V}_{\chi b} = -H V_{\chi b} - \left(1 + \frac{f_{\chi,\text{int}} \rho_\text{DM}}{\rho_b}\right) \sum_j \frac{m_j n_j \sigma_{0,j}}{m_\chi + m_j} \frac{F(r_j)}{V_{\chi b}^2}.
    \label{eqn:V_chi_b}
\end{alignat}
When $f_{\chi, \text{int}} < 1$, Eq.~(\ref{eqn:T_chi_cooling}) assumes that the interacting component of DM has a temperature that is separate from the rest of the dark sector; relaxing this assumption would mean that the energy flow into the dark sector is distributed among more particles, with the exact effect on the thermal history determined by the masses of both the interacting and non-interacting components. 

To set constraints on DM energy injection in the presence of this scattering process, the full set of rate equations which should be integrated are Eqs.~(\ref{eqn:T_chi_cooling}), ~(\ref{eqn:V_chi_b}), together with
\begin{alignat}{1}
    \dot{T}_m &= \dot{T}_m^{(0)} + \dot{T}_m^\chi + \dot{T}_m^c, \nonumber \\
    \dot{x}_e &= \dot{x}_e^{(0)} + \dot{x}_e^\chi.
    \label{eqn:cooling_full_eqns}
\end{alignat}
For simplicity, we will restrict our discussion to the case of DM-hydrogen scattering (both neutral and ionized, with no scattering on helium or free electrons) until we discuss the millicharged DM model, where scattering occurs between DM and free charged particles. 

\subsection{Weak and Strong Coupling Regimes}

The magnitude of $\sigma_0$ determines how tightly coupled the interacting DM and baryon fluids are. In the weakly coupled regime, the DM temperature $T_\chi$ remains well below the matter temperature $T_m$, and the interacting DM component is able to collapse into structures well before recombination, leading to a non-zero bulk relative velocity $V_{\chi b}$. However, for a sufficiently large $\sigma_0$, the temperature of the interacting DM becomes close to the matter temperature, and collapse into structures becomes impossible. For $f_{\chi,\text{int}} = 1$, i.e. all of the DM interacts with the SM, this scenario is highly constrained by the damping effect this would have on the CMB power spectrum \cite{McDermott:2010pa,Dvorkin:2013cea,Xu:2018efh,Slatyer:2018aqg}. However, a subdominant component ($f_{\chi,\text{int}} \lesssim 0.01$) of DM can have significant interactions with the SM at recombination without contradicting precision CMB measurements: the interacting DM component would essentially be an additional, small contribution to the baryon fluid, while leaving structure formation due to the bulk of DM unaffected \cite{Dolgov:2013una}. 

In the weak-coupling regime, the interacting component of DM remains cold and collapses efficiently, and $V_{\chi b}$ is expected to have an rms velocity of \SI{29}{km \, s^{-1}} at photon decoupling, $z = 1010$, the value expected for cold, non-interacting DM \cite{Ali-Haimoud:2013hpa}. From Eq.~(\ref{eqn:T_m_cooling}), while $T_m \gg T_\chi$, the effect of a non-zero value of $V_{\chi b}$ is generally to increase $T_m$. This additional source of heating forces the energy injection from DM annihilation or decay to be smaller than if we set $V_{\chi b} = 0$, leading to tighter cosmological constraints. For the rest of the results in this section, we will show only results with $V_{\chi b} = 0$, which leads to the most robust constraints: the effect of fully evolving $V_{\chi b}$ starting at a non-zero value at recombination will be shown in Appendix~\ref{app:systematics}. We integrate Eqs.~(\ref{eqn:T_chi_cooling}) to~(\ref{eqn:cooling_full_eqns}), with $T_\chi = 0$, $T_m = T_\text{CMB}$ and $x_e = 1$ starting from before recombination.

In the strong-coupling regime, the interacting component of DM is in thermal equilibrium with baryons and the CMB, and cannot collapse into structures. In this case, $V_{\chi b} = 0$ at the point of recombination, and the strong coupling between the two sectors ensures $T_m = T_\chi$ throughout. We can therefore integrate Eq.~(\ref{eqn:T_chi_cooling}) and~(\ref{eqn:cooling_full_eqns}), with $T_\chi = T_m = T_\text{CMB}$ and $x_e = 1$ starting from before recombination, with $V_{\chi b} = 0$. 

We delineate the two regimes by requiring the rate of DM heating due to DM-baryon scattering to be larger than $H T_\chi$ at recombination in the strong-coupling limit, so that DM and baryons remain at the same temperature at this point. This leads to the criterion
\begin{alignat}{1}
    \sigma_0^\text{strong} \gtrsim \frac{H}{n_H} \frac{(m_\chi + m_p)^2}{m_\chi m_p} \left(\frac{T_\text{CMB}}{m_p} + \frac{T_\text{CMB}}{m_\chi}\right)^{3/2}
\end{alignat}
at recombination for strong coupling to be valid, and we take the weak-coupling regime to be $\sigma_0 < 0.1 \sigma_0^\text{strong}$. 

\subsection{CMB Power Spectrum Limits}

\begin{figure}
    \includegraphics[scale=0.34]{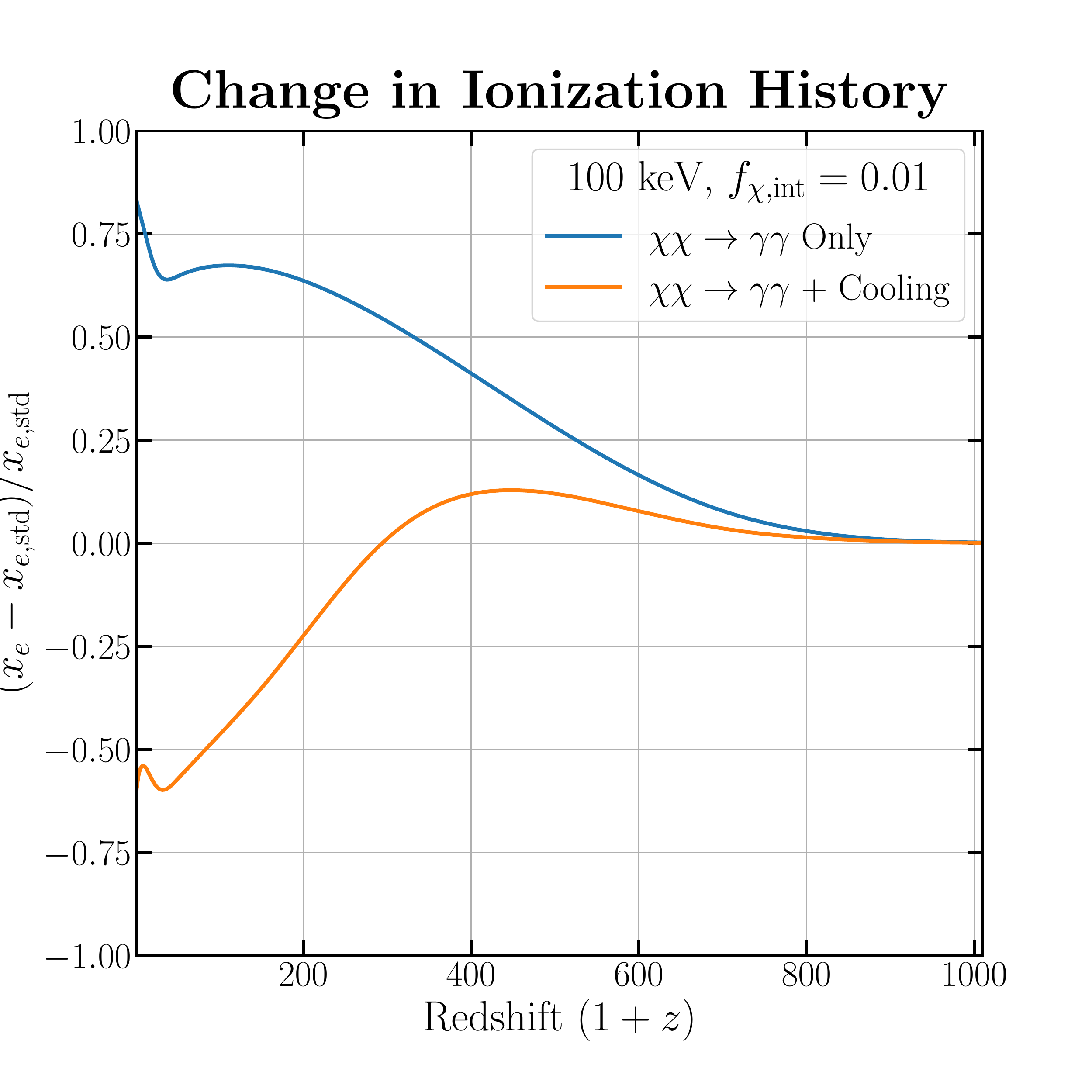}
    \caption{The change in ionization histories for $\chi \chi \to \gamma \gamma$ annihilation, with (yellow) and without (blue) Rutherford cooling, with respect to the standard ionization history (with no DM energy injection), $x_{e,\text{std}}$. Here, $m_\chi$ = 100 keV and $f_{\chi,\text{int}} = 0.01$. The standard history (blue, dotted), DM annihilation only with $\langle \sigma v \rangle = 6.6 \times 10^{-32} \, \text{cm}^3 \text{ s}^{-1}$ (orange), and DM annihilation with DM-baryon Rutherford scattering, $\sigma_0 = 10^{-33} \, \text{cm}^2$ (green) are shown. The chosen value of $\langle \sigma v \rangle$ is the maximum allowed from the Planck CMB limits in the absence of scattering; this scenario with scattering may evade these limits.}
    \label{fig:altered_CMB_const_history}
\end{figure}

DM annihilation and decay during the cosmic dark ages increase the residual ionization of the universe after recombination as compared to the standard history, and this change to the ionization history can be constrained by considering its impact on the CMB power spectrum. The presence of an additional source of cooling of the matter temperature, however, also modifies the ionization history during this time. If the rate of cooling is sufficiently large to decouple baryons from the CMB at a time earlier than $(1+z)_{\text{td},0}$ given in Eq.~(\ref{eqn:td_redshift}), then $T_m$ becomes smaller than expected, which in turn increases the recombination rate, decreasing the residual ionization. 

\begin{figure*}[t!]
    \subfloat[]{
        \label{fig:weak_coupling_Tm_history}
        \includegraphics[scale=0.34]{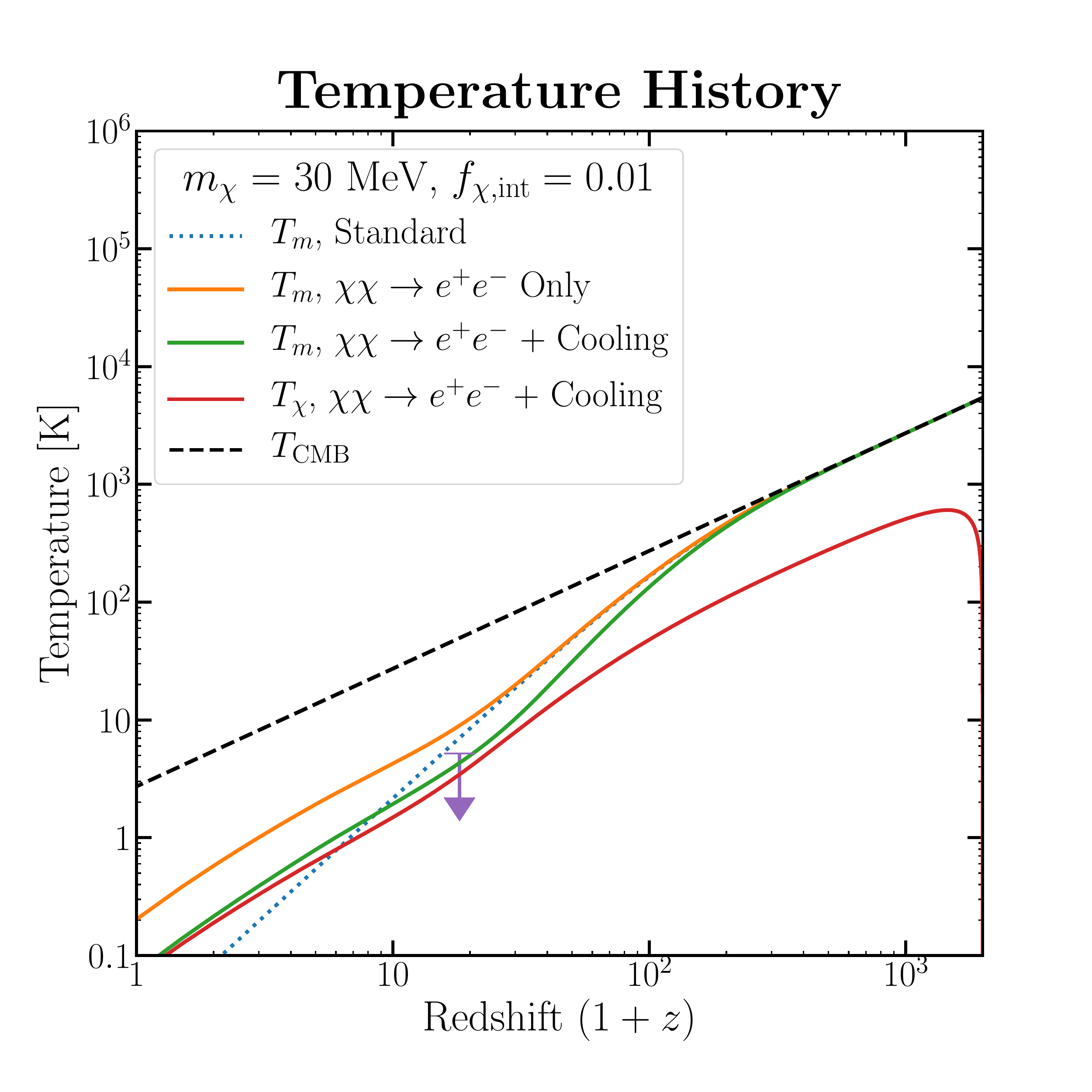}
    }
    \hfil
    \subfloat[]{
        \label{fig:weak_coupling_xe_history}
        \includegraphics[scale=0.34]{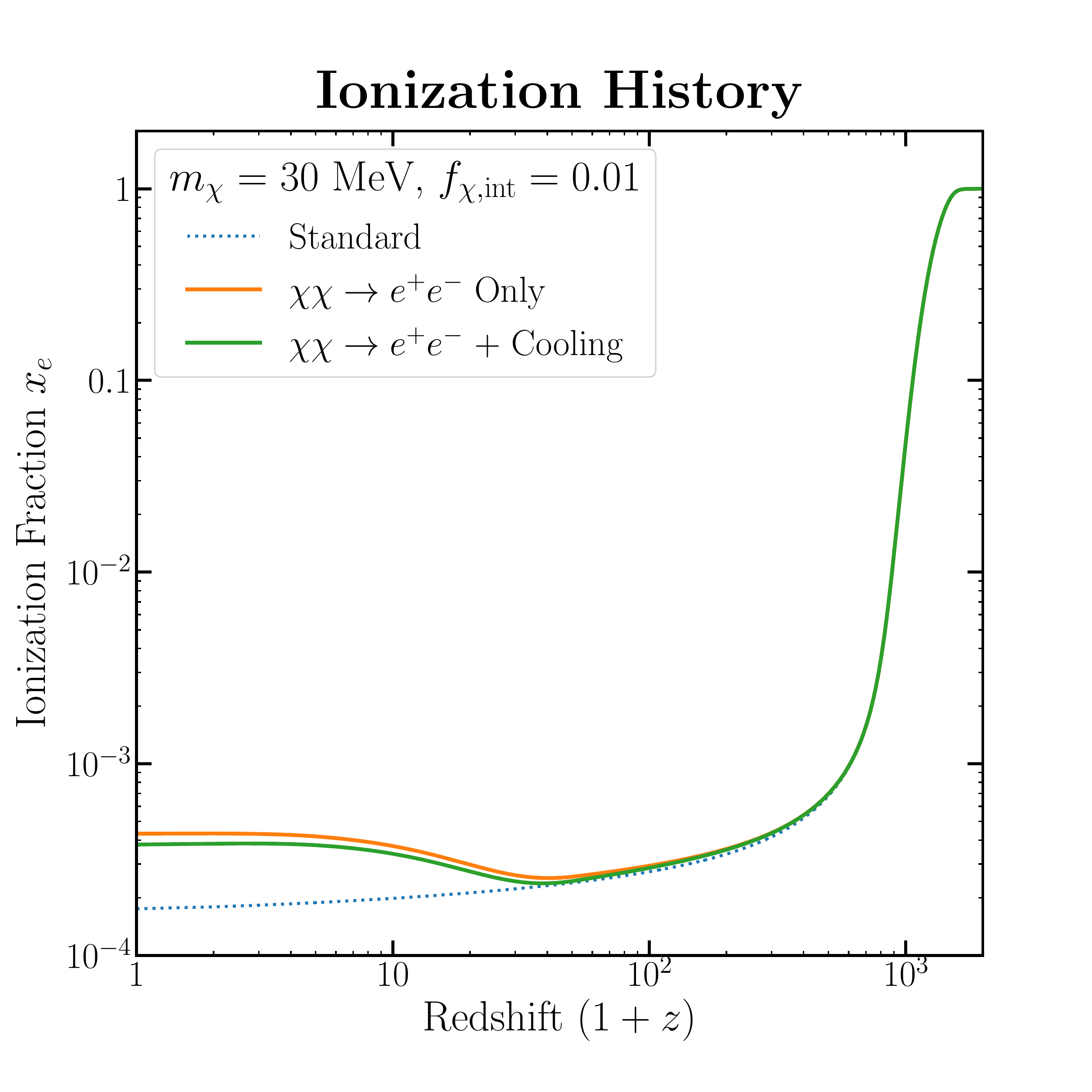}
    }
    \caption{Thermal (left) and ionization (right) histories with $\chi \chi \to e^+e^-$ annihilation and Rutherford cooling, $m_\chi$ = 30 MeV and $f_{\chi,\text{int}} = 0.01$. The standard history (blue, dotted), DM annihilation only with $\langle \sigma v \rangle = 10^{-30} \, \text{cm}^3 \text{ s}^{-1}$ (orange), and DM annihilation and DM-baryon Rutherford scattering with $\sigma_0 = 10^{-38} \, \text{cm}^2$ (green) are shown. The DM temperature evolution (red), CMB temperature (black, dashed) and the EDGES temperature limit (purple arrow) are also shown.}
    \label{fig:weak_coupling_history}
\end{figure*}

Fig.~\ref{fig:altered_CMB_const_history} shows an example of the change in ionization history with respect to the standard ionization history. We have chosen an annihilation cross section for $\chi \chi \to \gamma \gamma$ that is naively ruled out by the Planck CMB limits. Due to the presence of Rutherford cooling, however, the ionization at $z \sim 600$ ($z \sim 300$) relative to the standard history is reduced by a factor of 2 (more than 10). Since the $s$-wave annihilation and decay constraints from the CMB power spectrum are most sensitive to energy injection at $z \sim 600$ and $z \sim 300$ respectively \cite{Slatyer:2016qyl}, we conclude that \emph{the CMB power spectrum constraints on energy injection during this epoch can be significantly relaxed if additional sources of cooling lead to thermal decoupling of baryons during or before these redshifts}. 

To estimate when thermal decoupling of baryons occurs in the presence of Rutherford cooling, we can compare the heat transfer rate due to cooling from DM to the Compton scattering term. For DM-hydrogen scattering, this gives the following condition for thermal decoupling to occur at $(1+z)_\text{td}$:

\begin{alignat}{1}
    \sigma_{0,\text{td}} \lesssim \sigma_T (1+z)_\text{td}^{5/2} \frac{x_e}{f_{\chi,\text{int}}} \left[\frac{ T_{\text{CMB},0}^{11/2} m_\chi^2}{\mu_{\chi p}^2 m_p^{1/2} m_e \rho_{\text{DM},0}}\right],
\end{alignat}
where $\mu_{\chi p}$ is the reduced mass of DM and protons, $T_{\text{CMB},0}$ is the CMB temperature today, and $\rho_{\text{DM},0}$ is the DM density today. Numerically, 
\begin{alignat}{1}
    \sigma_{0,\text{td}} \lesssim \left(\frac{m_\chi}{\mu_{\chi p}}\right)^2 \left(\frac{(1+z)_\text{td}}{600}\right)^{5/2} \frac{10^{-40}\,\text{cm}^2}{f_{\chi,\text{int}}},
    \label{eqn:thermal_decoupling_condition}
\end{alignat}
where we have taken $x_e \approx 3 \times 10^{-4}$. Thus, for $(1+z)_\text{td} = 300$ and $(1+z)_\text{td} = 600$, the CMB power spectrum constraints for decays and $s$-wave annihilation may become inapplicable for $\sigma_0 > \sigma_{0,\text{td}}$ due to the enhanced recombination from cooling at these redshifts. A sufficiently large $T_\chi$ can relax this condition, but we neglect this effect; CMB constraints on all plots are therefore only shown in regions where their validity is assured. A comprehensive study of how CMB constraints on DM annihilation relax under these circumstances is left to future work.

\subsection{Weak Coupling Results}

Fig.~\ref{fig:weak_coupling_history} shows a typical ionization and temperature history in the weak-coupling limit with both cooling and DM annihilation. Thermal decoupling of matter from the CMB occurs slightly earlier than $(1+z)_\text{td} \sim 155$ due to the additional cooling, but not significantly earlier. Since the matter temperature is locked to the radiation temperature until well after $z \sim 600$, the ionization history, even in the presence of DM annihilation, differs very little from the expected history without cooling. As a result, constraints on $s$-wave annihilation set by the CMB spectrum, which is most sensitive to energy injection at $z \sim 600$, are still applicable. 

\begin{figure*}[t!]
    \subfloat[]{
        \label{fig:cooling_elec_decay}
        \includegraphics[scale=0.34]{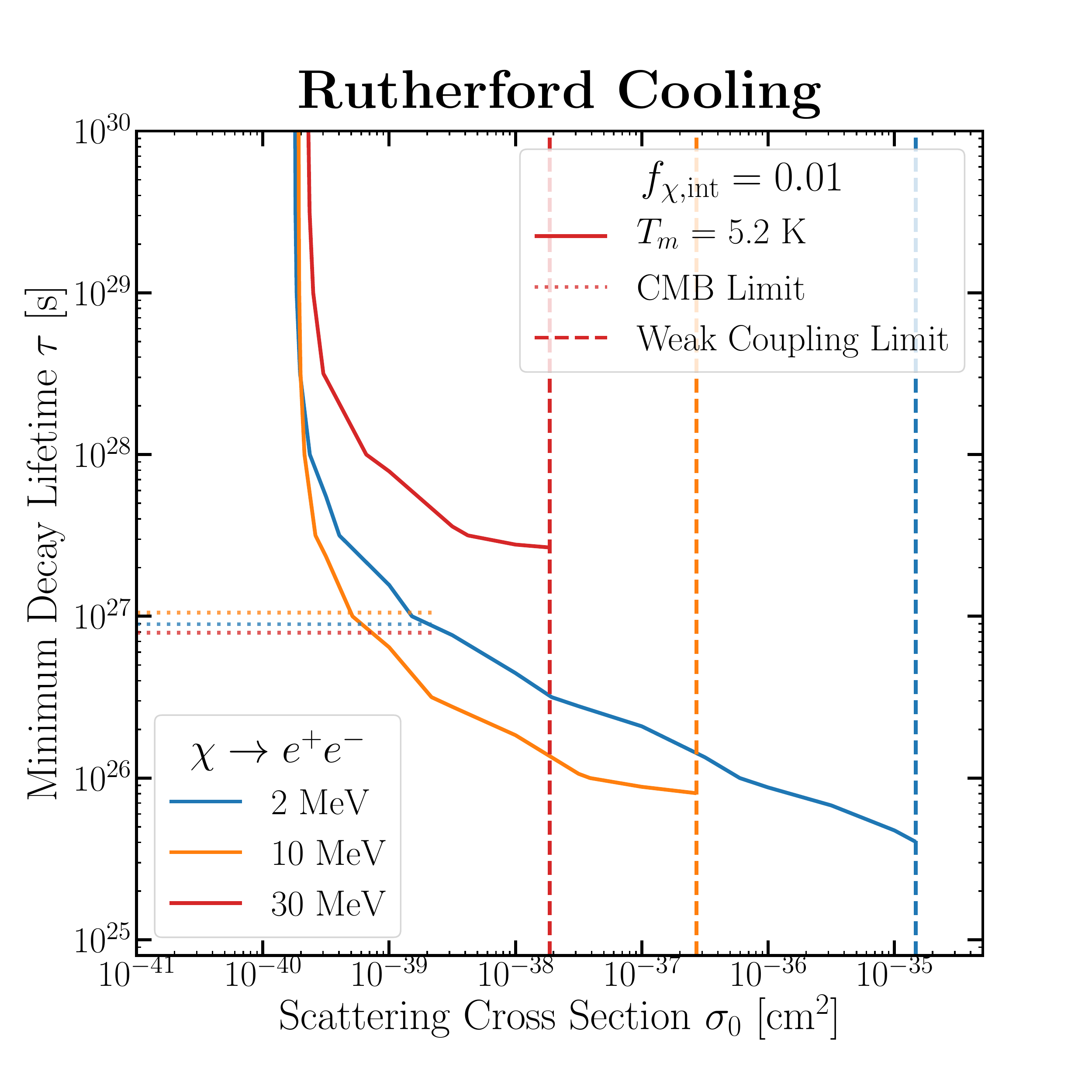}
    }
    \hfil
    \subfloat[]{
        \label{fig:cooling_phot_decay}
        \includegraphics[scale=0.34]{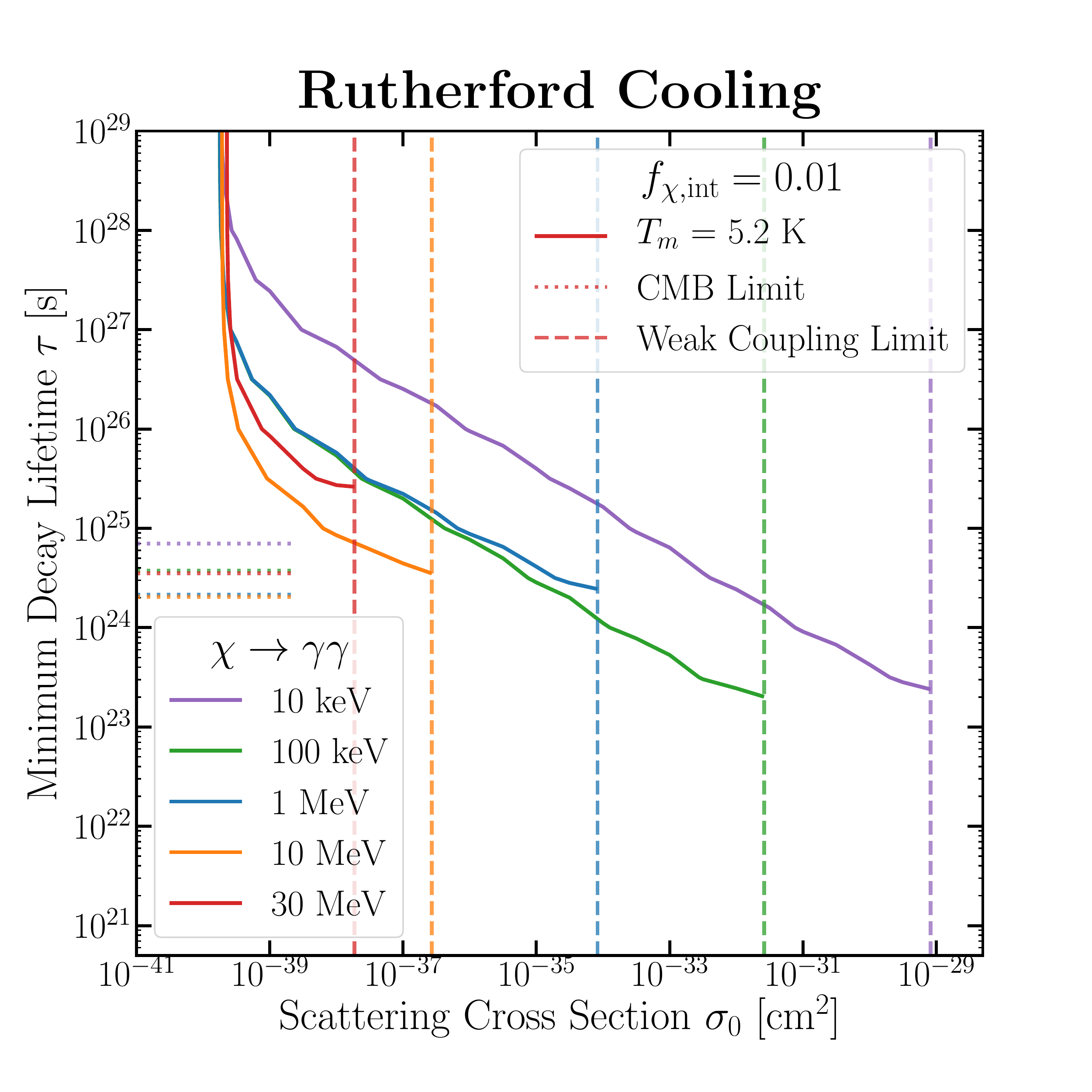}
    }
  \caption{Rutherford cooling constraints on the minimum decay lifetime for $\chi \to e^+e^-$ (left) and $\chi \to \gamma \gamma$ (right) from the matter temperature $T_m(z = 17.2) = $5.2 K (solid), $f_{\chi,\text{int}} = 0.01$. Limits from the Planck measurement of the CMB power spectrum are also shown up to $\sigma_0 = \sigma_{0,\text{td}}(z = 300)$ (dotted), together with the maximum scattering cross sections for the weak coupling limit to hold (dashed). 
  }
  \label{fig:cooling_decay}
\end{figure*}

Fig.~\ref{fig:cooling_decay} shows the constraints for DM decays to $e^+e^-$ and $\gamma \gamma$ respectively as a function of $\sigma_0$ for DM-hydrogen scattering in the weak coupling limit (set by the dashed lines), for the case where $f_{\chi,\text{int}} = 0.01$. The CMB power spectrum constraints are shown up to $\sigma_0 = \sigma_{0,\text{td}}$, after which the constraints may not be applicable. A minimum value of $\sigma_0 \sim 10^{-40} \text{ cm}^2$ is required for sufficient cooling to bring $T_m$ down to 5.2 K, absent any additional heat source. Over a large range of $\sigma_0$, the temperature constraint set by the EDGES $21$-cm measurement is more constraining than the CMB limits for parts of parameter space. For 10 - 100 keV DM decaying to photons, thermal decoupling as given in Eq.~(\ref{eqn:thermal_decoupling_condition}) occurs earlier than $z \sim 300$ even in the weak coupling regime, and at large scattering cross sections, only the temperature measurement can effectively constrain the decay lifetime.

\begin{figure*}
    \subfloat[]{
        \label{fig:cooling_elec_swave}
        \includegraphics[scale=0.34]{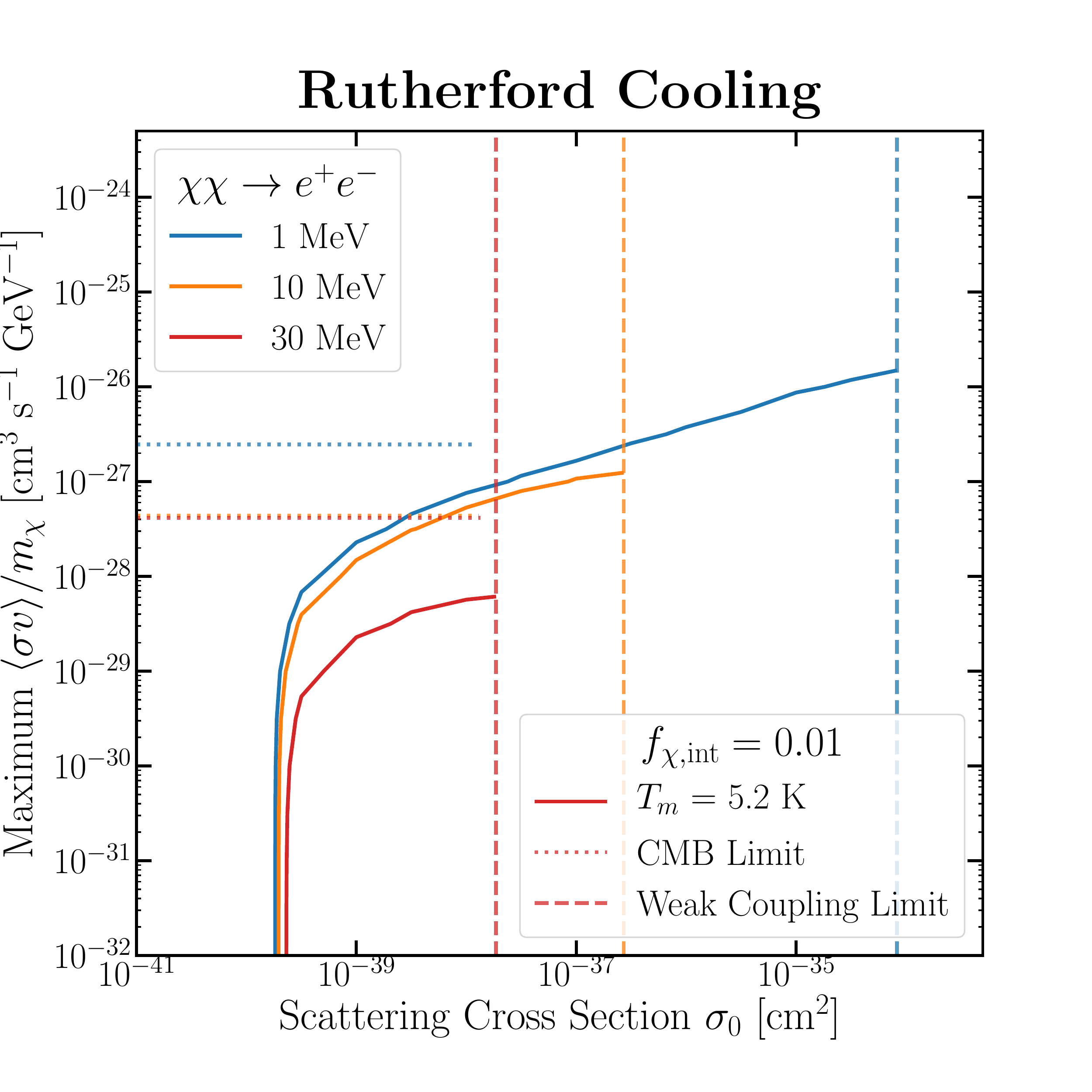}
    }
    \hfil
    \subfloat[]{
        \label{fig:cooling_phot_swave}
        \includegraphics[scale=0.34]{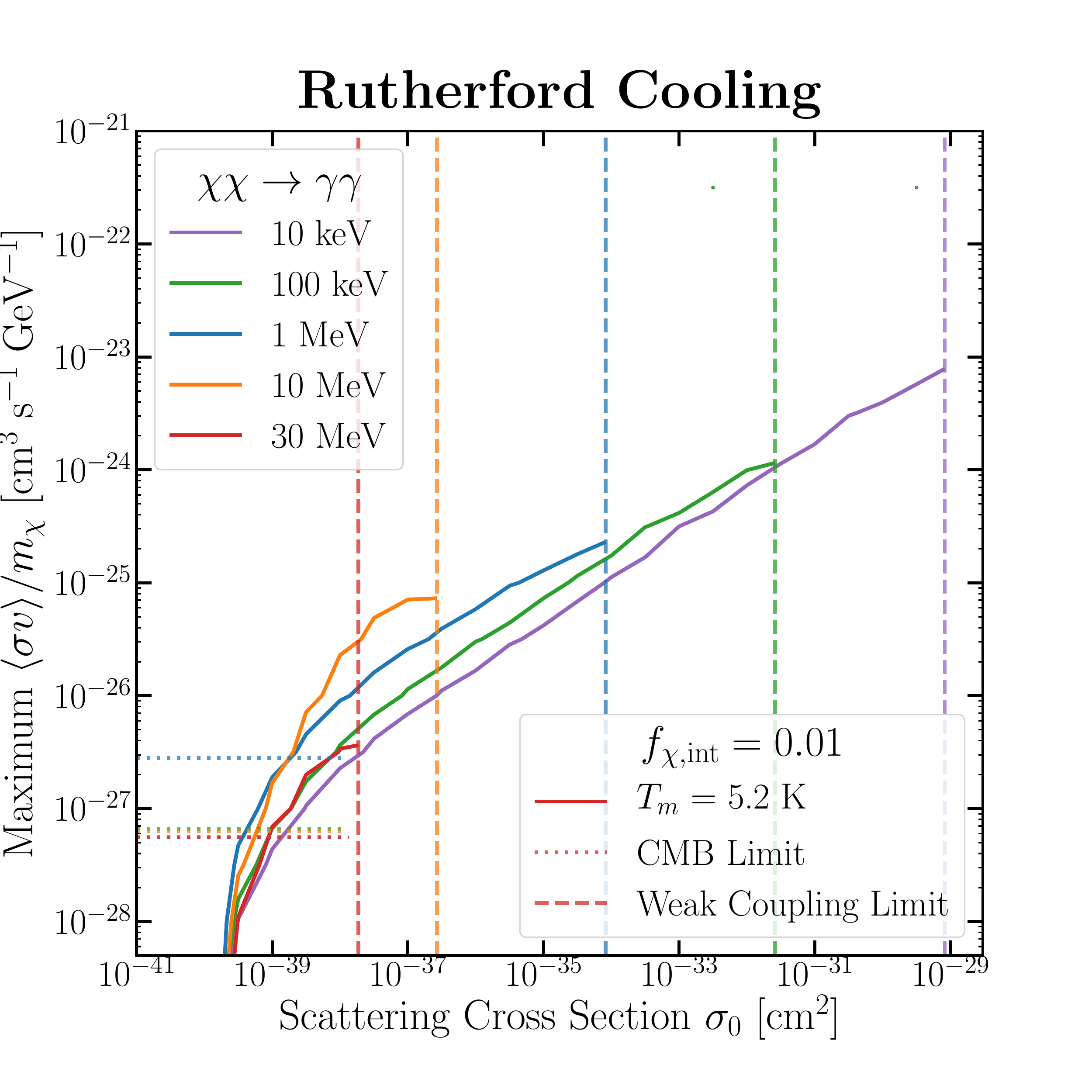}
    }
  \caption{Rutherford cooling $s$-wave annihilation constraints for $\chi \chi \to e^+e^-$ (left) and $\chi \chi \to \gamma \gamma$ (right) from the matter temperature $T_m(z = 17.2) = $ 5.2 K (solid), $f_{\chi,\text{int}} = 0.01$. Limits from the Planck measurement of the CMB power spectrum are also shown up to $\sigma_0 = \sigma_{0,\text{td}}(z = 600)$ (dotted), together with the maximum scattering cross section for the weak coupling limit to hold (dashed). 
  }
  \label{fig:cooling_swave}
\end{figure*}

Fig~\ref{fig:cooling_swave} shows similar constraints on the $s$-wave annihilation cross section. The temperature limits in both cases are relatively insensitive to the actual value of $T_m$ at $z \sim 20$.  The exact value of $T_m$ sets the minimum scattering cross section for cooling with no energy injection, but at higher cross sections/shorter decay lifetimes, the constraints are essentially set by having the large amount of heating almost entirely cancelled by Rutherford cooling.

Analogous plots for the case where $f_{\chi,\text{int}} = 1$ are shown in Figs.~\ref{fig:cooling_decay_f_1} and~\ref{fig:cooling_swave_f_1} in Appendix~\ref{app:supplemental_plots}.

\subsection{Strong Coupling Results}

The ionization history with both cooling and DM annihilation in the strong-coupling limit, on the other hand, exhibits important differences from the history with no cooling, especially when the interacting component of DM is light. Since the transfer of energy between baryons and the interacting DM is efficient, Compton heating from the CMB must also be able to efficiently heat all of the interacting DM particles in order to keep the matter temperature at the CMB temperature. The additional heating needed means that thermal decoupling between the CMB and DM can occur at a much higher redshift, if the DM mass is sufficiently light. After decoupling, since $T_m = T_\chi$, Eq.~(\ref{eqn:T_m_cooling}) shows that both of these temperatures simply evolve through adiabatic cooling in the absence of DM energy injections. The strong-coupling limit therefore reduces to a non-standard recombination history with early thermal decoupling, discussed in Sec.~\ref{sec:non_standard_recombination}. 

We can obtain the redshift of thermal decoupling between photons and the coupled baryon-DM fluid by replacing the Compton heating term in Eq.~(\ref{eqn:Tm}) and~(\ref{eqn:compton_rate}) by
 \begin{alignat}{1}
     \Gamma_C \to \Gamma_C \frac{n_\text{H}}{n_\text{H} + n_\chi},
 \end{alignat}
since energy from Compton heating must be redistributed into the dark sector as well. This gives
\begin{alignat}{1}
    (1 + z)_\text{td}^\text{strong} &\approx \left[\frac{45 m_e H_0 \sqrt{\Omega_m}}{4 \pi^2 \sigma_T x_e T_{\gamma,0}^4} \left(1 + \frac{f_{\chi,\text{int}} \rho_\chi}{n_\text{H} m_\chi}\right)\right]^{2/5} \nonumber \\
    &\approx 155 \left(1 + 5 f_{\chi,\text{int}} \frac{m_p}{m_\chi}\right)^{2/5},
\end{alignat}
with the redshift of thermal decoupling being independent of the scattering cross section. Note that we limit $(1+z)_\text{td}^\text{strong}$ to a maximum value of 1090, corresponding to the redshift of recombination, since thermal decoupling cannot occur before that, owing to the strong coupling between the fully ionized plasma and the CMB. In the limit of strong coupling, the exact details of how this coupling comes about is not important in determining the thermal and ionization history of the baryons.

For a given $\langle \sigma v \rangle$, the heating rate of baryons in the presence of a strongly-coupled DM is less than without DM, since some amount of the heating is transferred to the dark sector. If $x_e$ is small, we can account for this difference by replacing
\begin{alignat}{1}
    \langle \sigma v \rangle \to \frac{n_\text{H}}{n_\text{H} + n_\chi} \langle \sigma v \rangle,
\end{alignat}
and similarly for $1/\tau$ with decays. The constraints for the strong coupling limit can be easily determined from the non-standard recombination constraints: if $\langle \sigma v \rangle_{\max,(1+z)_\text{td}}$ is the maximum annihilation cross section from early thermal decoupling at redshift $(1+z)_\text{td}$, then the corresponding constraint from the strong coupling limit is
\begin{alignat}{1}
    \langle \sigma v \rangle_{\max, \text{strong}} &= \left(1 + \frac{f_{\chi,\text{int}} \rho_\chi}{n_\text{H} m_\chi} \right) \langle \sigma v \rangle_{\max,(1+z)_\text{td}^\text{strong}} \nonumber \\
    &\approx \left(1 + 5f_{\chi,\text{int}} \frac{m_p}{m_\chi}\right)\langle \sigma v \rangle_{\max,(1+z)_\text{td}^\text{strong}}.
\end{alignat}
This has been explicitly checked by directly solving Eqs.~(\ref{eqn:T_chi_cooling}) and~(\ref{eqn:cooling_full_eqns}) with $V_{\chi b} = 0$. 

\subsection{Millicharged DM}

We now turn our attention to the millicharged DM model discussed in \cite{Munoz:2018pzp,Berlin:2018sjs}, focusing on the case where $f_{\chi,\text{int}} = 0.01$, which evades the DM-baryon scattering CMB limits. 

 \begin{figure*}[t!]
    \subfloat[]{
        \label{fig:millicharged_elec_decay}
        \includegraphics[scale=0.34]{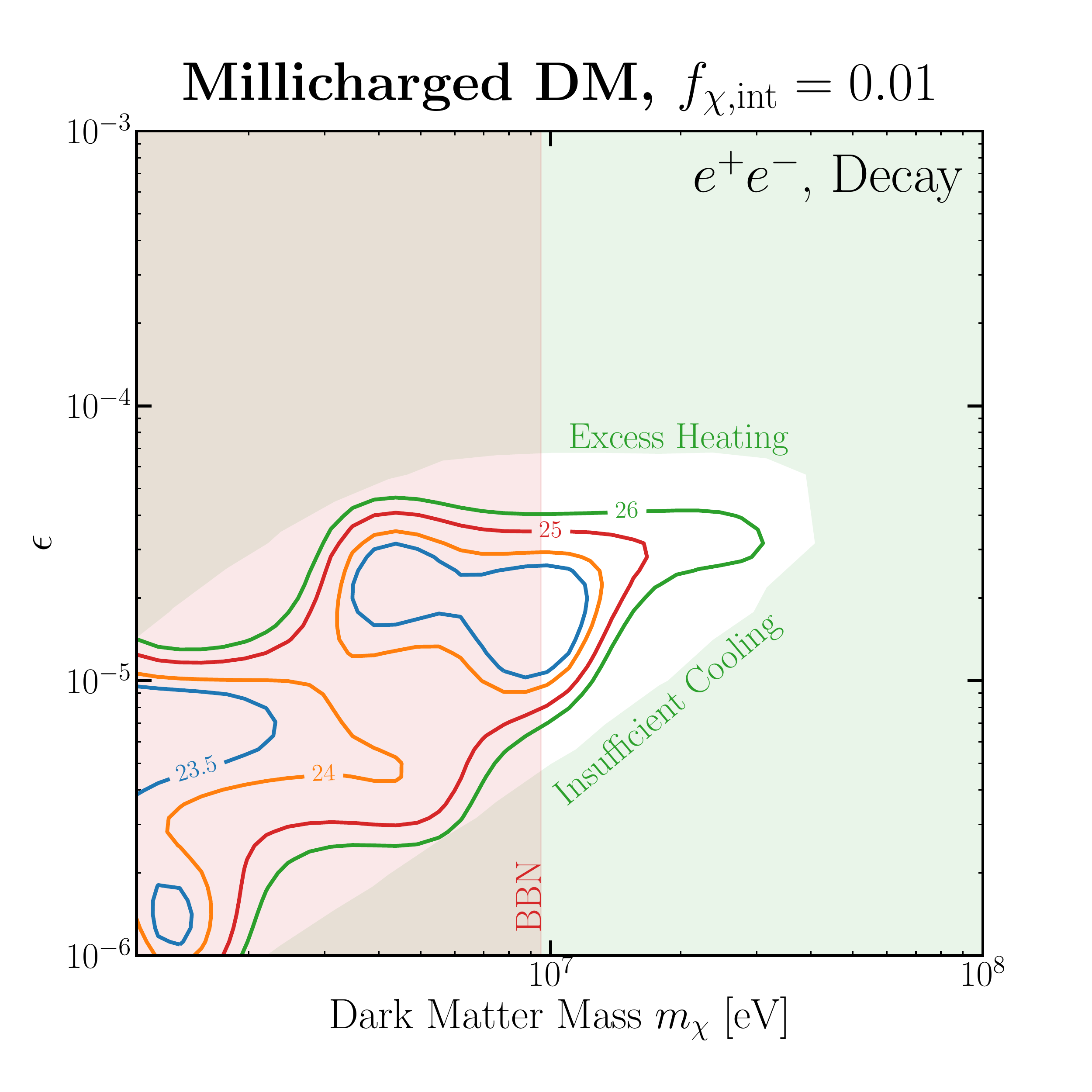}
    }
    \hfil
    \subfloat[]{
        \label{fig:millicharged_phot_decay}
        \includegraphics[scale=0.34]{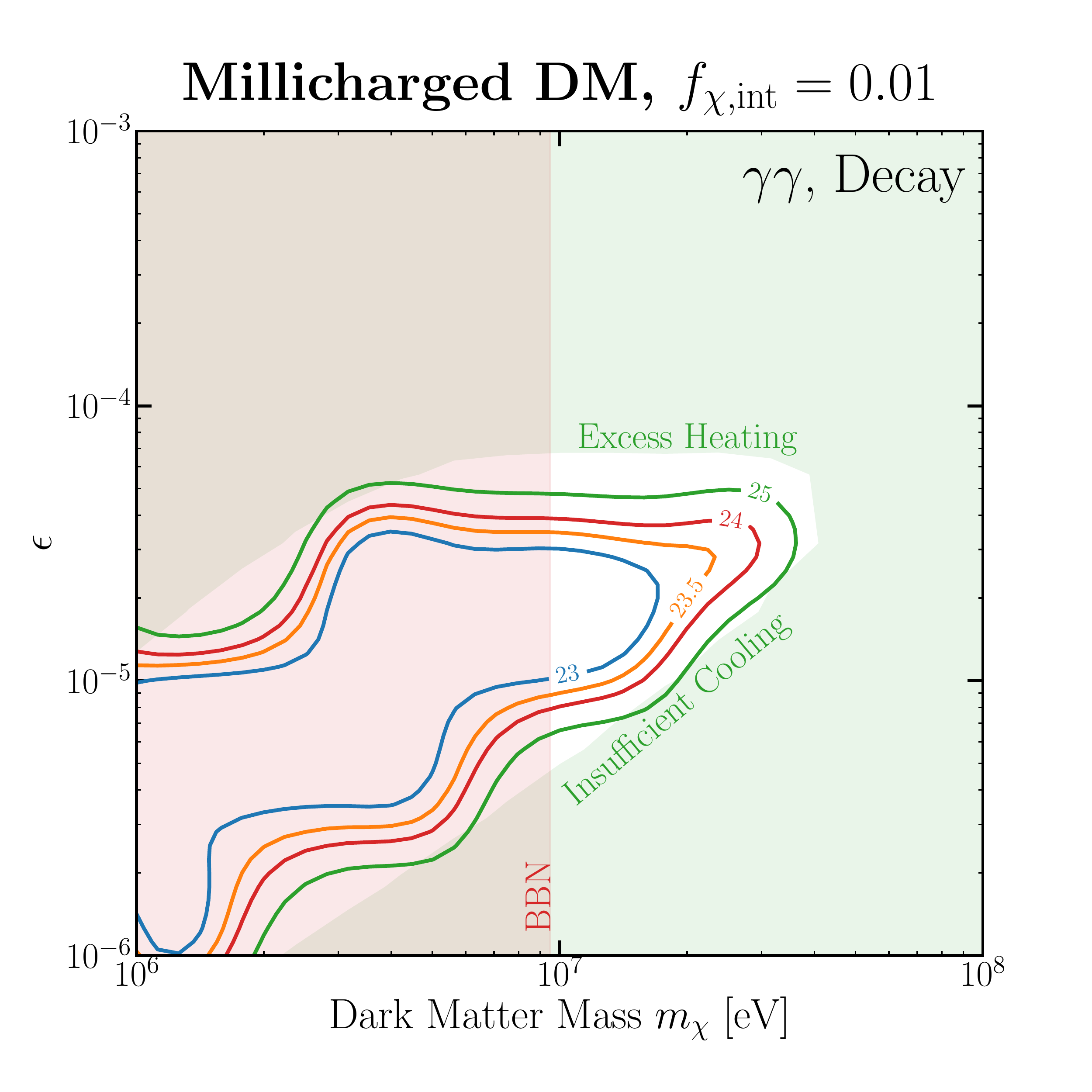}
    } \\
    \subfloat[]{
        \label{fig:millicharged_elec_swave}
        \includegraphics[scale=0.34]{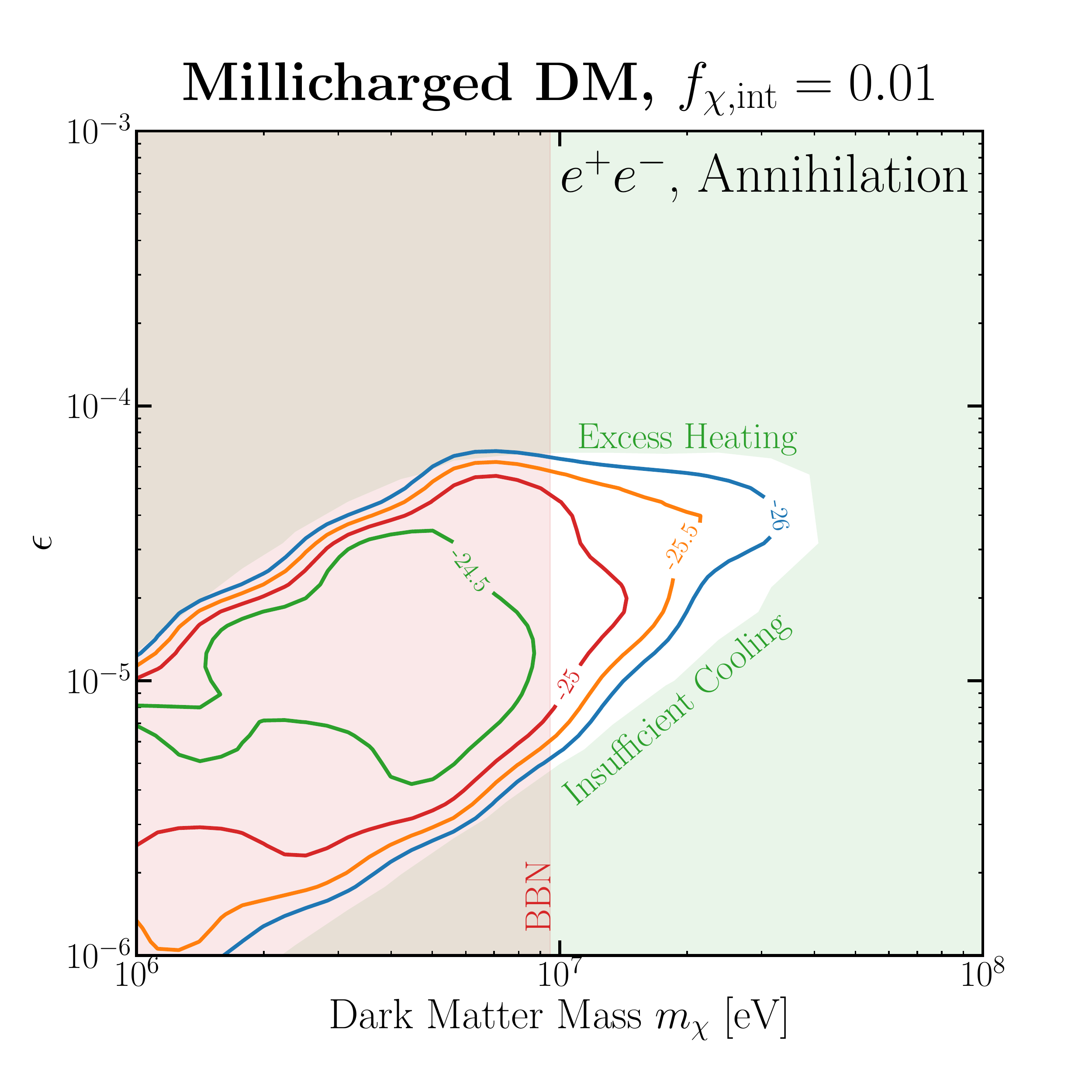}
    }
    \hfil
    \subfloat[]{
        \label{fig:millicharged_phot_swave}
        \includegraphics[scale=0.34]{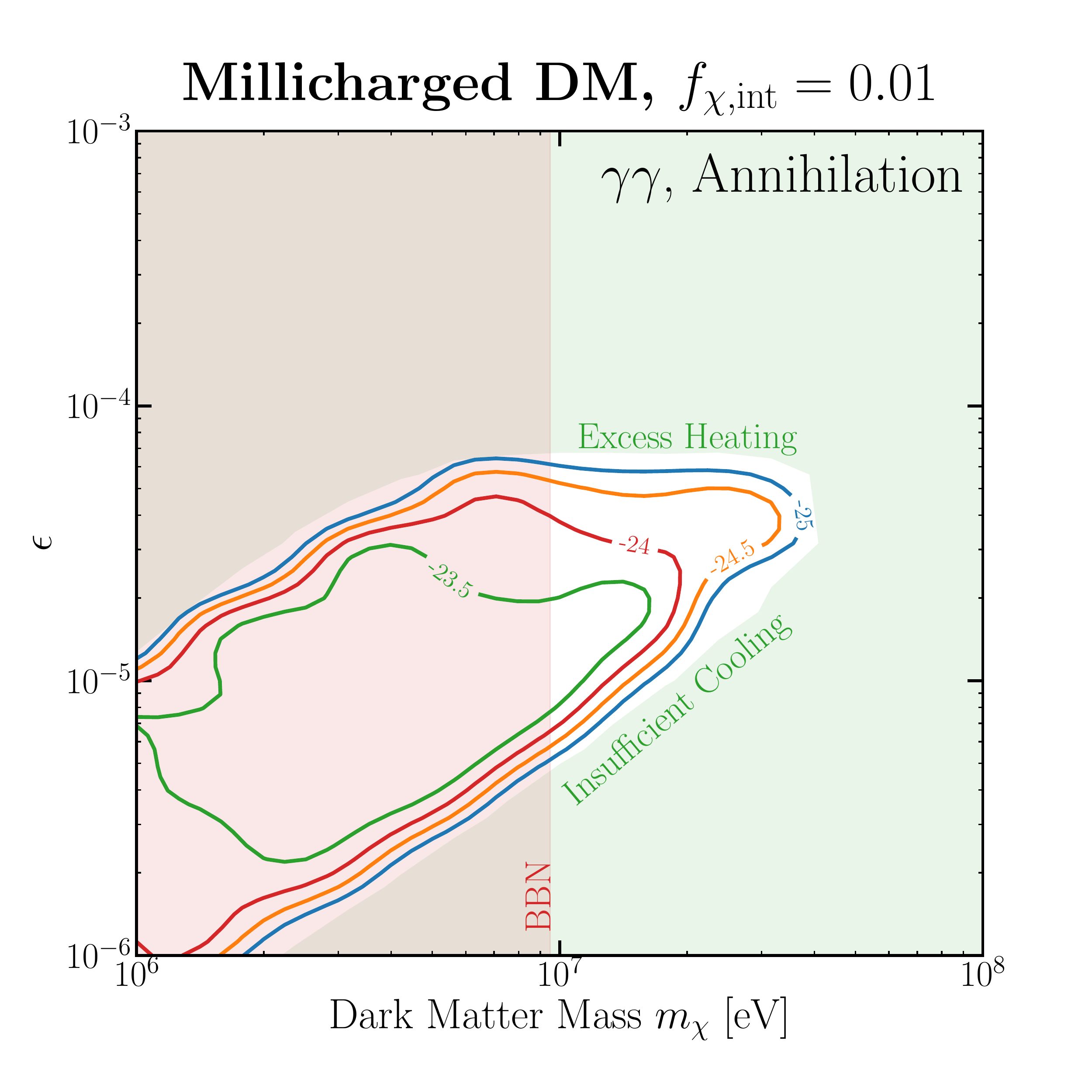}
    }
  \caption{Constraints on the millicharged DM, with an additional source of DM decay (upper panels) or annihilation (lower panels) to $e^+e^-$ (left panels) and $\gamma \gamma$ (right panels), with $f_{\chi,\text{dec}}$ and $f_{\chi,\text{ann}} = 0.01$ respectively. The region of parameter space ruled out by BBN (red) is shown. Charges $\epsilon$ that are not large enough for efficient cooling of baryons (green, below) or so large that excess heating occurs at $z \sim 20$ are excluded. Contours of constant minimum $\log_{10} \tau$ in seconds for decay and maximum $\log_{10} \langle \sigma v \rangle$ in $\text{cm}^3 \text{ s}^{-1}$ so that $T_m(z = 17.2) < 4$ K are drawn.
  }
  \label{fig:millicharged}
\end{figure*}

For DM masses of interest ($\gtrsim$ 1 MeV), a symmetric, Dirac fermion\footnote{The cross section for annihilation of complex scalars to $e^+e^-$ pairs is $p$-wave suppressed, and while $p$-wave annihilation also leads to significant heating at $z \sim 20$, we defer a proper treatment of this process to future work.} millicharged DM has an unavoidable $s$-wave annihilation channel into $e^+e^-$, with a velocity-averaged cross section given by
 \begin{alignat}{1}
     \langle \sigma v \rangle = \frac{\pi \alpha_\text{EM}^2 \epsilon^2}{m_\chi^2} \sqrt{1 - \frac{m_e^2}{m_\chi^2}} \left(1 + \frac{m_e^2}{2m_\chi^2}\right),
 \end{alignat}
 where $\epsilon$ is the charge of the millicharged DM, and $\alpha_{\text{EM}}$ is the electromagnetic fine-structure constant. Raising $\epsilon$ therefore both increases the rate of DM-baryon scattering and the rate of DM annihilation to electrons, and the opposing effects on $T_m$ should be properly taken into account when considering the viability of this model. Since annihilation takes place between Dirac millicharged particles, the annihilation rate given in Eq.~(\ref{eqn:energy_injection}) must have the additional factor of 1/2.

Fig.~\ref{fig:millicharged} show a plot of the $m_\chi - \epsilon$ parameter space of this model with several relevant constraints. Sufficiently light millicharged DM remains in thermal equilibrium with electrons and photons during Big Bang nucleosynthesis (BBN) \cite{Berlin:2018sjs}, and is excluded by the known light elemental abundances. Since the irreducible annihilation to $e^+e^-$ heats the baryons, by requiring $T_m(z = 17.2) \leq \SI{4}{K}$,\footnote{We choose 4 K in this section for consistency with existing results in the literature.} we can set an upper limit of $\epsilon \lesssim 5 \times 10^{-5}$, cutting the remaining parameter space down to a narrow window between $m_\chi \sim 10 - 100$ MeV and $\epsilon \sim 5 \times 10^{-6}$ to $5 \times 10^{-5}$. These limits are stronger than the conventional CMB power spectrum limits, since at large values of $\epsilon$, DM and baryons become strongly coupled early on, and the temperature evolution is mostly dominated by adiabatic cooling until structure formation; once structure formation starts, a small perturbation on the order of a few kelvins from millicharged DM annihilation to $e^+e^-$ is all that is required to raise the temperature above the EDGES measurement. Other experimental constraints set by the SLAC millicharge experiment \cite{Prinz:1998ua} and observations of the cooling of SN1987a \cite{Chang:2018rso} set limits that are already ruled out by a combination of the two limits shown, and have been left out.

We now consider an additional source of DM-related energy injection through $s$-wave annihilation or decay. This need not come from annihilation or decay of the millicharged DM itself; in principle, other particles in the dark sector could contribute such an energy injection. However, the existence of an additional annihilation channel for the millicharged DM could potentially allow it to obtain its correct relic abundance through thermal freezeout, since the cross section of the irreducible annihilation to $e^+e^-$ is too small in the allowed region for this to happen.

We set $f_{\chi,\text{ann}} = 0.01$ in Eq.~\ref{eqn:energy_injection} by convention for both annihilation and decay when discussing this new source of energy injection: the result for other values of $f_{\chi,\text{ann}}$ can be obtained by simple rescaling. For $s$-wave annihilation, we find that $\langle \sigma v \rangle \lesssim 2 \times 10^{-25} \text{ cm}^3 \text{ s}^{ -1}$ for $\chi \overline{\chi} \to e^+e^-$ and $\langle \sigma v \rangle \lesssim 7 \times 10^{-24} \text{ cm}^3 \text{ s}^{ -1}$ for $\chi \overline{\chi} \to \gamma \gamma$. Since the cross section to produce the correct relic abundance of the millicharged DM with $f_{\chi,\text{int}} = 0.01$ is $\langle \sigma v \rangle \sim 6 \times 10^{-24} \text{ cm}^3 \, \text{s}^{ -1}$, it is unlikely that any additional source of $s$-wave annihilation to $e^+e^-$ (on top of the irreducible $s$-channel annihilation through the SM photon) can produce the correct relic abundance while remaining consistent with the EDGES $T_m$ measurement at $z \sim 20$. There is a small parameter space allowed for annihilation to photons to get the correct relic abundance without late-time suppression, but this requires a small branching ratio to electrons at the same time. 

\section{Conclusion}
\label{sec:conclusion}

We have computed the constraints that can be set on annihilating/decaying DM by a measurement of the 21-cm line of neutral hydrogen from the end of the cosmic dark ages. The recent claimed observation of an absorption trough by EDGES motivates the inclusion of some mechanism beyond the simplest scenario to explain the unexpectedly low inferred gas temperature; however, even if a future experiment found a weaker absorption signal, such additional mechanisms could still potentially be present and should be included to obtain conservative constraints.

We have considered three general scenarios that could weaken constraints from 21-cm observations on exotic energy injection from heating in the cosmic dark ages: (1) additional radiation backgrounds in the frequency range surrounding 21-cm, (2) non-standard recombination allowing the gas to decouple thermally from the CMB earlier, and (3) cooling of the gas through DM-baryon scattering. We have demonstrated that the strong-coupling limit of scenario (3) implements scenario (2) as a corollary, and that scenario (3) can generically weaken previously studied constraints on exotic energy injections from modifications to the ionization history during the cosmic dark ages.

We have mapped out the constraints on DM annihilation/decay in these three scenarios. We have found that in cases (2) and (3), there is an asymptotic behavior where the constraints become nearly independent of the redshift of decoupling (in case (2)) or the interaction cross section (in case (3)) for sufficiently early decoupling/large cross sections (see Fig.~\ref{fig:recomb_decay} and~\ref{fig:recomb_swave}). In these scenarios, we can thus present robust constraints that do not depend on the exact redshift of decoupling in case (2) or the size of the cross section in case (3). 

In the case where a small fraction of light DM (below 100 MeV) is millicharged and scatterings on this component are responsible for cooling of the gas, we have demonstrated that if this component has additional annihilation channels sufficient to obtain its relic density through thermal freezeout, then the energy injection from those channels will generically overheat the gas. Thus such a component would likely need to possess a non-thermal origin, or if a thermal relic, have annihilation channels in the early universe that are suppressed at late times (or have a large branching ratio for annihilation directly to neutrinos).

\begin{figure*}[hp]
    \subfloat[]{
        \label{fig:summary_elec_decay}
        \includegraphics[scale=0.34]{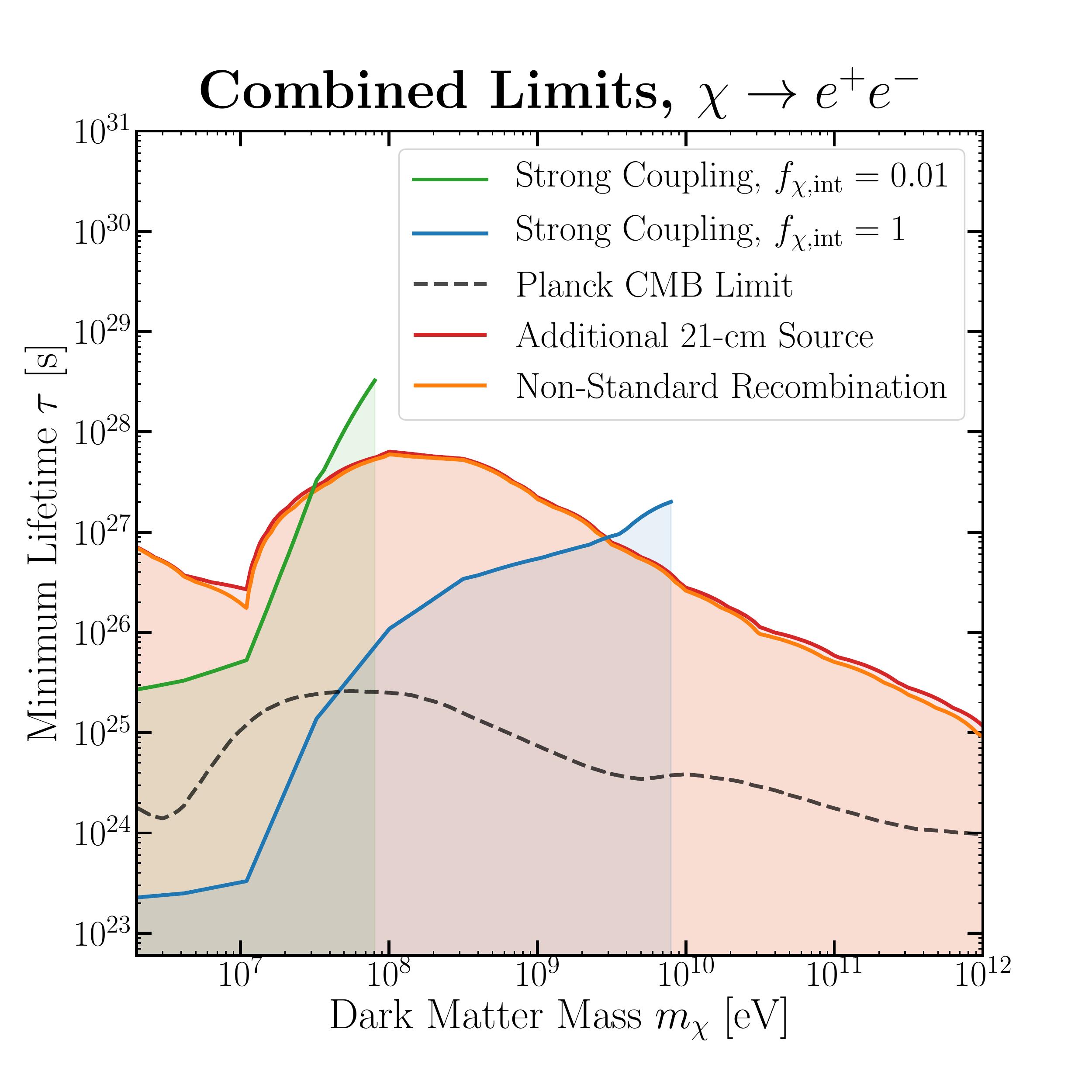}
    }
    \hfil
    \subfloat[]{
        \label{fig:summary_phot_decay}
        \includegraphics[scale=0.34]{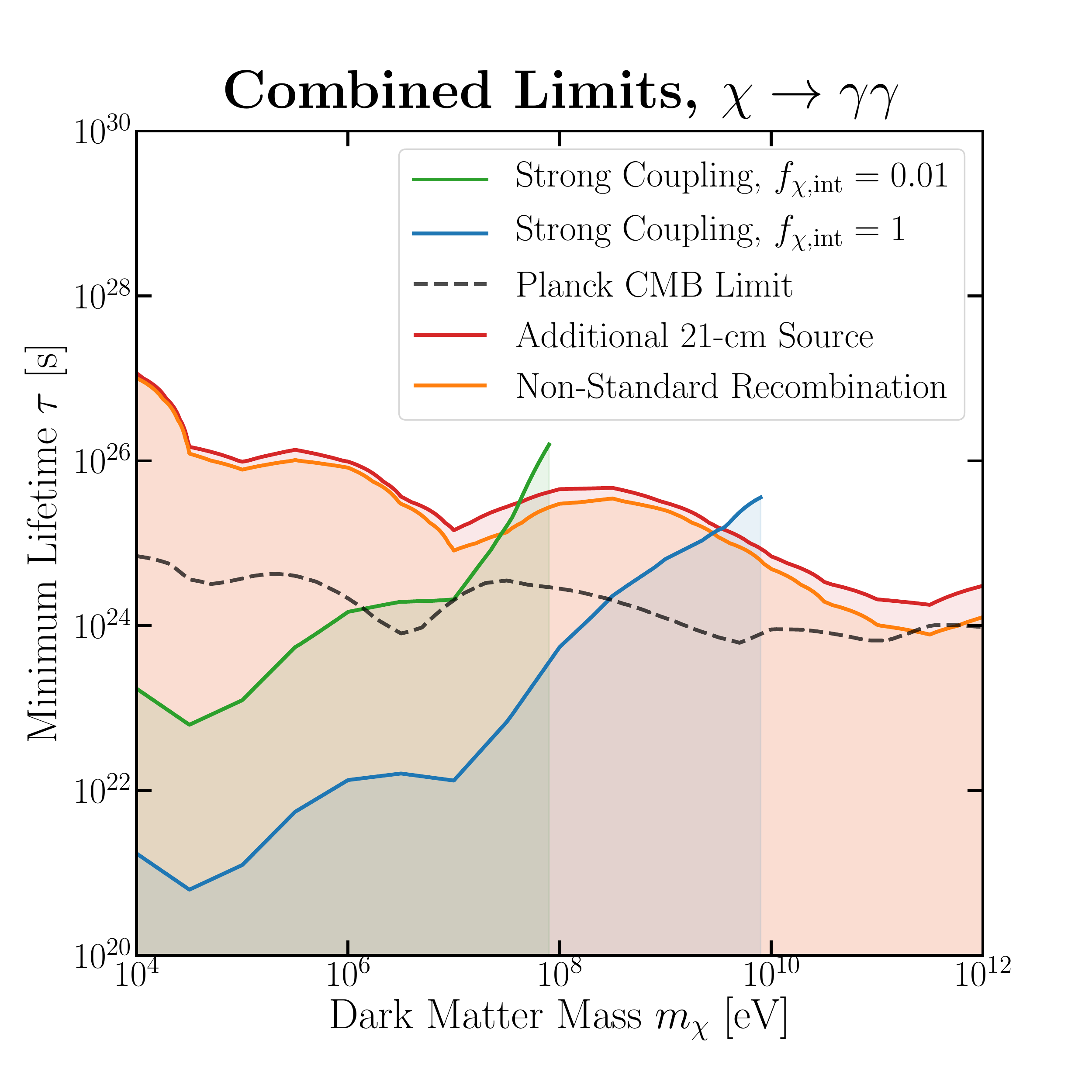}
    } \\
    \subfloat[]{
        \label{fig:summary_elec_swave}
        \includegraphics[scale=0.34]{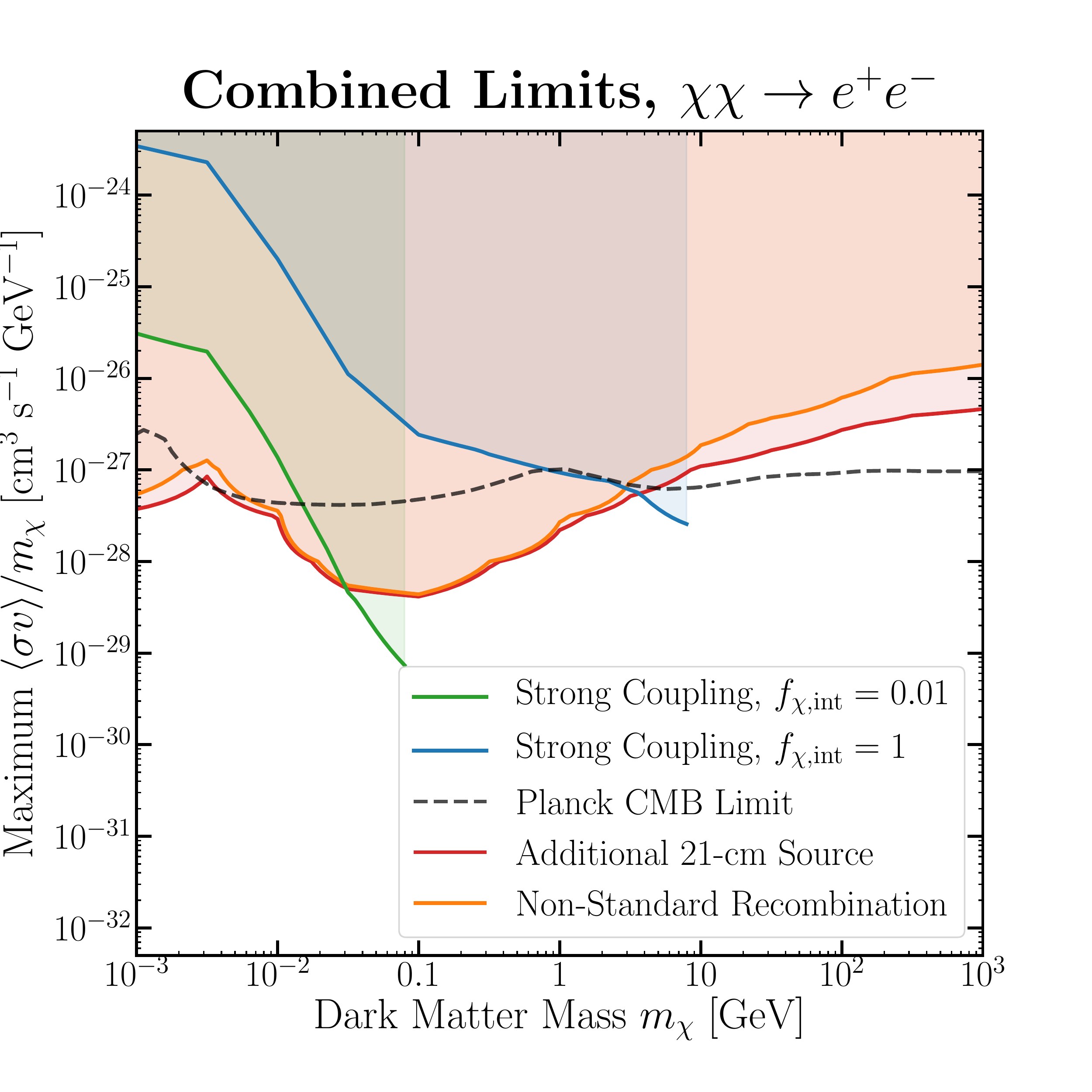}
    }
    \hfil
    \subfloat[]{
        \label{fig:summary_phot_swave}
        \includegraphics[scale=0.34]{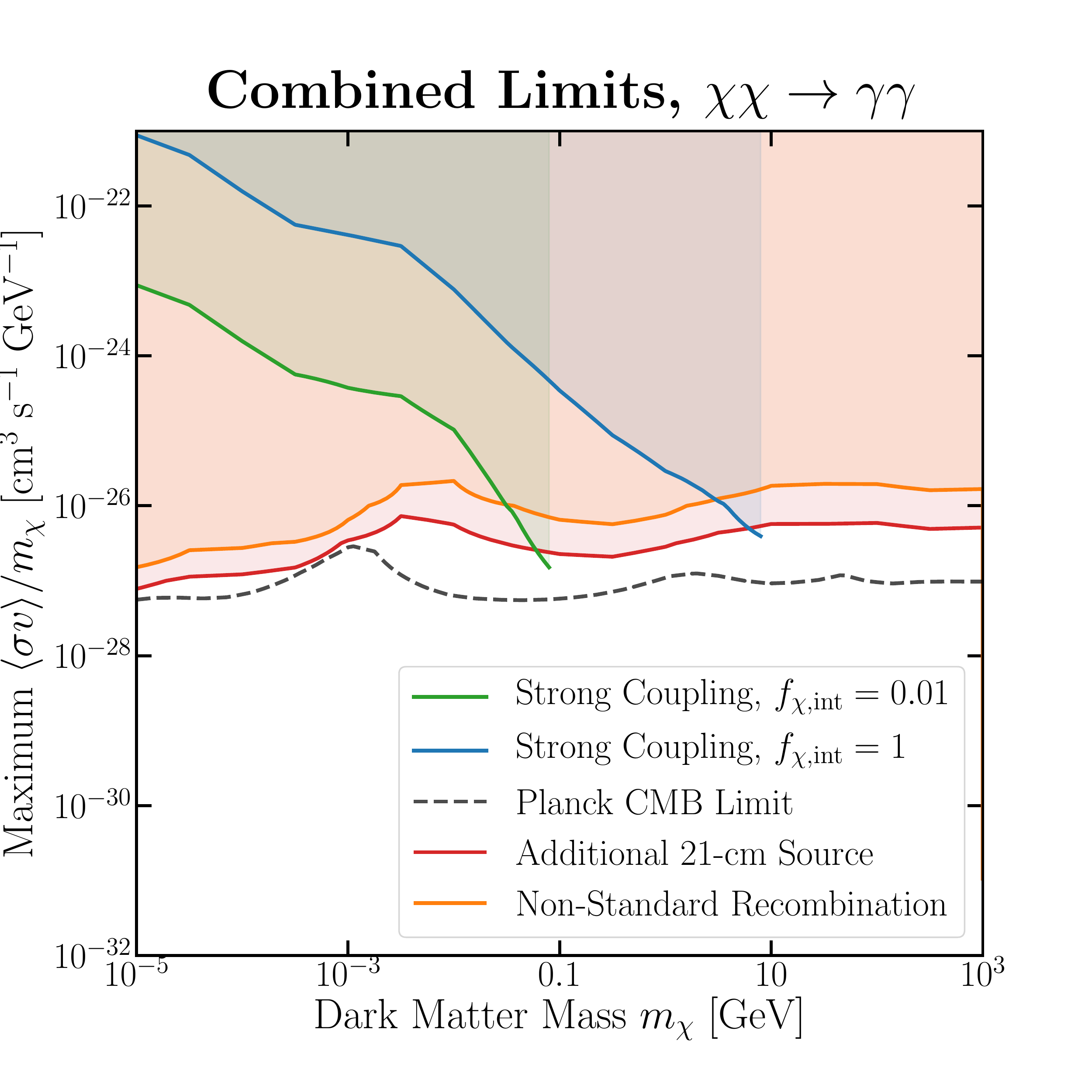}
    }
  \caption{Lower limits on the DM decay lifetime (upper panels) and upper limits of annihilation cross section (lower panels) from requiring $T_m/T_R(z=17.2) \le 0.105$ \cite{Bowman:2018yin}, for decay to $e^+e^-$ pairs (left panels) and photons (right panels). In the presence of an additional 21-cm radiation source with number density (at that frequency) smaller than or equal to that of the CMB number density, constraints are shown by the red solid line. In the limit of early baryon-photon decoupling, constraints are shown by the orange solid line. The solid green and solid blue lines  represent the constraints in the presence of DM-baryon scattering, in the limit of large cross section, for respectively 1\% and 100\% of the DM interacting with the baryons (these mechanisms cannot sufficiently cool baryons to match the data above critical mass scales, represented by the vertical cutoffs on the right-hand-side of the green/blue regions). The black dashed line represents previously derived constraints on the decay lifetime  \cite{Slatyer:2016qyl} (upper panels) or annihilation cross section (lower panels) \cite{Slatyer:2015jla} from measurements of the CMB.
  }
  \label{fig:summary}
\end{figure*}

In Fig.~\ref{fig:summary} we summarize the constraints that can be obtained on keV-TeV DM annihilation or decay into $e^+ e^-$ pairs or photons, in these three scenarios, if the EDGES result is confirmed, and compare these limits with the Planck CMB constraints; other limits from indirect detection also exist for both channels, and may be more constraining at higher DM masses (e.g. \cite{Cohen:2016uyg,Fermi-LAT:2016uux}). These particles are the main stable, electromagnetically interacting byproducts of more general annihilation/decay channels (other than annihilation/decay directly to neutrinos), and consequently the constraints on more general channels can be estimated by combining these results. To set a limit in the case of additional radiation backgrounds, we assume that the effective radiation temperature $T_R$ at the 21-cm wavelength is not more than twice the temperature of the CMB at $z = 17.2$. 

In the case of DM-baryon scattering (scenario (3)), the cooling is only sufficient to reduce the gas temperature below 5.2K if the DM mass is below a certain critical scale (depending on the fraction of the DM that is interacting); consequently, the constraints cut off above a certain mass scale because even for zero energy injection from annihilation/decay, the proposed mechanism cannot explain the data. The other two scenarios are in principle viable at all DM mass scales. We find that in these scenarios, for decaying DM, these constraints would generically be stronger than previously derived early-universe bounds, and in the case of decay primarily to electrons (as is expected for sub-100-MeV DM), these limits are stronger by up to two orders of magnitude. For DM annihilating to electrons, the constraints in these scenarios are generally stronger than CMB-based limits for sub-GeV DM (without taking into account that the CMB constraints may weaken due to a modified ionization history).

Simultaneous with and slightly after the release of this work, several other authors also studied the sensitivity of 21-cm measurements to DM annihilation or decays~\cite{Cheung:2018vww,Clark:2018ghm,Mitridate:2018iag}. In particular, the authors of~\cite{Clark:2018ghm,Mitridate:2018iag} set decay lifetime limits in a similar manner as~\cite{DAmico:2018sxd}, assuming an absorption signal that is smaller than the EDGES signal, with either $T_{21} = \SI{100}{mK}$ or \SI{50}{mK}. We reiterate that these limits are equivalent to our additional 21-cm source limits, with $(T_S/T_R)_\text{obs}= 0.26$ and 0.41 respectively, and setting $T_R = T_\text{CMB}$ at $z = 17.2$. Our work is more general than these other studies as we consider new effects that must be present to account for the large negative value of $T_{21}$ for the EDGES measurement. Consequently, our results are not merely a sensitivity study, and are immediately applicable to the various scenarios that have been suggested to explain the claimed EDGES detection. Even if future 21-cm measurements report a less negative value of $T_{21}$, the effects that we study here could potentially be present and are important to consider in setting future limits on DM annihilation and decay.

\acknowledgments

We thank Rebecca Leane, Julian Mu\~{n}oz, Nicholas Rodd, Yotam Soreq, Jesse Thaler and Chih-Liang Wu for helpful discussions and feedback. We also thank the authors of~\cite{Venumadhav:2018uwn} for pointing out typographical errors and omissions. This work was supported by the Office of High Energy Physics of the U.S. Department of Energy under grant Contract Numbers DE-SC00012567 and DE-SC0013999, and by the MIT Research Support Committee.

\appendix

\section{Astrophysical Systematics}
\label{app:systematics}

\renewcommand{\thefigure}{A\arabic{figure}}

\setcounter{figure}{0}

\subsection{Uncertainties from annihilation in DM halos}

\begin{figure*}[t!]
    \subfloat[]{
        \label{fig:cooling_struct_form_sys_erfc}
        \includegraphics[scale=0.34]{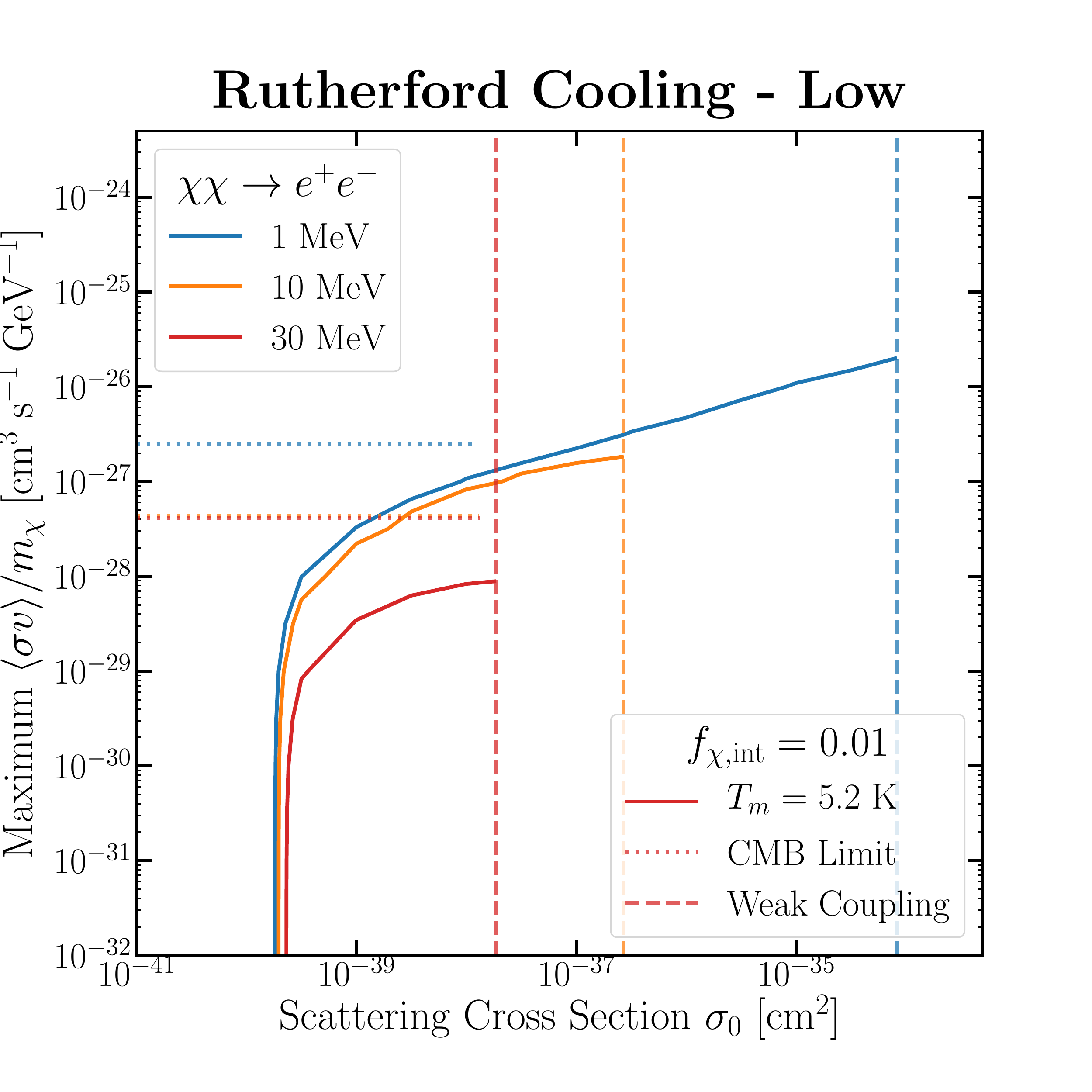}
    }
    \hfil
    \subfloat[]{
        \label{fig:cooling_struct_form_sys_NFW}
        \includegraphics[scale=0.34]{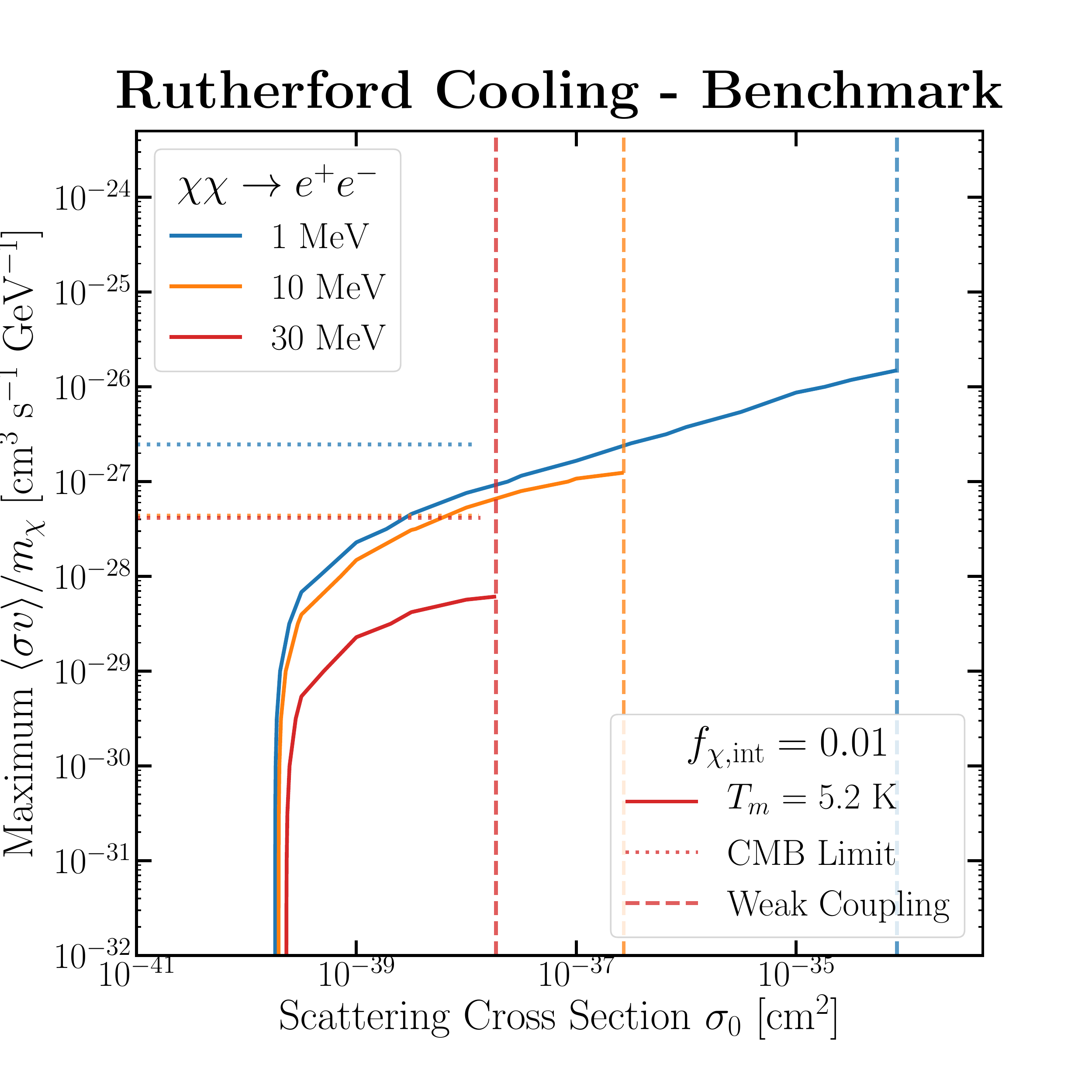}
    } \\
    \subfloat[]{
        \label{fig:cooling_struct_form_sys_einasto}
        \includegraphics[scale=0.34]{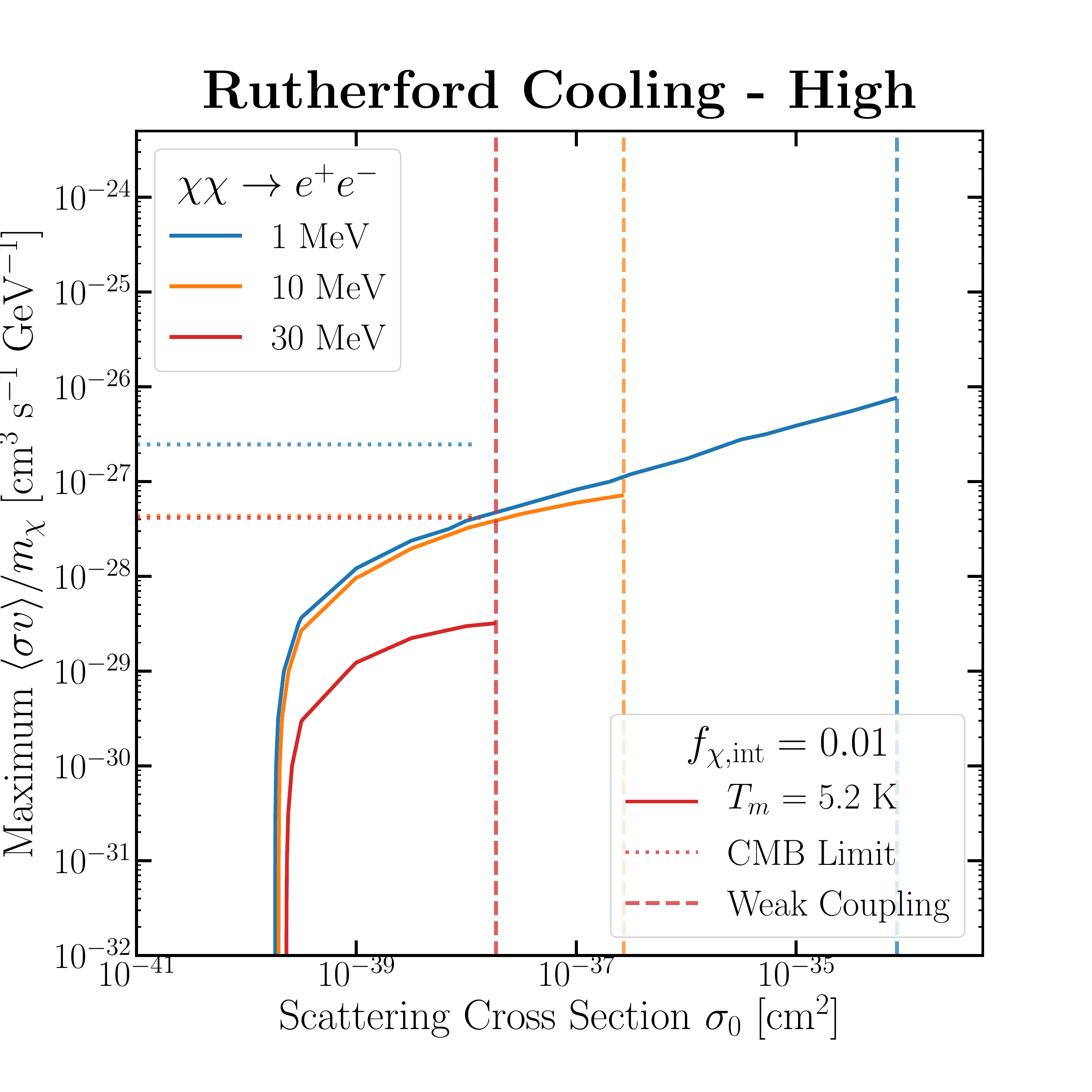}
    }
    \hfil
    \subfloat[]{
        \label{fig:cooling_struct_form_sys_no_boost
        }
        \includegraphics[scale=0.34]{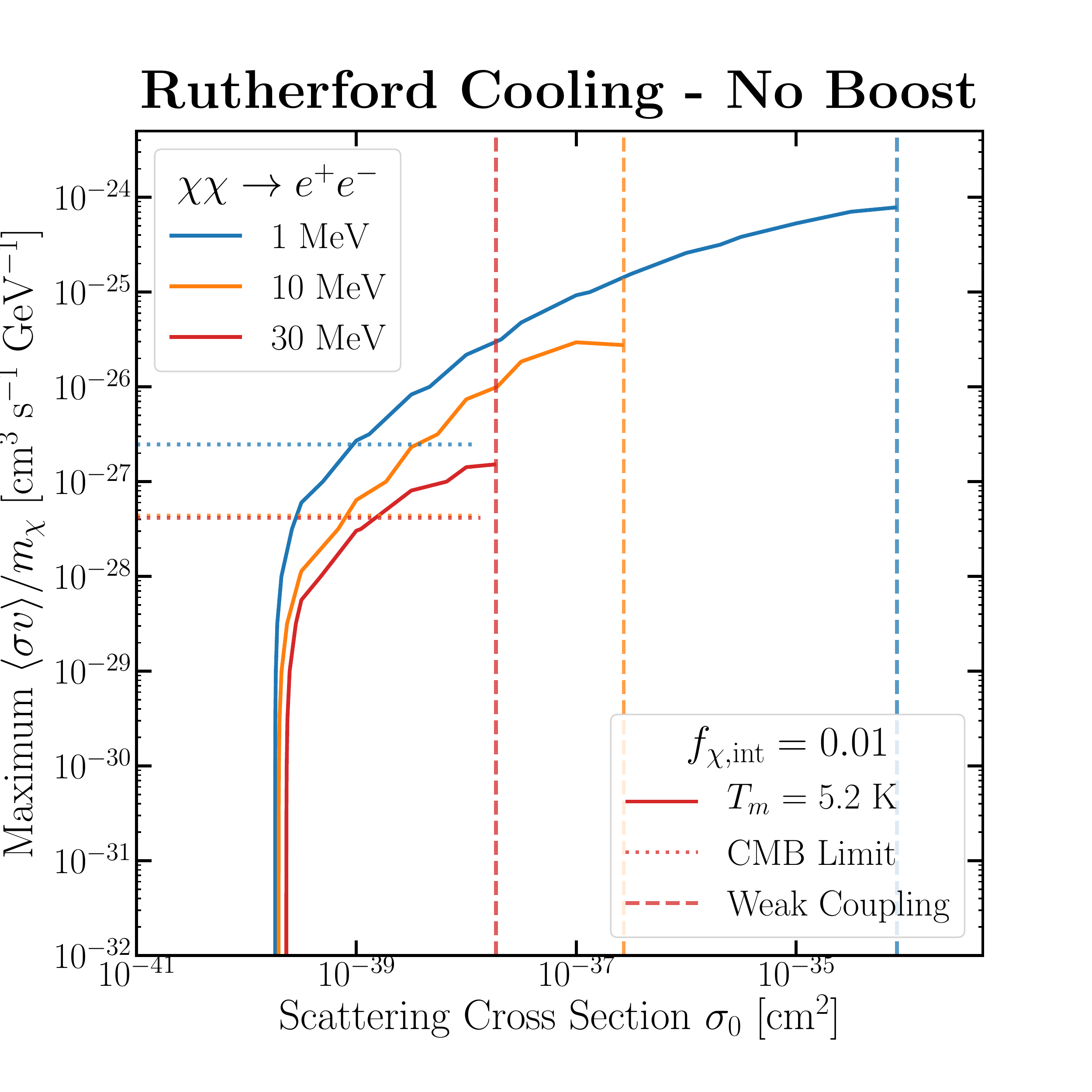}
    }
  \caption{Comparison of DM annihilation constraints when the gas is cooled by Rutherford scattering, where the HIGH (upper left), BENCHMARK (upper right) and LOW (lower left) and NO BOOST (lower right) models for the DM structure formation history are employed (see text for details).}
  \label{fig:cooling_swave_struct_form_sys}
\end{figure*}

\begin{figure*}[t!]
    \subfloat[]{
        \label{fig:cooling_V_pec_0}
        \includegraphics[scale=0.34]{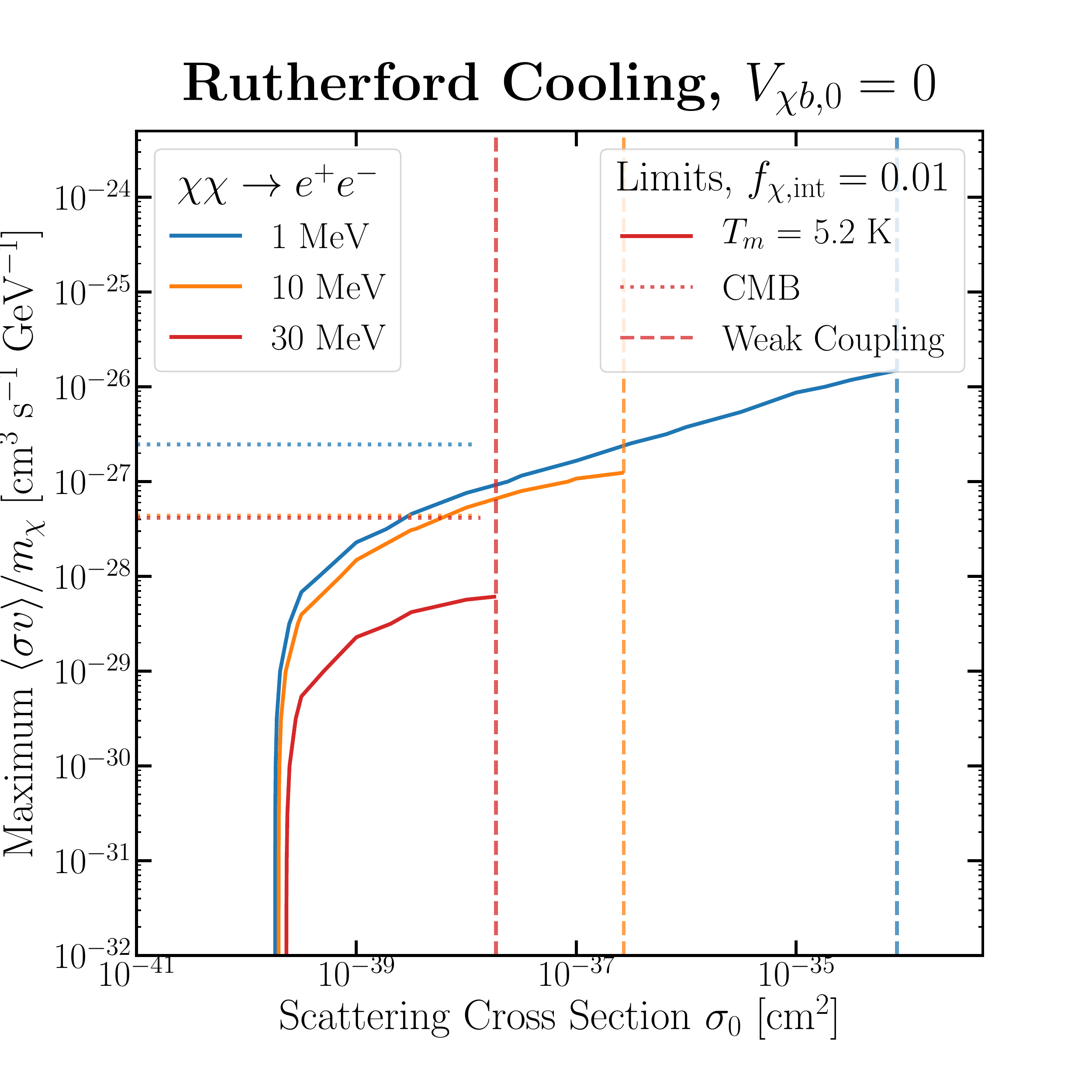}
    }
    \hfil
    \subfloat[]{
        \label{fig:cooling_V_pec_rms}
        \includegraphics[scale=0.34]{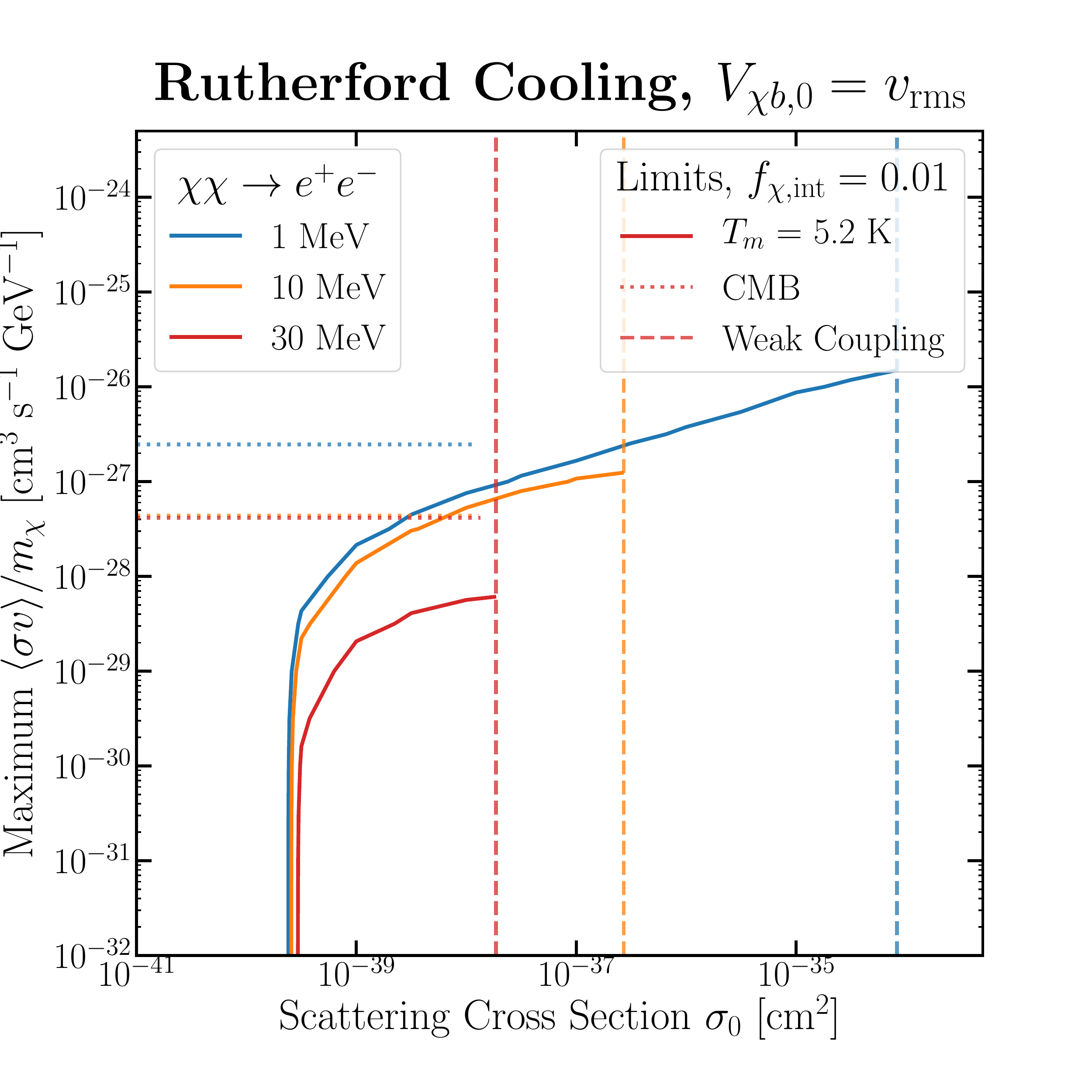}
    }
  \caption{Comparison of DM annihilation constraints when the gas is cooled by Rutherford scattering, where the bulk DM-baryon relative velocity at recombination is taken to be $V_{\chi b,0} = 0$ (left) or $V_{\chi b,0} = v_\text{rms}$ (right).}
  \label{fig:cooling_swave_vchib_init_sys}
\end{figure*}

Once structure formation begins, the DM annihilation rate is no longer set purely by the well-measured cosmological average DM density, but instead becomes dominated by annihilation in over-dense regions, which can enhance $\langle \rho^2\rangle$ greatly over $\langle \rho \rangle^2$. A model for this ``boost factor'' $\langle\rho^2(z)\rangle/\langle\rho(z)\rangle^2$ is included in the $f_c(z)$ factors that determine the amount of heating and ionization from $s$-wave DM annihilation (as defined in Eq.~(\ref{eqn:f_c_definition})), as discussed in \cite{Liu:2016cnk}.

On one hand, the size of this enhancement is quite uncertain, due mostly to large uncertainties in the abundance and concentration of low-mass halos that cannot be resolved by cosmological simulations. On the other hand, at $z\sim 17$ the enhancement factor is expected to still be relatively modest; furthermore the heating and ionization at that epoch are determined by the integral over DM annihilation at all previous times, not only at $z\sim 17$, which also dilutes the effect of late-time overdensities (e.g. \cite{Slatyer:2012yq}). This last effect is stronger for secondary particles that take longer to cool and deposit their energy; in particular, for most energies of injection, photons take longer to deposit their energy than electrons, and thus the systematic uncertainty in $f_c(z)$ due to structure formation is expected to be smaller (as the typical photon contributing to late-time ionization/heating was injected at an earlier epoch where structure formation was less important).

To quantitatively estimate the uncertainty in our quoted annihilation limits due to uncertainties in the contribution from DM overdensities, we repeat the analysis using three different models from the literature for the boost factor. The first two are limiting cases from \cite{Liu:2016cnk}: they correspond to (1) DM halos with Einasto density profiles \cite{Einasto}, and including an estimate of substructure within main halos, and (2) DM halos with Navarro-Frenk-White profiles \cite{Navarro:1995iw}, without substructure included. The second model, with the lowest boost factor of those considered in \cite{Liu:2016cnk}, is the benchmark we use for the plots in the main text. Finally, the third model is (3) a simple analytic form proposed as a conservative model for the boost factor by \cite{DAmico:2018sxd}. We label these models as HIGH, BENCHMARK and LOW respectively.

We show the limits on the annihilation cross section for DM annihilating to $e^+ e^-$ in the presence of Rutherford cooling, for these three models, in Fig.~\ref{fig:cooling_swave_struct_form_sys}. Since the boost factor is approximately degenerate with the annihilation cross section (changing it can also lead to a slight modification of the redshift dependence of the annihilation rate), we expect the changes in the limits on the cross section to be similar for the other scenarios (early decoupling and additional radiation). 

We find that, as expected, the constraints are more stringent for the HIGH model, by roughly a factor of 2. However, the BENCHMARK and LOW models agree closely, with the constraints differing on the $15 - 20$\% level. We have performed the same check for DM annihilating to photons, and the difference between the models is even smaller. Thus we expect our benchmark constraints to be similar to others set using a conservative structure formation model. 

If we completely ignore structure formation and consider only annihilations in the smooth DM density, then the constraints weaken considerably, by about a factor of 50, as shown in the fourth panel of Fig.~\ref{fig:cooling_swave_struct_form_sys}. This is the maximally conservative case, and is probably unrealistic; we leave a detailed study of the minimum possible boost factor to further work.

We note these uncertainties do not apply to DM decay, which probes the average DM density rather than the average DM density-squared.

\subsection{Uncertainties in the initial value of $V_{\chi b}$}

In the scenario where Rutherford scattering cools the gas, the scattering rate depends on the relative velocity between the DM and baryons, and hence on the initial value of the bulk relative velocity $V_{\chi b,0}$ at recombination. As argued in Sec.~\ref{sec:rutherford_cooling}, we expect the cooling effect to be strongest for $V_{\chi b,0} = 0$, thus leaving the maximum amount of room for heating from annihilation/decay products.

We test this hypothesis in Fig.~\ref{fig:cooling_swave_vchib_init_sys}, comparing constraints on DM annihilating to $e^+e^-$ for $V_{\chi b,0} = 0$ and $V_{\chi b,0} = v_\text{rms} = 29$ km s$^{-1}$, in the case where a 1\% fraction of the DM participates in the scattering. We find that once the scattering cross section $\sigma_0$ is well above the value required to cool the gas to 5.2 K, the constraints on annihilation are unaffected by this change in the initial conditions, because the large baryon-DM scattering cross section induces a drag force that drives $V_{\chi b}$ to zero (for the interacting DM component). However, the minimum $\sigma_0$ needed to cool the baryons to that temperature does increase modestly (by a few tens of percent) when $V_{\chi b,0}=v_\text{rms}$.

Accordingly we conclude that in the regime where the constraints are not very rapidly varying as a function of $\sigma_0$, away from the minimum $\sigma_0$ needed to achieve the required cooling, the systematic error due to neglecting the distribution of initial relative velocities $V_{\chi b}$ is small.

\section{Supplemental Plots}
\label{app:supplemental_plots}

\renewcommand{\thefigure}{B\arabic{figure}}

\setcounter{figure}{0}

\begin{figure*}[b]
    \subfloat[]{
        \label{fig:source_elec_decay_zoom}
        \includegraphics[scale=0.34]{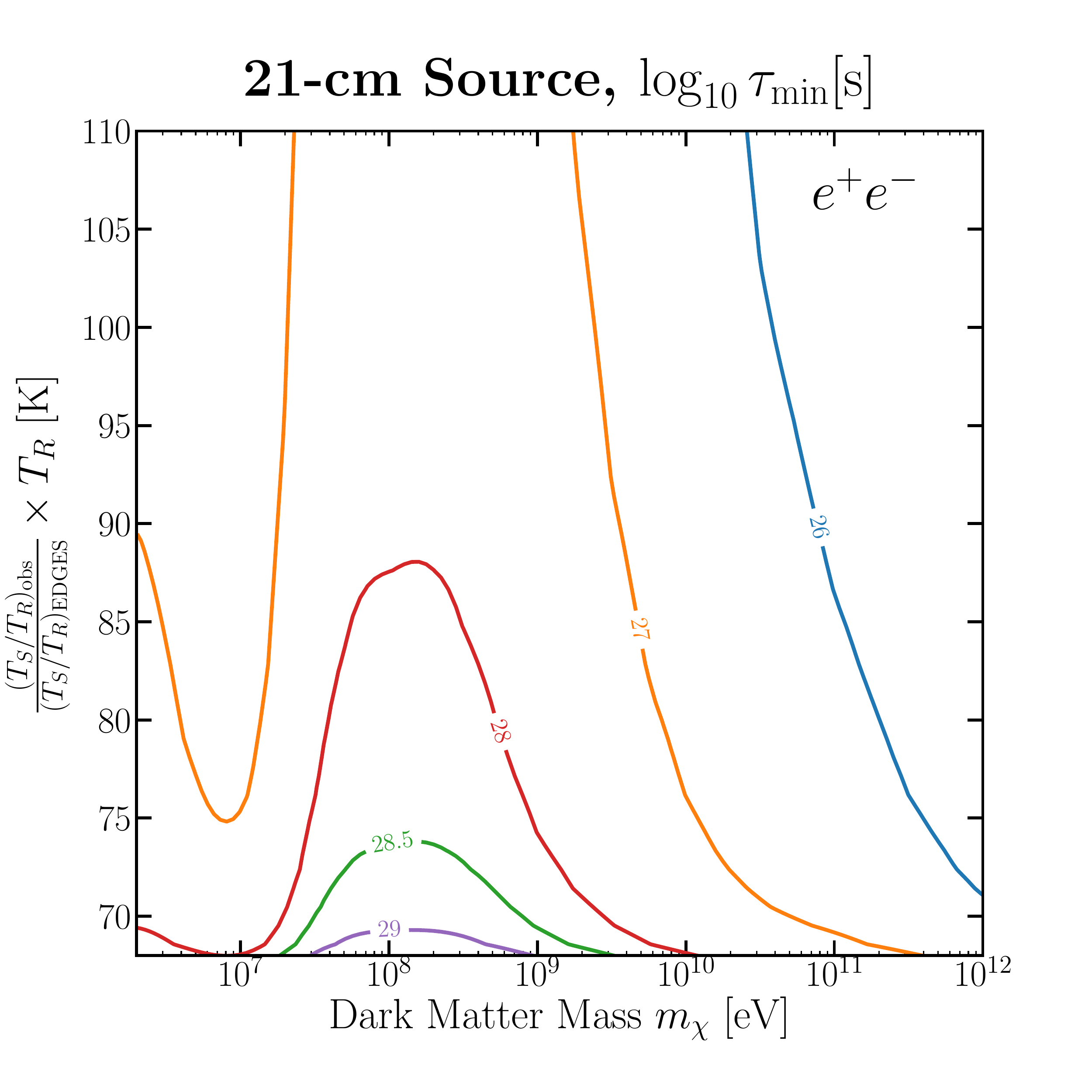}
    }
    \hfil
    \subfloat[]{
        \label{fig:source_phot_decay_zoom}
        \includegraphics[scale=0.34]{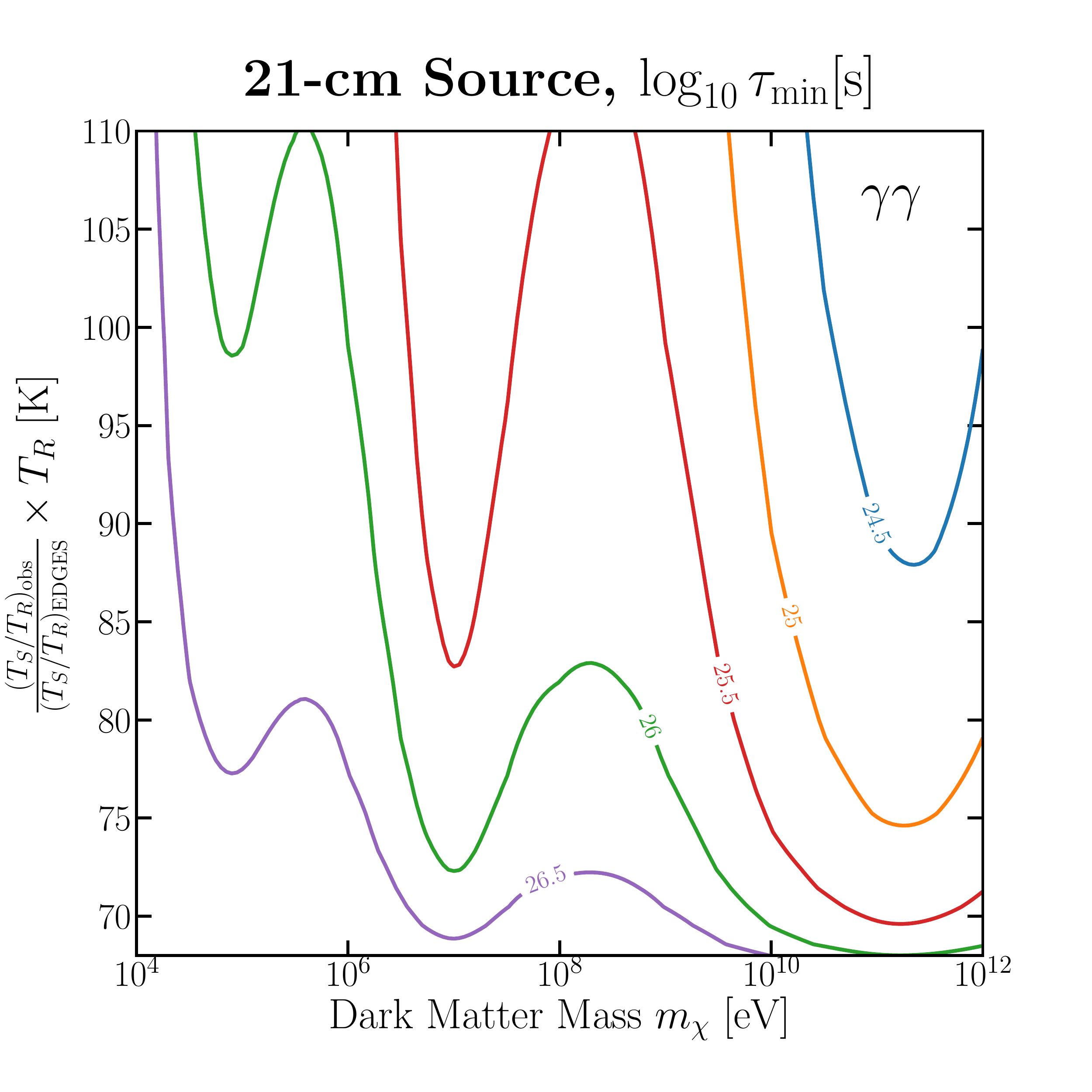}
    }
  \caption{Decay lifetime constraints with an additional 21-cm source with $\chi \to e^+e^-$ (left) and $\chi \to \gamma \gamma$ (right), as a function of $m_\chi$ and $(T_S/T_R)_\text{obs}/(T_S/T_R)_\text{EDGES} \times T_R$. This is a zoomed-in version of Fig.~\ref{fig:source_decay}. Contour lines of constant minimum $\log_{10}\tau$ (in seconds) are shown. 
  }
  \label{fig:source_decay_zoom}
\end{figure*}

\begin{figure*}[t]
    \subfloat[]{
        \label{fig:source_elec_swave_zoom}
        \includegraphics[scale=0.34]{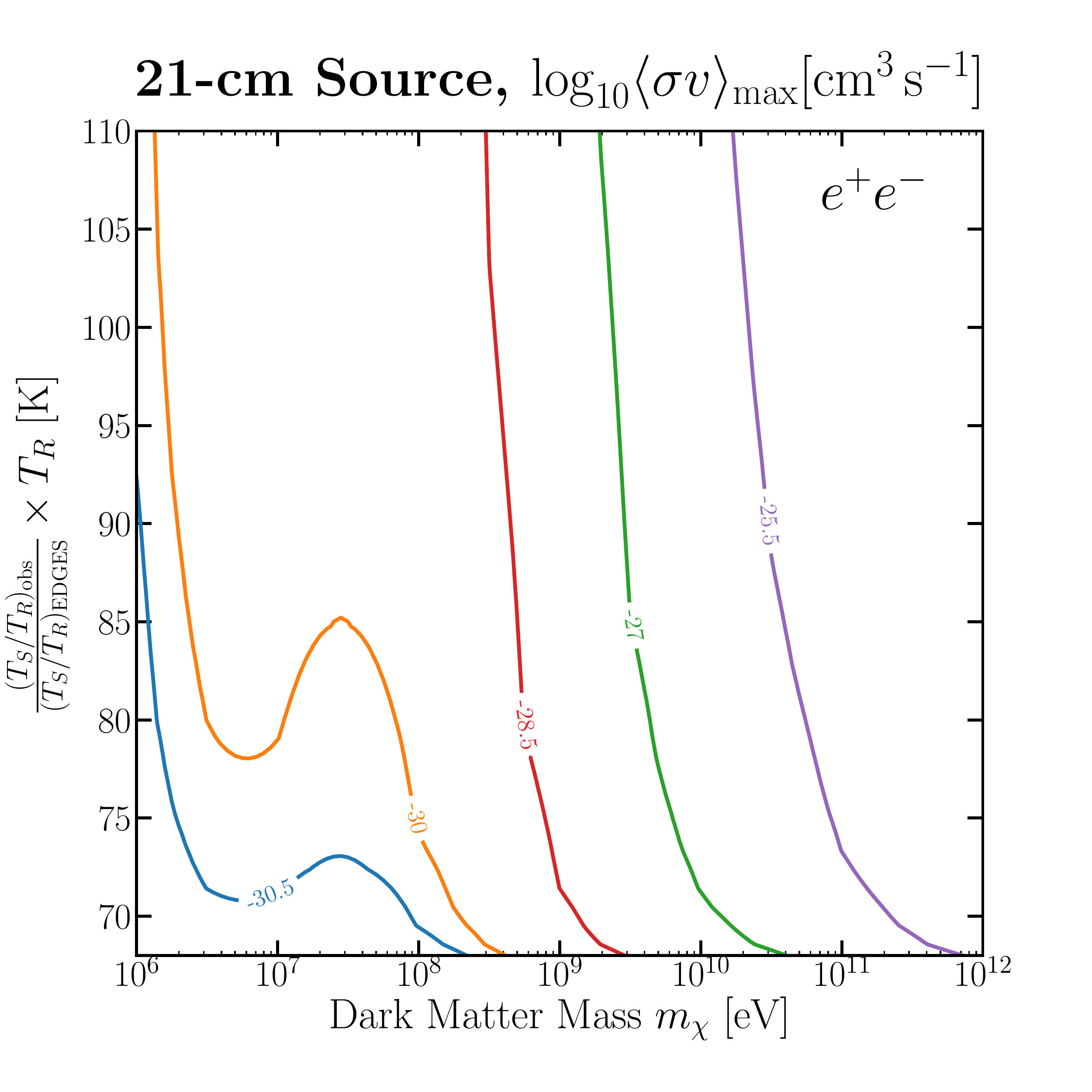}
    }
    \hfil
    \subfloat[]{
        \label{fig:source_phot_swave_zoom}
        \includegraphics[scale=0.34]{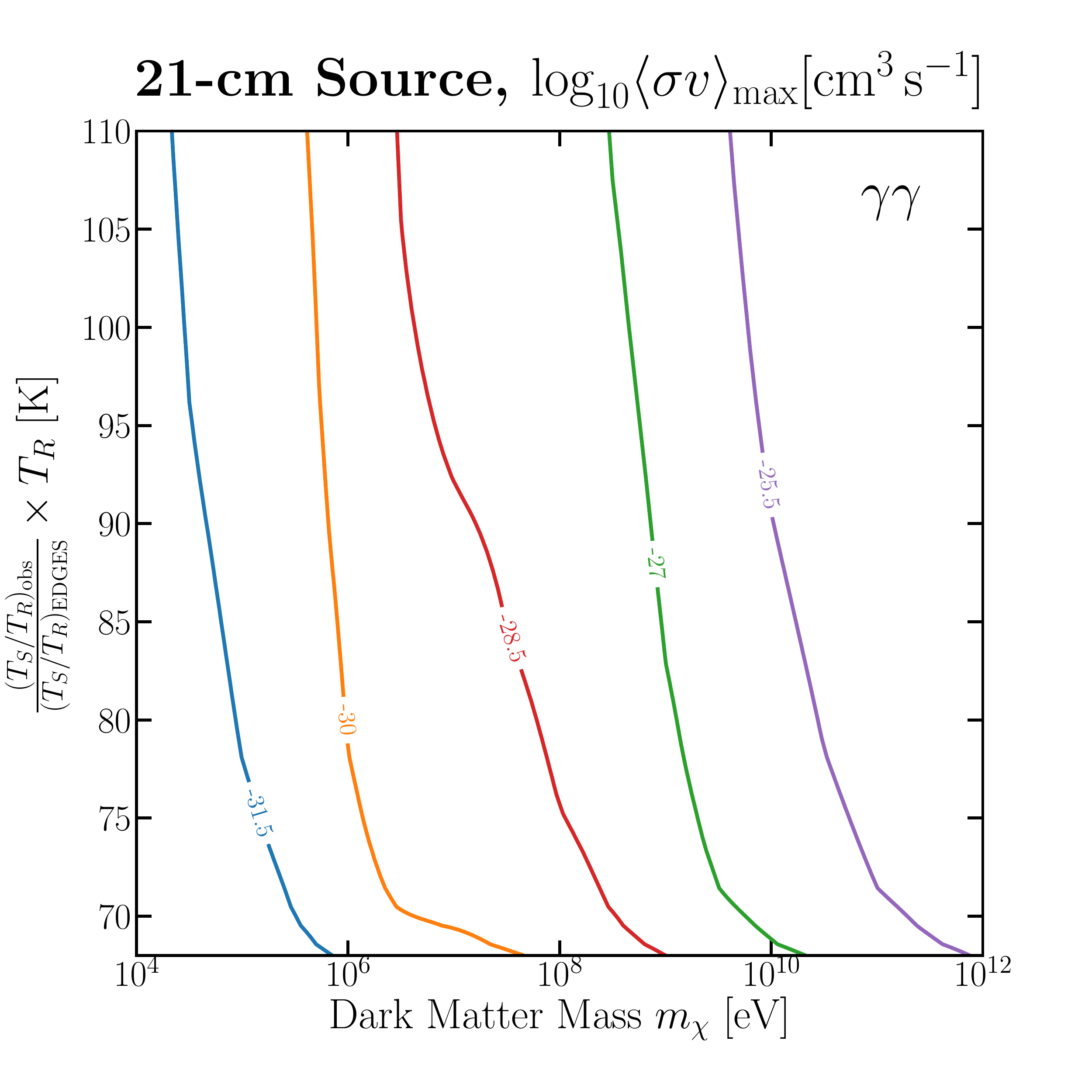}
    }
  \caption{Annihilation cross section constraints with an additional 21-cm source with $\chi \chi \to e^+e^-$ (left) and $\chi \chi \to \gamma \gamma$ (right), as a function of $m_\chi$ and $(T_S/T_R)_\text{obs}/(T_S/T_R)_\text{EDGES} \times T_R$. Contour lines of constant maximum $\log_{10}\langle \sigma v \rangle$ (in $\text{ cm}^3 \text{ s}^{-1}$) are shown. This is a zoomed-in version of Fig.~\ref{fig:source_swave} The green contour corresponds to the canonical relic abundance cross section of $\SI{3e-26}{cm^3 \, s^{-1}}$.
  }
  \label{fig:source_swave_zoom}
\end{figure*}

In Figs.~\ref{fig:source_decay_zoom} and~\ref{fig:source_swave_zoom}, we show zoomed-in versions of Figs.~\ref{fig:source_decay} and~\ref{fig:source_swave}, to highlight the region where the additional radiation source is comparable or smaller to the CMB, in terms of number density at a wavelength of 21 cm. 

\begin{figure*}[t!]
    \subfloat[]{
        \label{fig:recomb_elec_decay_low}
        \includegraphics[scale=0.34]{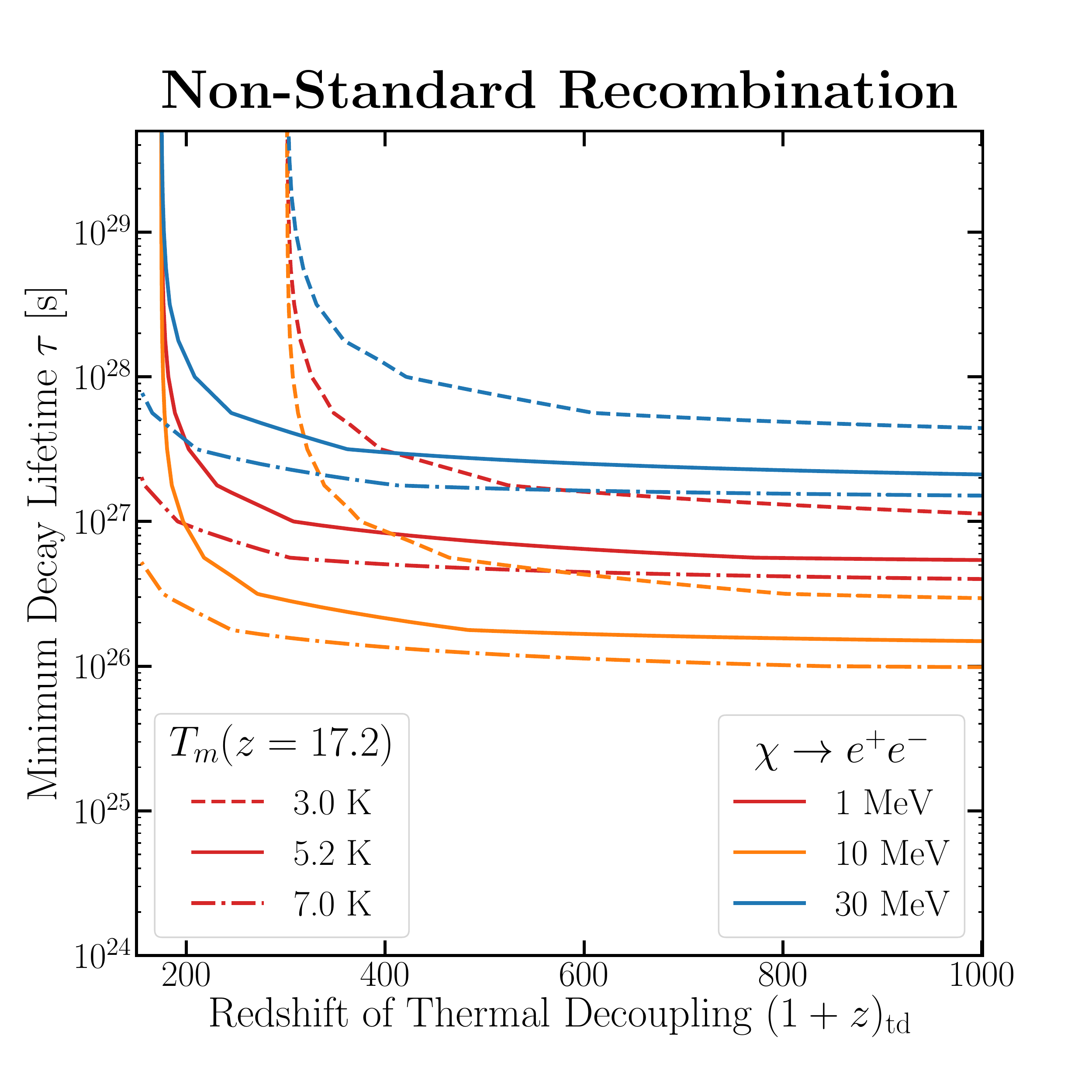}
    }
    \hfil
    \subfloat[]{
        \label{fig:recomb_phot_decay_low}
        \includegraphics[scale=0.34]{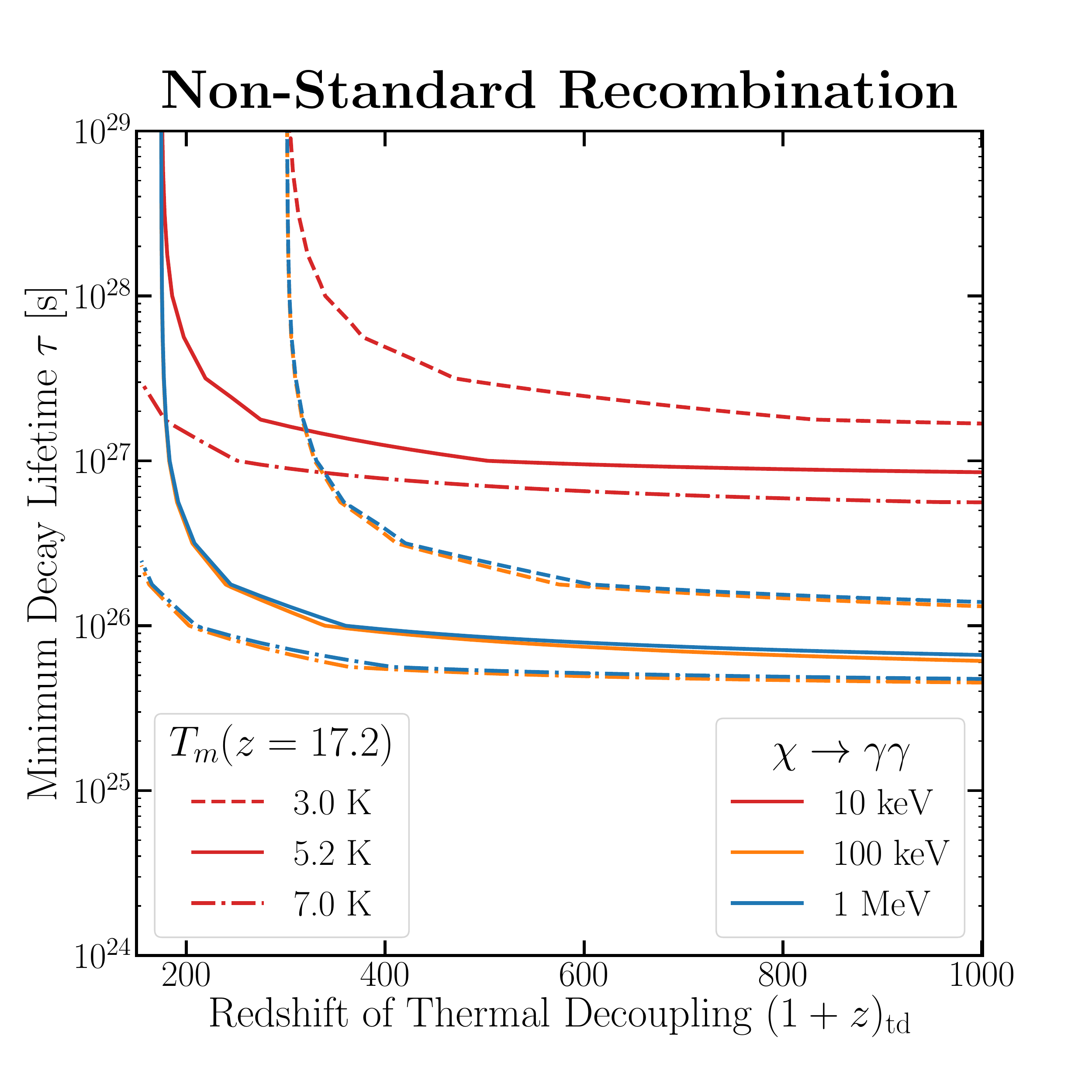}
    }
  \caption{As Fig.~\ref{fig:recomb_decay}, but extended to lower masses.}
  \label{fig:recomb_decay_low}
\end{figure*}

\begin{figure*}[b!]
    \subfloat[]{
        \label{fig:recomb_elec_swave_low}
        \includegraphics[scale=0.34]{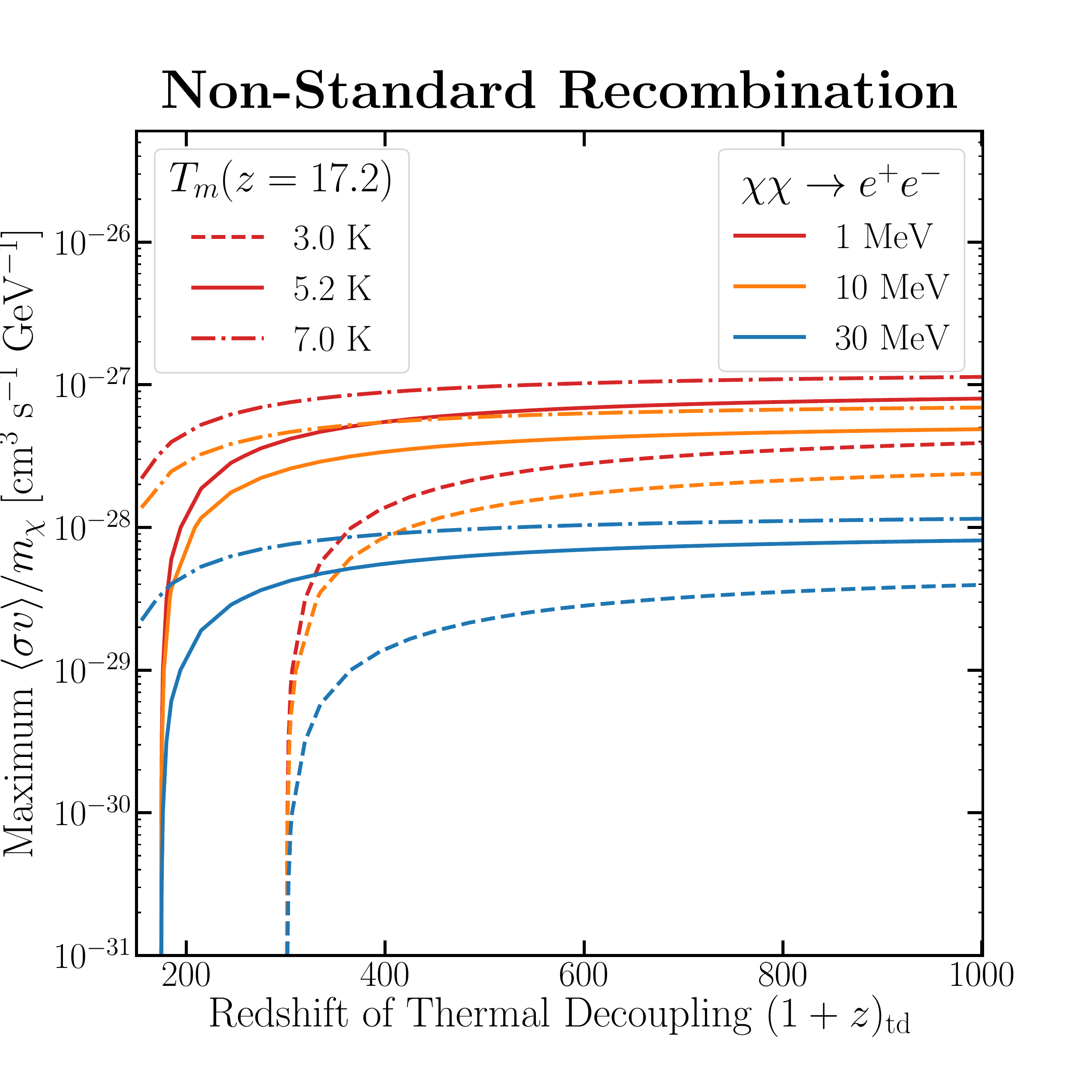}
    }
    \hfil
    \subfloat[]{
        \label{fig:recomb_phot_swave_low}
        \includegraphics[scale=0.34]{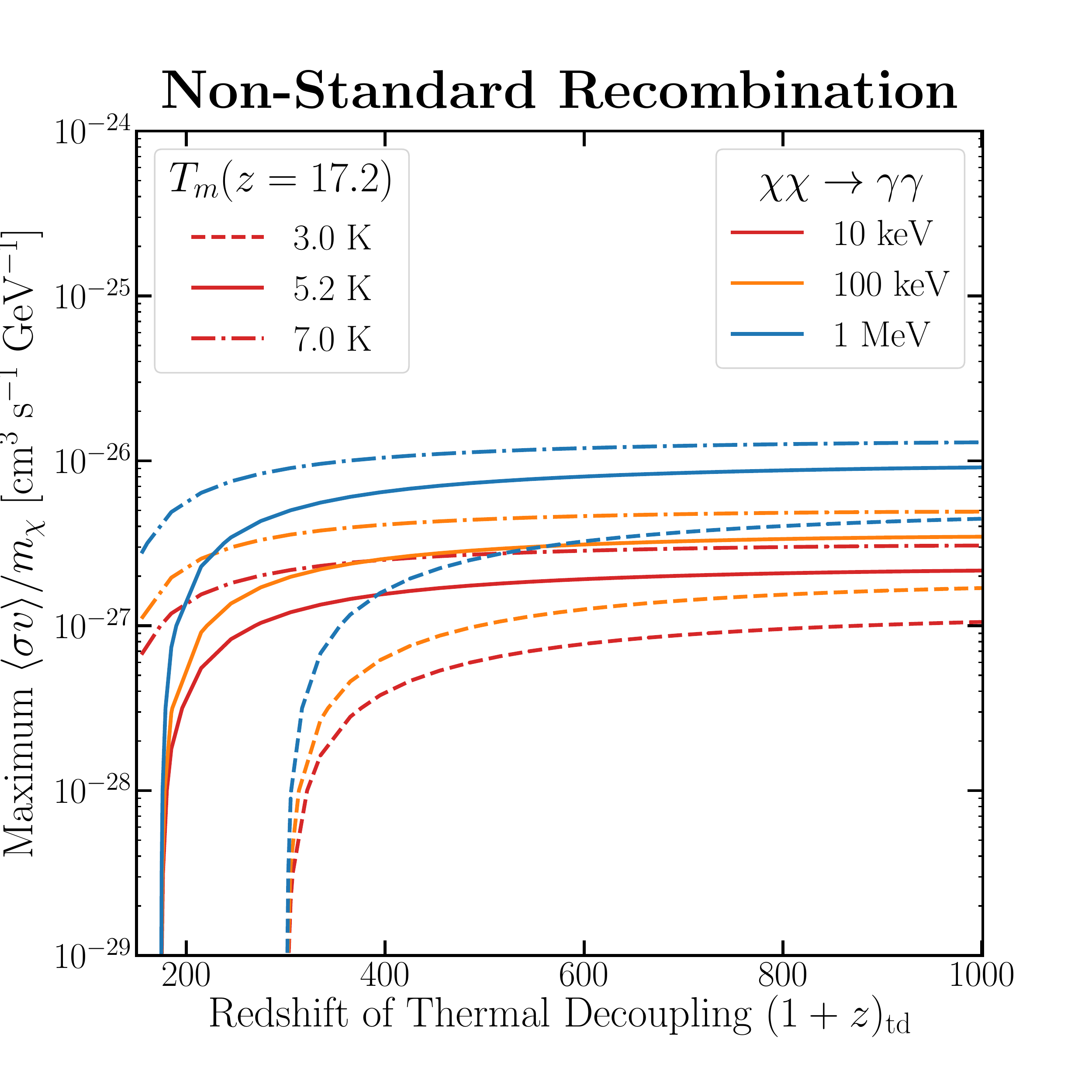}
    }
  \caption{As Fig.~\ref{fig:recomb_swave}, but extended to lower masses.}
  \label{fig:recomb_swave_low}
\end{figure*}

In Figs.~\ref{fig:recomb_decay_low} and~\ref{fig:recomb_swave_low}, we show constraint plots for DM masses below 100 MeV in the presence of non-standard recombination, for $s$-wave annihilation and decay respectively. These analyses are otherwise performed as discussed in the main text.

\begin{figure*}[t!]
    \subfloat[]{
        \label{fig:cooling_elec_decay_f_1}
        \includegraphics[scale=0.34]{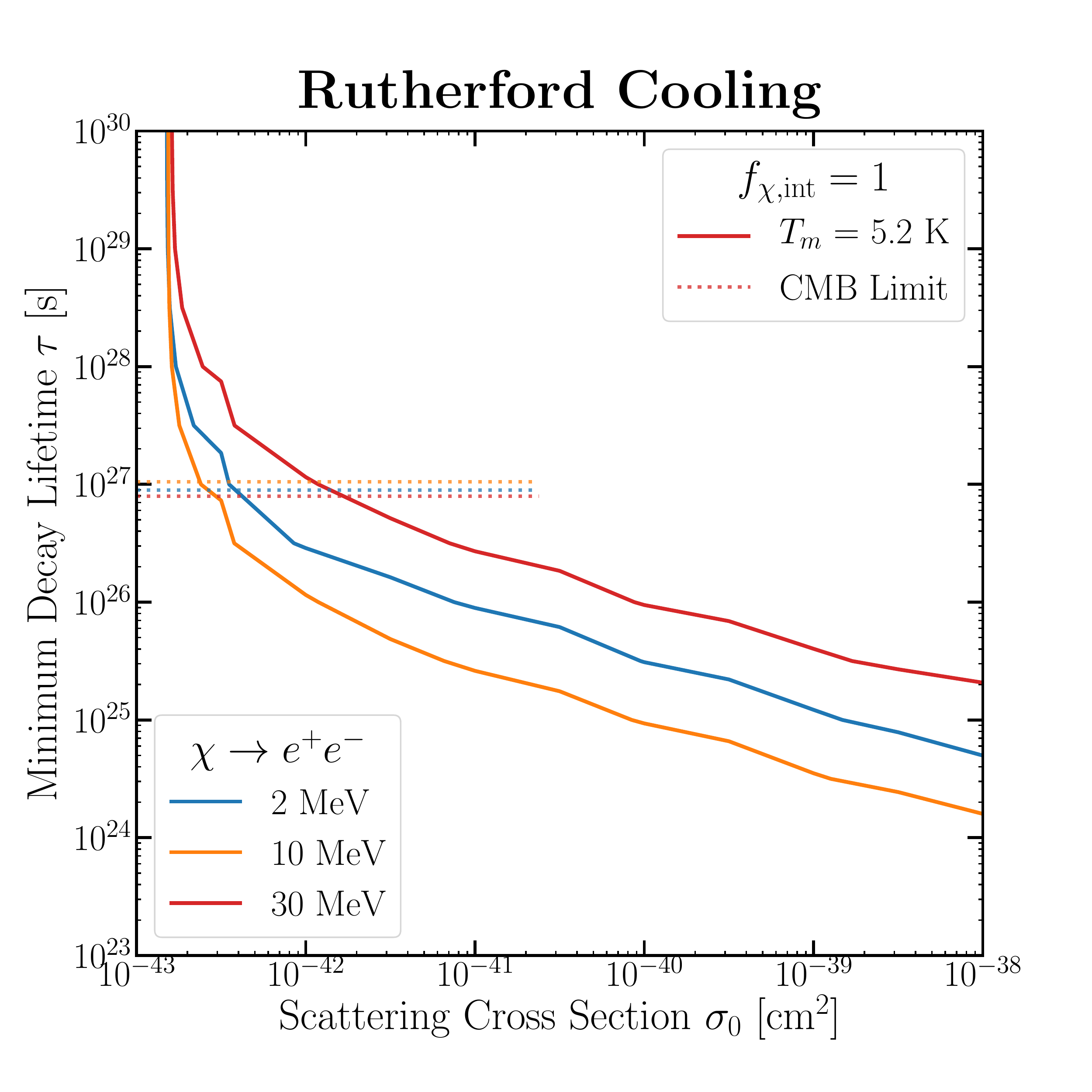}
    }
    \hfil
    \subfloat[]{
        \label{fig:cooling_phot_decay_f_1}
        \includegraphics[scale=0.34]{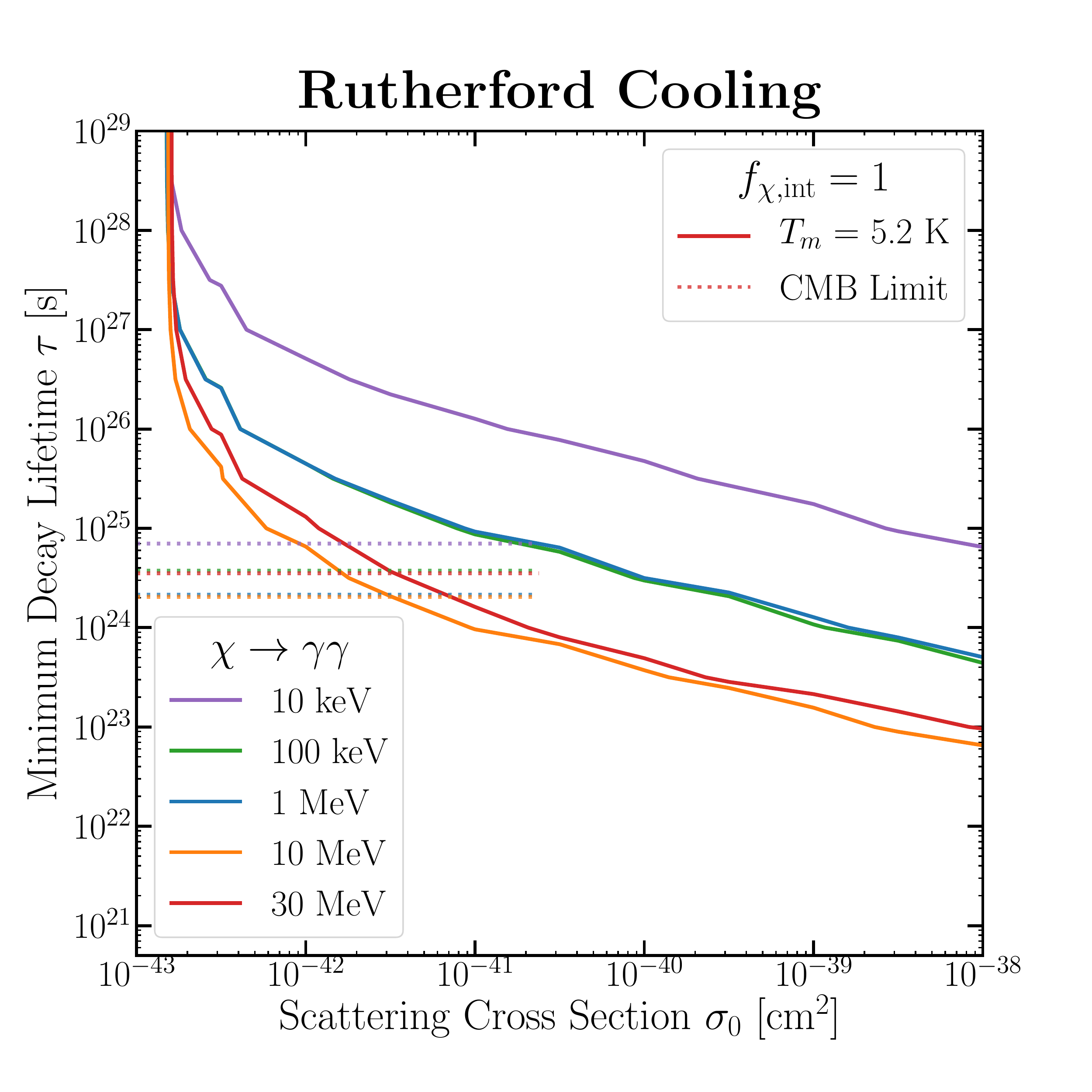}
    }
  \caption{Rutherford cooling constraints on the minimum decay lifetime for $\chi \to e^+e^-$ (left) and $\chi \to \gamma \gamma$ (right) from the matter temperature $T_m(z = 17.2) = $5.2 K (solid), $f_{\chi,\text{int}} = 1$. Limits from the Planck measurement of the CMB power spectrum are also shown up to $\sigma_0 = \sigma_{0,\text{td}}(z = 300)$ (dotted). 
  }
  \label{fig:cooling_decay_f_1}
\end{figure*}

\begin{figure*}[t!]
    \subfloat[]{
        \label{fig:cooling_elec_swave_f_1}
        \includegraphics[scale=0.34]{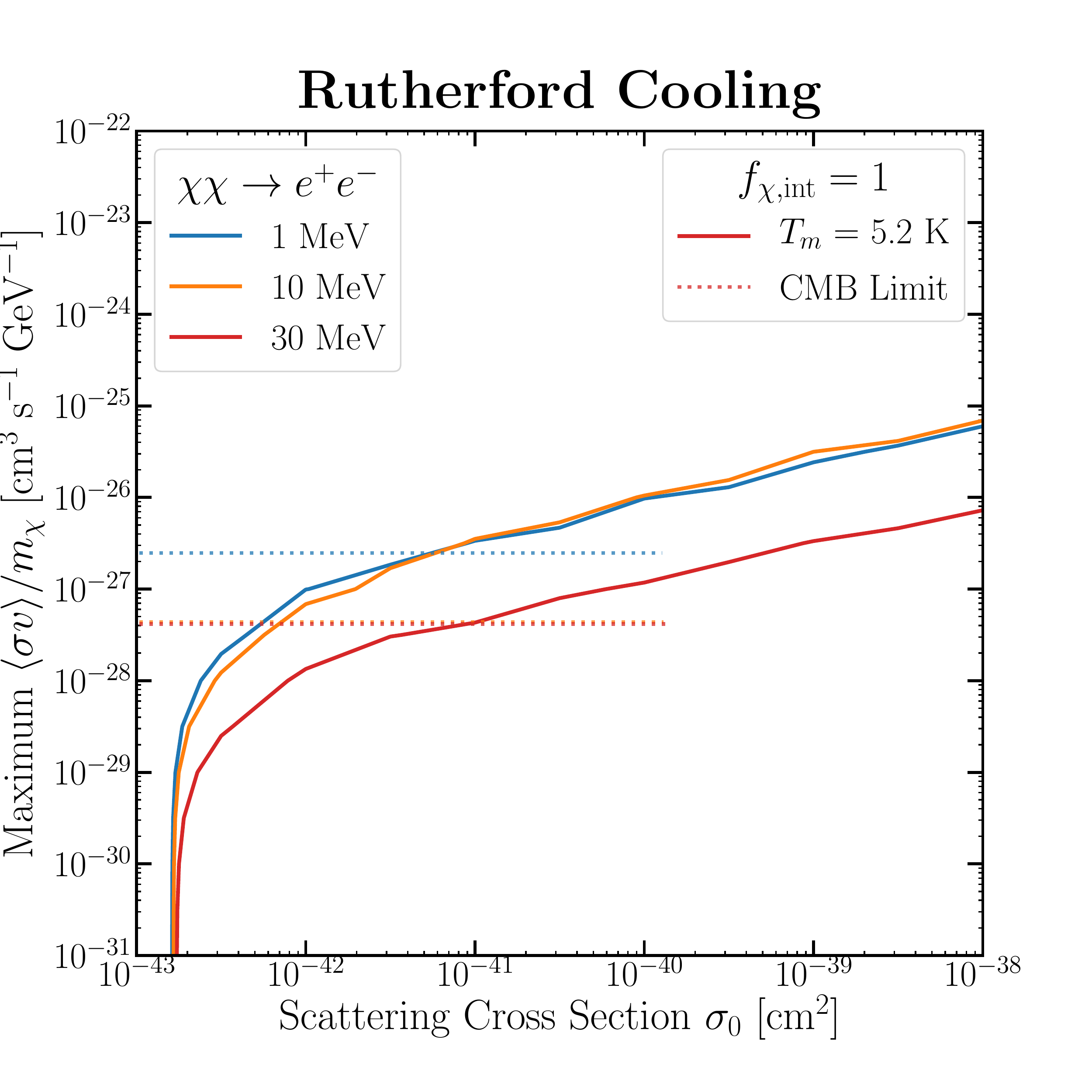}
    }
    \hfil
    \subfloat[]{
        \label{fig:cooling_phot_swave_f_1}
        \includegraphics[scale=0.34]{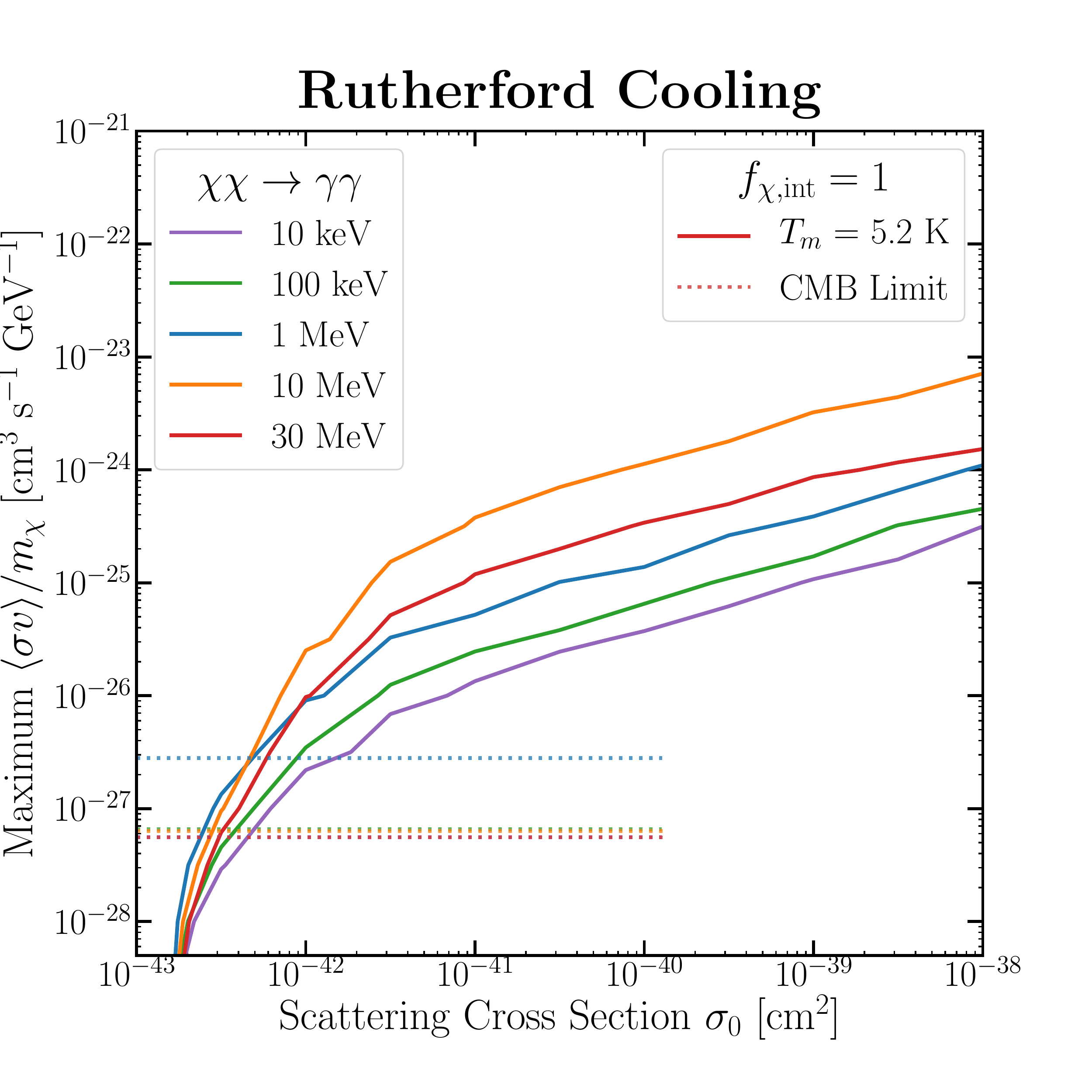}
    }
  \caption{Rutherford cooling $s$-wave annihilation constraints for $\chi \chi \to e^+e^-$ (left) and $\chi \chi \to \gamma \gamma$ (right) from the matter temperature $T_m(z = 17.2) = $ 5.2 K (solid), $f_{\chi,\text{int}} = 1$. Limits from the Planck measurement of the CMB power spectrum are also shown up to $\sigma_0 = \sigma_{0,\text{td}}(z = 600)$ (dotted). 
  }
  \label{fig:cooling_swave_f_1}
\end{figure*}

Finally, Figs.~\ref{fig:cooling_decay_f_1} and~\ref{fig:cooling_swave_f_1} show the limits on the minimum decay lifetime and maximum annihilation cross section for Rutherford cooling with $f_{\chi,\text{int}} = 1$. Values of $\sigma_0$ exceeding $\sim 10^{-42} \text{ cm}^2$ affect the CMB power spectrum significantly and are ruled out by Planck \cite{Slatyer:2018aqg}; for models that are consistent with this limit, the value of $T_m$ at $z \sim 20$ is a more powerful constraint on additional energy injection on models than the high-redshift CMB limits on annihilation and decay. 

\bibliography{21cm_annihilation_decay}

\begin{thebibliography}{53}
\expandafter\ifx\csname natexlab\endcsname\relax\def\natexlab#1{#1}\fi
\expandafter\ifx\csname bibnamefont\endcsname\relax
  \def\bibnamefont#1{#1}\fi
\expandafter\ifx\csname bibfnamefont\endcsname\relax
  \def\bibfnamefont#1{#1}\fi
\expandafter\ifx\csname citenamefont\endcsname\relax
  \def\citenamefont#1{#1}\fi
\expandafter\ifx\csname url\endcsname\relax
  \def\url#1{\texttt{#1}}\fi
\expandafter\ifx\csname urlprefix\endcsname\relax\def\urlprefix{URL }\fi
\providecommand{\bibinfo}[2]{#2}
\providecommand{\eprint}[2][]{\url{#2}}

\bibitem[{\citenamefont{Ali-Haimoud and Hirata}(2011)}]{AliHaimoud:2010dx}
\bibinfo{author}{\bibfnamefont{Y.}~\bibnamefont{Ali-Haimoud}} \bibnamefont{and}
  \bibinfo{author}{\bibfnamefont{C.~M.} \bibnamefont{Hirata}},
  \bibinfo{journal}{Phys. Rev.} \textbf{\bibinfo{volume}{D83}},
  \bibinfo{pages}{043513} (\bibinfo{year}{2011}), \eprint{1011.3758}.

\bibitem[{\citenamefont{Chluba and Thomas}(2011)}]{Chluba:2010ca}
\bibinfo{author}{\bibfnamefont{J.}~\bibnamefont{Chluba}} \bibnamefont{and}
  \bibinfo{author}{\bibfnamefont{R.~M.} \bibnamefont{Thomas}},
  \bibinfo{journal}{Mon. Not. Roy. Astron. Soc.}
  \textbf{\bibinfo{volume}{412}}, \bibinfo{pages}{748} (\bibinfo{year}{2011}),
  \eprint{1010.3631}.

\bibitem[{\citenamefont{Bowman et~al.}(2018)\citenamefont{Bowman, Rogers,
  Monsalve, Mozdzen, and Mahesh}}]{Bowman:2018yin}
\bibinfo{author}{\bibfnamefont{J.~D.} \bibnamefont{Bowman}},
  \bibinfo{author}{\bibfnamefont{A.~E.~E.} \bibnamefont{Rogers}},
  \bibinfo{author}{\bibfnamefont{R.~A.} \bibnamefont{Monsalve}},
  \bibinfo{author}{\bibfnamefont{T.~J.} \bibnamefont{Mozdzen}},
  \bibnamefont{and} \bibinfo{author}{\bibfnamefont{N.}~\bibnamefont{Mahesh}},
  \bibinfo{journal}{Nature} \textbf{\bibinfo{volume}{555}}, \bibinfo{pages}{67}
  (\bibinfo{year}{2018}).

\bibitem[{\citenamefont{Zaldarriaga et~al.}(2004)\citenamefont{Zaldarriaga,
  Furlanetto, and Hernquist}}]{Zaldarriaga:2003du}
\bibinfo{author}{\bibfnamefont{M.}~\bibnamefont{Zaldarriaga}},
  \bibinfo{author}{\bibfnamefont{S.~R.} \bibnamefont{Furlanetto}},
  \bibnamefont{and}
  \bibinfo{author}{\bibfnamefont{L.}~\bibnamefont{Hernquist}},
  \bibinfo{journal}{Astrophys. J.} \textbf{\bibinfo{volume}{608}},
  \bibinfo{pages}{622} (\bibinfo{year}{2004}), \eprint{astro-ph/0311514}.

\bibitem[{\citenamefont{Furlanetto
  et~al.}(2006{\natexlab{a}})\citenamefont{Furlanetto, Oh, and
  Briggs}}]{Furlanetto:2006jb}
\bibinfo{author}{\bibfnamefont{S.}~\bibnamefont{Furlanetto}},
  \bibinfo{author}{\bibfnamefont{S.~P.} \bibnamefont{Oh}}, \bibnamefont{and}
  \bibinfo{author}{\bibfnamefont{F.}~\bibnamefont{Briggs}},
  \bibinfo{journal}{Phys. Rept.} \textbf{\bibinfo{volume}{433}},
  \bibinfo{pages}{181} (\bibinfo{year}{2006}{\natexlab{a}}),
  \eprint{astro-ph/0608032}.

\bibitem[{\citenamefont{{Wouthuysen}}(1952)}]{1952AJ.....57R..31W}
\bibinfo{author}{\bibfnamefont{S.~A.} \bibnamefont{{Wouthuysen}}},
  \bibinfo{journal}{Astron. J.} \textbf{\bibinfo{volume}{57}},
  \bibinfo{pages}{31} (\bibinfo{year}{1952}).

\bibitem[{\citenamefont{{Field}}(1959{\natexlab{a}})}]{1959ApJ...129..536F}
\bibinfo{author}{\bibfnamefont{G.~B.} \bibnamefont{{Field}}},
  \bibinfo{journal}{Astrophys. J.} \textbf{\bibinfo{volume}{129}},
  \bibinfo{pages}{536} (\bibinfo{year}{1959}{\natexlab{a}}).

\bibitem[{\citenamefont{{Field}}(1959{\natexlab{b}})}]{1959ApJ...129..551F}
\bibinfo{author}{\bibfnamefont{G.~B.} \bibnamefont{{Field}}},
  \bibinfo{journal}{Astrophys. J.} \textbf{\bibinfo{volume}{129}},
  \bibinfo{pages}{551} (\bibinfo{year}{1959}{\natexlab{b}}).

\bibitem[{\citenamefont{Barkana}(2018)}]{Barkana:2018lgd}
\bibinfo{author}{\bibfnamefont{R.}~\bibnamefont{Barkana}},
  \bibinfo{journal}{Nature} \textbf{\bibinfo{volume}{555}}, \bibinfo{pages}{71}
  (\bibinfo{year}{2018}).

\bibitem[{\citenamefont{Mu\~{n}oz and Loeb}(2018)}]{Munoz:2018pzp}
\bibinfo{author}{\bibfnamefont{J.~B.} \bibnamefont{Mu\~{n}oz}}
  \bibnamefont{and} \bibinfo{author}{\bibfnamefont{A.}~\bibnamefont{Loeb}}
  (\bibinfo{year}{2018}), \eprint{1802.10094}.

\bibitem[{\citenamefont{Berlin et~al.}(2018)\citenamefont{Berlin, Hooper,
  Krnjaic, and McDermott}}]{Berlin:2018sjs}
\bibinfo{author}{\bibfnamefont{A.}~\bibnamefont{Berlin}},
  \bibinfo{author}{\bibfnamefont{D.}~\bibnamefont{Hooper}},
  \bibinfo{author}{\bibfnamefont{G.}~\bibnamefont{Krnjaic}}, \bibnamefont{and}
  \bibinfo{author}{\bibfnamefont{S.~D.} \bibnamefont{McDermott}}
  (\bibinfo{year}{2018}), \eprint{1803.02804}.

\bibitem[{\citenamefont{Fraser et~al.}(2018)}]{Fraser:2018acy}
\bibinfo{author}{\bibfnamefont{S.}~\bibnamefont{Fraser}} \bibnamefont{et~al.}
  (\bibinfo{year}{2018}), \eprint{1803.03245}.

\bibitem[{\citenamefont{Barkana et~al.}(2018)\citenamefont{Barkana,
  Outmezguine, Redigolo, and Volansky}}]{Barkana:2018qrx}
\bibinfo{author}{\bibfnamefont{R.}~\bibnamefont{Barkana}},
  \bibinfo{author}{\bibfnamefont{N.~J.} \bibnamefont{Outmezguine}},
  \bibinfo{author}{\bibfnamefont{D.}~\bibnamefont{Redigolo}}, \bibnamefont{and}
  \bibinfo{author}{\bibfnamefont{T.}~\bibnamefont{Volansky}}
  (\bibinfo{year}{2018}), \eprint{1803.03091}.

\bibitem[{\citenamefont{Falkowski and Petraki}(2018)}]{Falkowski:2018qdj}
\bibinfo{author}{\bibfnamefont{A.}~\bibnamefont{Falkowski}} \bibnamefont{and}
  \bibinfo{author}{\bibfnamefont{K.}~\bibnamefont{Petraki}}
  (\bibinfo{year}{2018}), \eprint{1803.10096}.

\bibitem[{\citenamefont{Hill and Baxter}(2018)}]{Hill:2018lfx}
\bibinfo{author}{\bibfnamefont{J.~C.} \bibnamefont{Hill}} \bibnamefont{and}
  \bibinfo{author}{\bibfnamefont{E.~J.} \bibnamefont{Baxter}}
  (\bibinfo{year}{2018}), \eprint{1803.07555}.

\bibitem[{\citenamefont{Costa et~al.}(2018)\citenamefont{Costa, Landim, Wang,
  and Abdalla}}]{Costa:2018aoy}
\bibinfo{author}{\bibfnamefont{A.~A.} \bibnamefont{Costa}},
  \bibinfo{author}{\bibfnamefont{R.~C.~G.} \bibnamefont{Landim}},
  \bibinfo{author}{\bibfnamefont{B.}~\bibnamefont{Wang}}, \bibnamefont{and}
  \bibinfo{author}{\bibfnamefont{E.}~\bibnamefont{Abdalla}}
  (\bibinfo{year}{2018}), \eprint{1803.06944}.

\bibitem[{\citenamefont{Pospelov et~al.}(2018)\citenamefont{Pospelov, Pradler,
  Ruderman, and Urbano}}]{Pospelov:2018kdh}
\bibinfo{author}{\bibfnamefont{M.}~\bibnamefont{Pospelov}},
  \bibinfo{author}{\bibfnamefont{J.}~\bibnamefont{Pradler}},
  \bibinfo{author}{\bibfnamefont{J.~T.} \bibnamefont{Ruderman}},
  \bibnamefont{and} \bibinfo{author}{\bibfnamefont{A.}~\bibnamefont{Urbano}}
  (\bibinfo{year}{2018}), \eprint{1803.07048}.

\bibitem[{\citenamefont{Gong and Kitajima}(2018)}]{Gong:2018sos}
\bibinfo{author}{\bibfnamefont{J.-O.} \bibnamefont{Gong}} \bibnamefont{and}
  \bibinfo{author}{\bibfnamefont{N.}~\bibnamefont{Kitajima}}
  (\bibinfo{year}{2018}), \eprint{1803.02745}.

\bibitem[{\citenamefont{Ewall-Wice et~al.}(2018)\citenamefont{Ewall-Wice,
  Chang, Lazio, Dore, Seiffert, and Monsalve}}]{Ewall-Wice:2018bzf}
\bibinfo{author}{\bibfnamefont{A.}~\bibnamefont{Ewall-Wice}},
  \bibinfo{author}{\bibfnamefont{T.~C.} \bibnamefont{Chang}},
  \bibinfo{author}{\bibfnamefont{J.}~\bibnamefont{Lazio}},
  \bibinfo{author}{\bibfnamefont{O.}~\bibnamefont{Dore}},
  \bibinfo{author}{\bibfnamefont{M.}~\bibnamefont{Seiffert}}, \bibnamefont{and}
  \bibinfo{author}{\bibfnamefont{R.~A.} \bibnamefont{Monsalve}}
  (\bibinfo{year}{2018}), \eprint{1803.01815}.

\bibitem[{\citenamefont{Poulin et~al.}(2017)\citenamefont{Poulin, Lesgourgues,
  and Serpico}}]{Poulin:2016anj}
\bibinfo{author}{\bibfnamefont{V.}~\bibnamefont{Poulin}},
  \bibinfo{author}{\bibfnamefont{J.}~\bibnamefont{Lesgourgues}},
  \bibnamefont{and} \bibinfo{author}{\bibfnamefont{P.~D.}
  \bibnamefont{Serpico}}, \bibinfo{journal}{JCAP}
  \textbf{\bibinfo{volume}{1703}}, \bibinfo{pages}{043} (\bibinfo{year}{2017}),
  \eprint{1610.10051}.

\bibitem[{\citenamefont{Furlanetto
  et~al.}(2006{\natexlab{b}})\citenamefont{Furlanetto, Oh, and
  Pierpaoli}}]{Furlanetto:2006wp}
\bibinfo{author}{\bibfnamefont{S.~R.} \bibnamefont{Furlanetto}},
  \bibinfo{author}{\bibfnamefont{S.~P.} \bibnamefont{Oh}}, \bibnamefont{and}
  \bibinfo{author}{\bibfnamefont{E.}~\bibnamefont{Pierpaoli}},
  \bibinfo{journal}{Phys. Rev.} \textbf{\bibinfo{volume}{D74}},
  \bibinfo{pages}{103502} (\bibinfo{year}{2006}{\natexlab{b}}),
  \eprint{astro-ph/0608385}.

\bibitem[{\citenamefont{Lopez-Honorez et~al.}(2016)\citenamefont{Lopez-Honorez,
  Mena, Molin\'{e}, Palomares-Ruiz, and Vincent}}]{Lopez-Honorez:2016sur}
\bibinfo{author}{\bibfnamefont{L.}~\bibnamefont{Lopez-Honorez}},
  \bibinfo{author}{\bibfnamefont{O.}~\bibnamefont{Mena}},
  \bibinfo{author}{\bibfnamefont{A.}~\bibnamefont{Molin\'{e}}},
  \bibinfo{author}{\bibfnamefont{S.}~\bibnamefont{Palomares-Ruiz}},
  \bibnamefont{and} \bibinfo{author}{\bibfnamefont{A.~C.}
  \bibnamefont{Vincent}}, \bibinfo{journal}{JCAP}
  \textbf{\bibinfo{volume}{1608}}, \bibinfo{pages}{004} (\bibinfo{year}{2016}),
  \eprint{1603.06795}.

\bibitem[{\citenamefont{Valdes et~al.}(2007)\citenamefont{Valdes, Ferrara,
  Mapelli, and Ripamonti}}]{Valdes:2007cu}
\bibinfo{author}{\bibfnamefont{M.}~\bibnamefont{Valdes}},
  \bibinfo{author}{\bibfnamefont{A.}~\bibnamefont{Ferrara}},
  \bibinfo{author}{\bibfnamefont{M.}~\bibnamefont{Mapelli}}, \bibnamefont{and}
  \bibinfo{author}{\bibfnamefont{E.}~\bibnamefont{Ripamonti}},
  \bibinfo{journal}{Mon. Not. Roy. Astron. Soc.}
  \textbf{\bibinfo{volume}{377}}, \bibinfo{pages}{245} (\bibinfo{year}{2007}),
  \eprint{astro-ph/0701301}.

\bibitem[{\citenamefont{Evoli et~al.}(2014)\citenamefont{Evoli, Mesinger, and
  Ferrara}}]{Evoli:2014pva}
\bibinfo{author}{\bibfnamefont{C.}~\bibnamefont{Evoli}},
  \bibinfo{author}{\bibfnamefont{A.}~\bibnamefont{Mesinger}}, \bibnamefont{and}
  \bibinfo{author}{\bibfnamefont{A.}~\bibnamefont{Ferrara}},
  \bibinfo{journal}{JCAP} \textbf{\bibinfo{volume}{1411}}, \bibinfo{pages}{024}
  (\bibinfo{year}{2014}), \eprint{1408.1109}.

\bibitem[{\citenamefont{Ade et~al.}(2016)}]{Ade:2015xua}
\bibinfo{author}{\bibfnamefont{P.~A.~R.} \bibnamefont{Ade}}
  \bibnamefont{et~al.} (\bibinfo{collaboration}{Planck}),
  \bibinfo{journal}{Astron. Astrophys.} \textbf{\bibinfo{volume}{594}},
  \bibinfo{pages}{A13} (\bibinfo{year}{2016}), \eprint{1502.01589}.

\bibitem[{\citenamefont{Peebles}(1968)}]{Peebles:1968ja}
\bibinfo{author}{\bibfnamefont{P.~J.~E.} \bibnamefont{Peebles}},
  \bibinfo{journal}{Astrophys. J.} \textbf{\bibinfo{volume}{153}},
  \bibinfo{pages}{1} (\bibinfo{year}{1968}).

\bibitem[{\citenamefont{Zeldovich et~al.}(1969)\citenamefont{Zeldovich, Kurt,
  and Sunyaev}}]{Zeldovich:1969en}
\bibinfo{author}{\bibfnamefont{{\relax Ya}.~B.} \bibnamefont{Zeldovich}},
  \bibinfo{author}{\bibfnamefont{V.~G.} \bibnamefont{Kurt}}, \bibnamefont{and}
  \bibinfo{author}{\bibfnamefont{R.~A.} \bibnamefont{Sunyaev}},
  \bibinfo{journal}{Sov. Phys. JETP} \textbf{\bibinfo{volume}{28}},
  \bibinfo{pages}{146} (\bibinfo{year}{1969}), \bibinfo{note}{[Zh. Eksp. Teor.
  Fiz.55,278(1968)]}.

\bibitem[{\citenamefont{Slatyer}(2016{\natexlab{a}})}]{Slatyer:2015kla}
\bibinfo{author}{\bibfnamefont{T.~R.} \bibnamefont{Slatyer}},
  \bibinfo{journal}{Phys. Rev.} \textbf{\bibinfo{volume}{D93}},
  \bibinfo{pages}{023521} (\bibinfo{year}{2016}{\natexlab{a}}),
  \eprint{1506.03812}.

\bibitem[{\citenamefont{Liu et~al.}(2016)\citenamefont{Liu, Slatyer, and
  Zavala}}]{Liu:2016cnk}
\bibinfo{author}{\bibfnamefont{H.}~\bibnamefont{Liu}},
  \bibinfo{author}{\bibfnamefont{T.~R.} \bibnamefont{Slatyer}},
  \bibnamefont{and} \bibinfo{author}{\bibfnamefont{J.}~\bibnamefont{Zavala}},
  \bibinfo{journal}{Phys. Rev.} \textbf{\bibinfo{volume}{D94}},
  \bibinfo{pages}{063507} (\bibinfo{year}{2016}), \eprint{1604.02457}.

\bibitem[{\citenamefont{Venumadhav et~al.}(2018)\citenamefont{Venumadhav, Dai,
  Kaurov, and Zaldarriaga}}]{Venumadhav:2018uwn}
\bibinfo{author}{\bibfnamefont{T.}~\bibnamefont{Venumadhav}},
  \bibinfo{author}{\bibfnamefont{L.}~\bibnamefont{Dai}},
  \bibinfo{author}{\bibfnamefont{A.}~\bibnamefont{Kaurov}}, \bibnamefont{and}
  \bibinfo{author}{\bibfnamefont{M.}~\bibnamefont{Zaldarriaga}}
  (\bibinfo{year}{2018}), \eprint{1804.02406}.

\bibitem[{\citenamefont{Chluba}(2014)}]{Chluba:2014sma}
\bibinfo{author}{\bibfnamefont{J.}~\bibnamefont{Chluba}}, in
  \emph{\bibinfo{booktitle}{{Proceedings, 49th Rencontres de Moriond on
  Cosmology: La Thuile, Italy, March 15-22, 2014}}} (\bibinfo{year}{2014}), pp.
  \bibinfo{pages}{327--334}, \eprint{1405.6938},
  \urlprefix\url{https://inspirehep.net/record/1298250/files/arXiv:1405.6938.pdf}.

\bibitem[{\citenamefont{D'Amico et~al.}(2018)\citenamefont{D'Amico, Panci, and
  Strumia}}]{DAmico:2018sxd}
\bibinfo{author}{\bibfnamefont{G.}~\bibnamefont{D'Amico}},
  \bibinfo{author}{\bibfnamefont{P.}~\bibnamefont{Panci}}, \bibnamefont{and}
  \bibinfo{author}{\bibfnamefont{A.}~\bibnamefont{Strumia}}
  (\bibinfo{year}{2018}), \eprint{1803.03629}.

\bibitem[{\citenamefont{Slatyer}(2013)}]{Slatyer:2012yq}
\bibinfo{author}{\bibfnamefont{T.~R.} \bibnamefont{Slatyer}},
  \bibinfo{journal}{Phys. Rev.} \textbf{\bibinfo{volume}{D87}},
  \bibinfo{pages}{123513} (\bibinfo{year}{2013}), \eprint{1211.0283}.

\bibitem[{\citenamefont{Slatyer and Wu}(2017)}]{Slatyer:2016qyl}
\bibinfo{author}{\bibfnamefont{T.~R.} \bibnamefont{Slatyer}} \bibnamefont{and}
  \bibinfo{author}{\bibfnamefont{C.-L.} \bibnamefont{Wu}},
  \bibinfo{journal}{Phys. Rev.} \textbf{\bibinfo{volume}{D95}},
  \bibinfo{pages}{023010} (\bibinfo{year}{2017}), \eprint{1610.06933}.

\bibitem[{\citenamefont{De~Bernardis et~al.}(2009)\citenamefont{De~Bernardis,
  Bean, Galli, Melchiorri, Silk, and Verde}}]{DeBernardis:2008tk}
\bibinfo{author}{\bibfnamefont{F.}~\bibnamefont{De~Bernardis}},
  \bibinfo{author}{\bibfnamefont{R.}~\bibnamefont{Bean}},
  \bibinfo{author}{\bibfnamefont{S.}~\bibnamefont{Galli}},
  \bibinfo{author}{\bibfnamefont{A.}~\bibnamefont{Melchiorri}},
  \bibinfo{author}{\bibfnamefont{J.~I.} \bibnamefont{Silk}}, \bibnamefont{and}
  \bibinfo{author}{\bibfnamefont{L.}~\bibnamefont{Verde}},
  \bibinfo{journal}{Phys. Rev.} \textbf{\bibinfo{volume}{D79}},
  \bibinfo{pages}{043503} (\bibinfo{year}{2009}), \eprint{0812.3557}.

\bibitem[{\citenamefont{Slatyer}(2016{\natexlab{b}})}]{Slatyer:2015jla}
\bibinfo{author}{\bibfnamefont{T.~R.} \bibnamefont{Slatyer}},
  \bibinfo{journal}{Phys. Rev.} \textbf{\bibinfo{volume}{D93}},
  \bibinfo{pages}{023527} (\bibinfo{year}{2016}{\natexlab{b}}),
  \eprint{1506.03811}.

\bibitem[{\citenamefont{Cirelli et~al.}(2011)\citenamefont{Cirelli, Corcella,
  Hektor, Hutsi, Kadastik, Panci, Raidal, Sala, and Strumia}}]{Cirelli:2010xx}
\bibinfo{author}{\bibfnamefont{M.}~\bibnamefont{Cirelli}},
  \bibinfo{author}{\bibfnamefont{G.}~\bibnamefont{Corcella}},
  \bibinfo{author}{\bibfnamefont{A.}~\bibnamefont{Hektor}},
  \bibinfo{author}{\bibfnamefont{G.}~\bibnamefont{Hutsi}},
  \bibinfo{author}{\bibfnamefont{M.}~\bibnamefont{Kadastik}},
  \bibinfo{author}{\bibfnamefont{P.}~\bibnamefont{Panci}},
  \bibinfo{author}{\bibfnamefont{M.}~\bibnamefont{Raidal}},
  \bibinfo{author}{\bibfnamefont{F.}~\bibnamefont{Sala}}, \bibnamefont{and}
  \bibinfo{author}{\bibfnamefont{A.}~\bibnamefont{Strumia}},
  \bibinfo{journal}{JCAP} \textbf{\bibinfo{volume}{1103}}, \bibinfo{pages}{051}
  (\bibinfo{year}{2011}), \bibinfo{note}{[Erratum: JCAP1210,E01(2012)]},
  \eprint{1012.4515}.

\bibitem[{\citenamefont{Mu\~{n}oz et~al.}(2015)\citenamefont{Mu\~{n}oz, Kovetz,
  and Ali-Haïmoud}}]{Munoz:2015bca}
\bibinfo{author}{\bibfnamefont{J.~B.} \bibnamefont{Mu\~{n}oz}},
  \bibinfo{author}{\bibfnamefont{E.~D.} \bibnamefont{Kovetz}},
  \bibnamefont{and}
  \bibinfo{author}{\bibfnamefont{Y.}~\bibnamefont{Ali-Haïmoud}},
  \bibinfo{journal}{Phys. Rev.} \textbf{\bibinfo{volume}{D92}},
  \bibinfo{pages}{083528} (\bibinfo{year}{2015}), \eprint{1509.00029}.

\bibitem[{\citenamefont{McDermott et~al.}(2011)\citenamefont{McDermott, Yu, and
  Zurek}}]{McDermott:2010pa}
\bibinfo{author}{\bibfnamefont{S.~D.} \bibnamefont{McDermott}},
  \bibinfo{author}{\bibfnamefont{H.-B.} \bibnamefont{Yu}}, \bibnamefont{and}
  \bibinfo{author}{\bibfnamefont{K.~M.} \bibnamefont{Zurek}},
  \bibinfo{journal}{Phys. Rev.} \textbf{\bibinfo{volume}{D83}},
  \bibinfo{pages}{063509} (\bibinfo{year}{2011}), \eprint{1011.2907}.

\bibitem[{\citenamefont{Dvorkin et~al.}(2014)\citenamefont{Dvorkin, Blum, and
  Kamionkowski}}]{Dvorkin:2013cea}
\bibinfo{author}{\bibfnamefont{C.}~\bibnamefont{Dvorkin}},
  \bibinfo{author}{\bibfnamefont{K.}~\bibnamefont{Blum}}, \bibnamefont{and}
  \bibinfo{author}{\bibfnamefont{M.}~\bibnamefont{Kamionkowski}},
  \bibinfo{journal}{Phys. Rev.} \textbf{\bibinfo{volume}{D89}},
  \bibinfo{pages}{023519} (\bibinfo{year}{2014}), \eprint{1311.2937}.

\bibitem[{\citenamefont{Xu et~al.}(2018)\citenamefont{Xu, Dvorkin, and
  Chael}}]{Xu:2018efh}
\bibinfo{author}{\bibfnamefont{W.~L.} \bibnamefont{Xu}},
  \bibinfo{author}{\bibfnamefont{C.}~\bibnamefont{Dvorkin}}, \bibnamefont{and}
  \bibinfo{author}{\bibfnamefont{A.}~\bibnamefont{Chael}}
  (\bibinfo{year}{2018}), \eprint{1802.06788}.

\bibitem[{\citenamefont{Slatyer and Wu}(2018)}]{Slatyer:2018aqg}
\bibinfo{author}{\bibfnamefont{T.~R.} \bibnamefont{Slatyer}} \bibnamefont{and}
  \bibinfo{author}{\bibfnamefont{C.-L.} \bibnamefont{Wu}}
  (\bibinfo{year}{2018}), \eprint{1803.09734}.

\bibitem[{\citenamefont{Dolgov et~al.}(2013)\citenamefont{Dolgov, Dubovsky,
  Rubtsov, and Tkachev}}]{Dolgov:2013una}
\bibinfo{author}{\bibfnamefont{A.~D.} \bibnamefont{Dolgov}},
  \bibinfo{author}{\bibfnamefont{S.~L.} \bibnamefont{Dubovsky}},
  \bibinfo{author}{\bibfnamefont{G.~I.} \bibnamefont{Rubtsov}},
  \bibnamefont{and} \bibinfo{author}{\bibfnamefont{I.~I.}
  \bibnamefont{Tkachev}}, \bibinfo{journal}{Phys. Rev.}
  \textbf{\bibinfo{volume}{D88}}, \bibinfo{pages}{117701}
  (\bibinfo{year}{2013}), \eprint{1310.2376}.

\bibitem[{\citenamefont{Ali-Haïmoud et~al.}(2014)\citenamefont{Ali-Haïmoud,
  Meerburg, and Yuan}}]{Ali-Haimoud:2013hpa}
\bibinfo{author}{\bibfnamefont{Y.}~\bibnamefont{Ali-Haïmoud}},
  \bibinfo{author}{\bibfnamefont{P.~D.} \bibnamefont{Meerburg}},
  \bibnamefont{and} \bibinfo{author}{\bibfnamefont{S.}~\bibnamefont{Yuan}},
  \bibinfo{journal}{Phys. Rev.} \textbf{\bibinfo{volume}{D89}},
  \bibinfo{pages}{083506} (\bibinfo{year}{2014}), \eprint{1312.4948}.

\bibitem[{\citenamefont{Prinz et~al.}(1998)}]{Prinz:1998ua}
\bibinfo{author}{\bibfnamefont{A.~A.} \bibnamefont{Prinz}}
  \bibnamefont{et~al.}, \bibinfo{journal}{Phys. Rev. Lett.}
  \textbf{\bibinfo{volume}{81}}, \bibinfo{pages}{1175} (\bibinfo{year}{1998}),
  \eprint{hep-ex/9804008}.

\bibitem[{\citenamefont{Chang et~al.}(2018)\citenamefont{Chang, Essig, and
  McDermott}}]{Chang:2018rso}
\bibinfo{author}{\bibfnamefont{J.~H.} \bibnamefont{Chang}},
  \bibinfo{author}{\bibfnamefont{R.}~\bibnamefont{Essig}}, \bibnamefont{and}
  \bibinfo{author}{\bibfnamefont{S.~D.} \bibnamefont{McDermott}}
  (\bibinfo{year}{2018}), \eprint{1803.00993}.

\bibitem[{\citenamefont{Cohen et~al.}(2017)\citenamefont{Cohen, Murase, Rodd,
  Safdi, and Soreq}}]{Cohen:2016uyg}
\bibinfo{author}{\bibfnamefont{T.}~\bibnamefont{Cohen}},
  \bibinfo{author}{\bibfnamefont{K.}~\bibnamefont{Murase}},
  \bibinfo{author}{\bibfnamefont{N.~L.} \bibnamefont{Rodd}},
  \bibinfo{author}{\bibfnamefont{B.~R.} \bibnamefont{Safdi}}, \bibnamefont{and}
  \bibinfo{author}{\bibfnamefont{Y.}~\bibnamefont{Soreq}},
  \bibinfo{journal}{Phys. Rev. Lett.} \textbf{\bibinfo{volume}{119}},
  \bibinfo{pages}{021102} (\bibinfo{year}{2017}), \eprint{1612.05638}.

\bibitem[{\citenamefont{Albert et~al.}(2017)}]{Fermi-LAT:2016uux}
\bibinfo{author}{\bibfnamefont{A.}~\bibnamefont{Albert}} \bibnamefont{et~al.}
  (\bibinfo{collaboration}{DES, Fermi-LAT}), \bibinfo{journal}{Astrophys. J.}
  \textbf{\bibinfo{volume}{834}}, \bibinfo{pages}{110} (\bibinfo{year}{2017}),
  \eprint{1611.03184}.

\bibitem[{\citenamefont{Cheung et~al.}(2018)\citenamefont{Cheung, Kuo, Ng, and
  Tsai}}]{Cheung:2018vww}
\bibinfo{author}{\bibfnamefont{K.}~\bibnamefont{Cheung}},
  \bibinfo{author}{\bibfnamefont{J.-L.} \bibnamefont{Kuo}},
  \bibinfo{author}{\bibfnamefont{K.-W.} \bibnamefont{Ng}}, \bibnamefont{and}
  \bibinfo{author}{\bibfnamefont{Y.-L.~S.} \bibnamefont{Tsai}}
  (\bibinfo{year}{2018}), \eprint{1803.09398}.

\bibitem[{\citenamefont{Clark et~al.}(2018)\citenamefont{Clark, Dutta, Gao, Ma,
  and Strigari}}]{Clark:2018ghm}
\bibinfo{author}{\bibfnamefont{S.}~\bibnamefont{Clark}},
  \bibinfo{author}{\bibfnamefont{B.}~\bibnamefont{Dutta}},
  \bibinfo{author}{\bibfnamefont{Y.}~\bibnamefont{Gao}},
  \bibinfo{author}{\bibfnamefont{Y.-Z.} \bibnamefont{Ma}}, \bibnamefont{and}
  \bibinfo{author}{\bibfnamefont{L.~E.} \bibnamefont{Strigari}}
  (\bibinfo{year}{2018}), \eprint{1803.09390}.

\bibitem[{\citenamefont{Mitridate and Podo}(2018)}]{Mitridate:2018iag}
\bibinfo{author}{\bibfnamefont{A.}~\bibnamefont{Mitridate}} \bibnamefont{and}
  \bibinfo{author}{\bibfnamefont{A.}~\bibnamefont{Podo}}
  (\bibinfo{year}{2018}), \eprint{1803.11169}.

\bibitem[{\citenamefont{Einasto}(1965)}]{Einasto}
\bibinfo{author}{\bibfnamefont{J.}~\bibnamefont{Einasto}},
  \bibinfo{journal}{Trudy Inst. Astrofiz. Alma-Ata}
  \textbf{\bibinfo{volume}{51, 87}} (\bibinfo{year}{1965}).

\bibitem[{\citenamefont{Navarro et~al.}(1996)\citenamefont{Navarro, Frenk, and
  White}}]{Navarro:1995iw}
\bibinfo{author}{\bibfnamefont{J.~F.} \bibnamefont{Navarro}},
  \bibinfo{author}{\bibfnamefont{C.~S.} \bibnamefont{Frenk}}, \bibnamefont{and}
  \bibinfo{author}{\bibfnamefont{S.~D.~M.} \bibnamefont{White}},
  \bibinfo{journal}{Astrophys. J.} \textbf{\bibinfo{volume}{462}},
  \bibinfo{pages}{563} (\bibinfo{year}{1996}), \eprint{astro-ph/9508025}.

\end{thebibliography}

\end{document}